\documentclass[aip,apr,twocolumn,showpacs,superscriptaddress,groupedaddress,nofootinbib,reprint]{revtex4-1}  

\usepackage{graphicx}  
\usepackage{dcolumn}   
\usepackage{bm}        
\usepackage{amssymb}   
\usepackage{epstopdf}
\usepackage{amsmath}
\usepackage{mathtools}
\usepackage{xcolor}
\usepackage[capitalize]{cleveref}
\usepackage{graphicx}
\usepackage{graphics}
\usepackage{enumerate}
\usepackage{epstopdf}
\usepackage[symbol]{footmisc}

\renewcommand{\eqref}[1]{\mbox{Eq.~(\ref{#1})}}

\newcommand{\be}{\begin{equation}}
\newcommand{\ee}{\end{equation}}
\newcommand{\bea}{\begin{eqnarray}}
\newcommand{\eea}{\end{eqnarray}}

\usepackage{lmodern}
\usepackage{circuitikz}
\usepackage{qcircuit}
\usepackage{dsfont}
\usepackage{xfrac}
\usepackage{wasysym}
\usepackage{mathrsfs}
\usepackage[capitalize]{cleveref}

\newcommand{\Id}{\mathds{1}}
\newcommand{\X}[1]{\textsf{X}_{#1}}
\newcommand{\Y}[1]{\textsf{Y}_{#1}}
\newcommand{\Z}[1]{\textsf{Z}_{#1}}
\newcommand{\T}{\textsf{T}}
\renewcommand{\S}{\textsf{S}}
\renewcommand{\H}{\textsf{H}}
\newcommand{\CNOT}{\textsf{CNOT}}
\newcommand{\iSWAP}{\emph{i}\textsf{SWAP}}
\newcommand{\bSWAP}{\emph{b}\textsf{SWAP}}
\newcommand{\CPHASE}{\textsf{CPHASE}}
\newcommand{\CZ}{\textsf{CZ}}
\newcommand{\ZX}[1]{\textsf{CR}_{#1}}
\newcommand{\CR}{\textsf{CR}}
\newcommand{\freq}{\omega_\text{q}}
\newcommand{\rf}{\text{rf}}
\newcommand{\omegad}{\delta\omega}

\newcommand{\etal}{\emph{et al.}}
\renewcommand{\d}{\text{d}}

\begin{document}



\title{A Quantum Engineer's Guide to Superconducting Qubits}

\author{P. Krantz$^{1,2,\dagger}$}
\author{M. Kjaergaard$^{1}$}
\author{F. Yan$^{1}$}
\author{T.P. Orlando$^{1}$}
\author{S. Gustavsson$^{1}$}
\author{W. D. Oliver$^{1,3,\ddagger}$}

\affiliation{$^1$Research Laboratory of Electronics, Massachusetts Institute of Technology, Cambridge, MA 02139, USA}
\affiliation{$^2$Wallenberg Centre for Quantum Technology (WACQT), Chalmers University of Technology, Gothenburg, SE-41296, Sweden}
\affiliation{$^3$MIT Lincoln Laboratory, 244 Wood Street, Lexington, MA 02420, USA}

\footnote{$^\dagger$philipk@mit.edu,$^\ddagger$william.oliver@mit.edu}

\date{\today}

\begin{abstract}

The aim of this review is to provide quantum engineers with an introductory guide to the central concepts and challenges in the rapidly accelerating field of superconducting quantum circuits. Over the past twenty years, the field has matured from a predominantly basic research endeavor to one that increasingly explores the engineering of larger-scale superconducting quantum systems. Here, we review several foundational elements --  qubit design, noise properties, qubit control, and readout techniques -- developed during this period, bridging fundamental concepts in circuit quantum electrodynamics (cQED) and contemporary, state-of-the-art applications in gate-model quantum computation.

\end{abstract}

\maketitle

\tableofcontents
\section{Introduction}
Quantum processors harness the intrinsic properties of quantum mechanical systems -- such as quantum parallelism and quantum interference -- to solve certain problems where classical computers fall short~\cite{Feynman1982,Lloyd1996,DiVincenzo2000,Dowling2003,Ladd2010,Nielsen2011}. Over the past two decades, rapid developments in the science and engineering of quantum systems have advanced the frontier in quantum computation, from the realm of scientific explorations on single isolated quantum systems toward the creation and manipulation of multi-qubit processors~\cite{Monroe2013,Bernien2017}. In particular, the requirements imposed by larger quantum processors have shifted of mindset within the community, from solely scientific discovery to the development of new, foundational engineering abstractions associated with the design, control, and readout of multi-qubit quantum systems. The result is the emergence of a new discipline termed \textit{quantum engineering}, which serves to bridge the basic sciences, mathematics, and computer science with fields generally associated with traditional engineering.

One prominent platform for constructing a multi-qubit quantum processor involves superconducting qubits, in which information is stored in the quantum degrees of freedom of nanofabricated, anharmonic oscillators constructed from superconducting circuit elements. In contrast to other platforms, e.g. electron spins in silicon~\cite{Loss1998,Kane1998,Vrijen2000,deSousa2004,Hollenberg2006,Morello2010} and quantum dots\cite{Imamoglu1999,Petta2005,Englund2005,Hanson2007}, trapped ions\cite{Cirac1995,Leibfried2003,Blatt2008,Haffner2008,Blatt2012}, ultracold atoms\cite{Jaksch2005,Lewenstein2007,Bloch2008,Gross2017}, nitrogen-vacancies in diamonds\cite{Hanson2006,Dutt2007}, and polarized photons\cite{Knill2001,Pittman2001,Franson2002,Pittman2003}, where the quantum information is encoded in natural microscopic quantum systems, superconducting qubits are macroscopic in size and lithographically defined.

One remarkable feature of superconducting qubits is that their energy-level spectra are governed by circuit element parameters and thus are configurable; they can be designed to exhibit ``atom-like'' energy spectra with desired properties. Therefore, superconducting qubits are also often referred to as \textit{artificial atoms}, offering a rich parameter space of possible qubit properties and operation regimes, with predictable performance in terms of transition frequencies, anharmonicity, and complexity.

While there are many other excellent reviews on superconducting qubits, see e.g. Refs. \onlinecite{Devoret2004,You2005b,Schoelkopf2008,Clarke2008,Girvin2009,You2011,Oliver2013,Gambetta2017,Wendin2017,Gu2017}, this work specifically aims to introduce new quantum engineers (academic and industrial alike) to the terminology and state-of-the-art practices used in the rapidly accelerating field of superconducting quantum computing. The reader is assumed to be familiar with basic concepts that span classical physics, quantum mechanics, and electrical engineering. In particular, readers will find it useful to have had previous exposure to classical mechanics, the Schr\"odinger equation, the Bloch sphere representation of qubit states, second quantization, basic concepts of superconductivity, electromagnetism, introductory circuit analysis, classical boolean logic, linear dynamical systems, analog and digital signal processing, and familiarity with microwave components such as transmission lines and mixers. These topics will be introduced as they arise, but having basic prior knowledge will be helpful.

\subsection{Organization of this article}
This review is organized in the following four sections; first, in Sec. \ref{sec:circuits}, we explore the parameter space available when designing superconducting circuits. In particular, we look at the promising capacitively-shunted planar qubit modalities and how these can be engineered with desired properties, such as transition frequency, anharmonicity, and reduced susceptibility to various sources of noise. In this section, we also introduce several ways in which interactions between qubits can be engineered, in order to implement two-qubit entangling operations, needed for a universal gate set.

In Sec.~\ref{sec:Noise}, we discuss systematic and stochastic noise, the concepts of noise strength and qubit noise susceptibility, and the common sources of noise which lead to decoherence in superconducting circuits. We introduce the Bloch-Redfield model of decoherence, characterized by longitudinal and transverse relaxation times $T_1$ and $T_2$, and discuss the implications of $1/f$ noise. We then define the noise power spectral density, which is commonly used to characterize noise processes, and describe how it drives decoherence. Finally, we close the section with a review of coherent control methods used to mitigate certain types of coherence, reversible noise.

In Sec. \ref{sec:QubitControl}, we provide a review of how single- and two-qubit operations are typically implemented in superconducing circuits, by using a combination of local magnetic flux control and microwave drives. In particular, we discuss the family of two-qubit gates arising from a capacitive coupling between qubits, and introduce several recent advances that have been demonstrated to achieve high-fidelity gates, as well as applications in quantum information processing that use these gates. The continued development of high-fidelity two-qubit gates in superconducting qubits is a highly active research area. For this reason, we include sufficient technical details that a reader may use this review as a starting point to critically assess the pros and cons of the various gates, as well as develop an appreciation for the types of gate-engineering already implemented in state-of-the-art superconducting quantum processors.

Finally, in Sec. \ref{sec:readout}, we discuss the physics and engineering associated with the dispersive readout technique, typically used to measure the individual qubit states in modern quantum processors. After a discussion of the theory behind dispersive coupling, we give an introduction to design of Purcell filters and the development of quantum-limited parametric amplifiers.

\section{\label{sec:circuits}Engineering quantum circuits}

In this section, we will demonstrate how quantum systems based on superconducting circuits can be engineered to achieve certain desired properties. Using the most common qubit modalities, we discuss how properties such as the qubit transition frequency, anharmonicity, and noise susceptibility can be tailored by the choice of circuit topology and element parameter values. We also discuss how to engineer the interactions between different quantum systems, in particular the cases of qubit-qubit and qubit-resonator couplings.

\subsection{\label{sec:HOtoTransmon}From quantum harmonic oscillator to the transmon qubit}

A quantum mechanical system is governed by the time-dependent Schr\"{o}dinger equation,
\begin{equation}
\hat{H} \vert \psi(t) \rangle = i \hbar \frac{\partial}{\partial t} \vert \psi(t) \rangle,
\end{equation}
\noindent where $\vert \psi(t) \rangle$ is the state of the quantum system at time $t$, $\hbar$ is the reduced Planck's constant $h/2\pi$, and $\hat{H}$ is the \textit{Hamiltonian} that describes the total energy of the system. The ``hat" is used to indicate that $\hat{H}$ is a quantum operator. As the Schr\"{o}dinger equation is a first-order linear differential equation, the temporal dynamics of the quantum system may be viewed as a straightforward example of a linear dynamical system with formal solution,
\begin{equation}
\vert \psi(t) \rangle = e^{-i \hat{H} t / \hbar} \vert \psi(0) \rangle.
\end{equation}
\noindent The time-independent Hamiltonian $\hat{H}$ governs the time evolution of the system through the operator $e^{-i \hat{H} t / \hbar}$. Thus, just as with classical systems, determining the Hamiltonian of a system -- whether the classical Hamiltonian $H$ or its quantum counterpart $\hat{H}$ -- is the first step to deriving its dynamical behavior. In Sec. \ref{sec:QubitControl}, we consider the case when the Hamiltonian is time-dependent in the context of qubit control.

\begin{figure}[t!]
\begin{center}
\includegraphics[width=8.6cm]{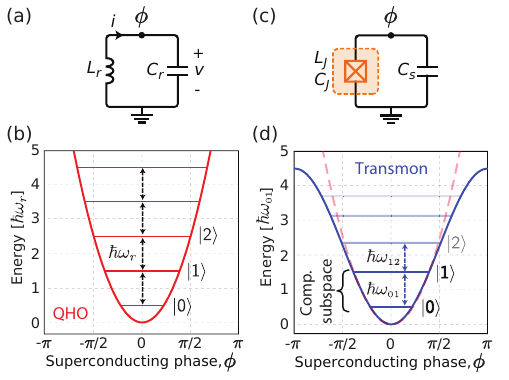}
\caption{\textbf{(a)} Circuit for a parallel LC-oscillator (quantum harmonic oscillator, QHO), with inductance $L$ in parallel with capacitance, $C$. The superconducting phase on the island is denoted $\phi$, referencing ground as zero. \textbf{(b)} Energy potential for the QHO, where energy levels are equidistantly spaced $\hbar \omega_{r}$ apart. \textbf{(c)} Josephson qubit circuit, where the nonlinear inductance $L_J$ (represented with the Josephson-subcircuit in the dashed orange box) is shunted by a capacitance, $C_s$. \textbf{(d)} The Josephson inductance reshapes the quadratic energy potential (dashed red) into sinusoidal (solid blue), which yields non-equidistant energy levels. This allows us to isolate the two lowest energy levels $|0\rangle$ and $|1\rangle$, forming a computational subspace with an energy separation $\hbar\omega_{01}$, which is different than $\hbar\omega_{12}$.}
\label{Fig:EnergyPotentials}
\end{center}
\end{figure}

To understand the dynamics of a superconducting qubit circuit, it is natural to start with the classical description of a linear LC resonant circuit [Fig. \ref{Fig:EnergyPotentials}(a)]. In this system, energy oscillates between electrical energy in the capacitor $C$ and magnetic energy in the inductor $L$. In the following, we will arbitrarily associate the electrical energy with the ``kinetic energy'' and the magnetic energy with the ``potential energy'' of the oscillator. The instantaneous, time-dependent energy in each element is derived from its current and voltage,
\begin{equation}
E(t) = \int_{-\infty}^{t} V(t')I(t')dt',
\label{Eq:ElementEnergy}
\end{equation}
\noindent where $V(t')$ and $I(t')$ denote the voltage and current of the capacitor or inductor.

To derive the classical Hamiltonian, we follow the standard approach used in classical mechanics: the Lagrange-Hamilton formulation. Here, we represent the circuit elements in terms of one of its generalized circuit coordinates, charge or flux. In the following, we pick flux, defined as the time integral of the voltage

\begin{equation}
\Phi(t) = \int_{-\infty}^{t} V(t')dt'.
\label{Eq:NodeFlux}
\end{equation}

\noindent In this example, the voltage at the node is also the branch voltage across the element. In this section, we will simply refer to these as node voltages and fluxes for convenience. For a more detailed discussion of nodes and branches in this context, we refer the reader to Ref.~\onlinecite{Devoret1997}.

Note that in the following, we could have exchanged our associations with kinetic energy (momentum coordinate) and potential energy (position coordinate), and instead start with the charge variable $Q(t)$, which is the time integral of the current $I(t)$.

By combining Eqs. (\ref{Eq:ElementEnergy}) and (\ref{Eq:NodeFlux}), using the relations $V=L \; dI/dt$ and $I=C \; dV/dt$, and applying the integration by parts formula, we can write down energy terms for the capacitor and inductor in terms of the node flux,

\begin{equation}
\mathcal{T}_C = \frac{1}{2}C\dot{\Phi}^2,
\label{Eq:KineticEnergy}
\end{equation}

\begin{equation}
\mathcal{U}_L = \frac{1}{2L}\Phi^2.
\label{Eq:PotentialEnergy}
\end{equation}

The Lagrangian is defined as the difference between the kinetic and potential energy terms and can thus be expressed in terms of Eqs. (\ref{Eq:KineticEnergy}) and (\ref{Eq:PotentialEnergy})

\begin{equation}
\mathcal{L} = \mathcal{T}_{C} - \mathcal{U}_L = \frac{1}{2}C\dot{\Phi}^2 - \frac{1}{2L}\Phi^2.
\label{Eq:Lagrangian}
\end{equation}

From the Lagrangian in Eq. (\ref{Eq:Lagrangian}), we can further derive the Hamiltonian using the Legendre transformation, for which we need to calculate the momentum conjugate to the flux, which in this case is the charge on the capacitor

\begin{equation}
Q = \frac{\partial \mathcal{L}}{\partial \dot{\Phi}} = C\dot{\Phi}.
\end{equation}

The Hamiltonian of the system is now defined as

\begin{equation}
H = Q\dot{\Phi} - \mathcal{L} = \frac{Q^2}{2C} + \frac{\Phi^2}{2L} \equiv
    \frac{1}{2} C V^2 + \frac{1}{2} L I^2,
\label{Eq:HamiltonianLC}
\end{equation}

\noindent as one would expect for an electrical LC circuit. Note that this Hamiltonian is analogous to that of a mechanical harmonic oscillator, with mass $m = C$, and resonant frequency $\omega = 1/\sqrt{LC}$, which expressed in position, $x$, and momentum, $p$, coordinates takes the form $H = p^2/2m + m\omega^2 x^2/2$.

The Hamiltonian described above is classical. In order to proceed to a quantum-mechanical description of the system, we need to promote the charge and flux coordinates to quantum operators. And, whereas the classical coordinates satisfy the Poisson bracket:
\begin{align}
\{f,g\}
    &= \frac{\delta f}{\delta \Phi} \frac{\delta g}{\delta Q} - \frac{\delta g}{\delta \Phi} \frac{\delta f}{\delta Q} \\
 \rightarrow \{\Phi,Q\}  &= \frac{\delta \Phi}{\delta \Phi} \frac{\delta Q}{\delta Q} - \frac{\delta Q}{\delta \Phi} \frac{\delta \Phi}{\delta Q} = 1-0 = 1,
\end{align}
the quantum operators similarly satisfy a \textit{commutation relation}:
\begin{equation}
[\hat{\Phi},\hat{Q}] = \hat{\Phi}\hat{Q} - \hat{Q}\hat{\Phi} = i\hbar,
\label{Eq:CommutationRelationPhiq}
\end{equation}
\noindent where the operators are indicated by hats. From this point forward, however, the hats on operators will be omitted for simplicity.

In a simple LC resonant circuit [Fig. \ref{Fig:EnergyPotentials}(a)], both the inductor $L$ and the capacitor $C$ are linear circuit elements. Defining the reduced flux $\phi \equiv 2 \pi \Phi / \Phi_0$ and the reduced charge $n=Q/2e$, we can write down the following quantum-mechanical Hamiltonian for the circuit,

\begin{equation}
H = 4E_C n^2 + \frac{1}{2}E_L \phi^2,
\label{Eq:Hqho}
\end{equation}

\noindent where $E_C = e^2/(2C)$ is the charging energy required to add {\em each} electron of the Cooper-pair to the island and $E_L = (\Phi_0/2\pi)^2/L$ is the inductive energy, where $\Phi_0 = h/(2e)$ is the superconducting magnetic flux quantum. Moreover, the quantum operator $n$ is the excess number of Cooper-pairs on the island, and $\phi$ -- the reduced flux -- is denoted the ``gauge-invariant phase'' across the inductor. These two operators form a canonical conjugate pair, obeying the commutation relation $[\phi,n] = i$. We note that the factor 4 in front of the charging energy $E_C$ is solely a historical artifact, namely, that this energy scale was first defined for single-electron systems and then adopted to two-electron Cooper-pair systems.

The Hamiltonian in Eq.$\,$(\ref{Eq:Hqho}) is identical to the one describing a particle in a one-dimensional quadratic potential, a quantum harmonic oscillator (QHO). We can treat $\phi$ as the generalized position coordinate, so that the first term is the kinetic energy and the second term is the potential energy. We emphasize that the functional form of the potential energy influences the eigensolutions. For example, the fact that this term is quadratic ($U_L \propto \phi^2$) in Eq. (\ref{Eq:Hqho}) gives rise to the shape of the potential in Fig. \ref{Fig:EnergyPotentials}(b). The solution to this eigenvalue problem gives an infinite series of eigenstates $|k\rangle$, ($k = 0,1,2,\ldots$), whose corresponding eigenenergies $E_k$ are all equidistantly spaced, i.e. $E_{k+1} - E_k = \hbar \omega_r$, where $\omega_r = \sqrt{8 E_L E_C}/\hbar = 1/\sqrt{L C}$ denotes the resonant frequency of the system, see Fig. \ref{Fig:EnergyPotentials}(b). We may represent these results in a more compact form (second quantization) for the quantum harmonic oscillator (QHO) Hamiltonian

\begin{equation}
H = \hbar\omega_{r}\left( a^{\dagger}a + \frac{1}{2}\right),
\label{Eq:HqhoSQ}
\end{equation}

\noindent where $a^{\dagger}(a)$ is the creation (annihilation) operator of a single excitation of the resonator. The Hamiltonian in Eq. (\ref{Eq:HqhoSQ}) is written as an energy. It is, however, often preferred to divide by $\hbar$ so that the expression has units of radian frequency, since we will later resonantly drive transitions at a particular frequency or reference the rate at which two systems interact with one another. Therefore, from here on, $\hbar$ will be omitted.

The original charge number and phase operators can be expressed as $n = n_{\text{zpf}}\times i(a - a^{\dagger})$ and $\phi = \phi_{\text{zpf}} \times (a + a^{\dagger})$, where $n_{\text{zpf}} = [E_L/(32E_C)]^{1/4}$ and $\phi_{\text{zpf}} = (2E_C/E_L)^{1/4}$ are the \textit{zero-point fluctuations} of the charge and phase variables, respectively. Quantum mechanically, the quantum states are represented as wavefunctions that are generally distributed over a range of values of $n$ and $\phi$ and, consequently, the wavefunctions have non-zero standard deviations. Such wavefunction distributions are referred to as ``quantum fluctuations,'' and they exist, even in the ground state, where they are called ``zero-point fluctuations''.

The linear characteristics of the QHO has a natural limitation in its applications for processing quantum information. Before the system can be used as a qubit, we need to be able to define a computational subspace consisting of only two energy states (usually the two-lowest energy eigenstates) in between which transitions can be driven without also exciting other levels in the system. Since many gate operations, such as single-qubit gates (Sec. \ref{sec:QubitControl}), depend on frequency selectivity, the equidistant level-spacing of the QHO, illustrated in Fig. \ref{Fig:EnergyPotentials}(b), poses a practical limitation\footnote{Even though linear resonant systems cannot be addressed properly, their long coherence times have proven them useful as quantum access memories for storing quantum information, where a nonlinear ancilla system is used as a quantum controller for feeding and extracting excitations to/from the resonant cavity modes\cite{Naik2017}.}.

To mitigate the problem of unwanted dynamics involving non-computational states, we need to add anharmonicity (or nonlinearity) into our system. In short, we require the transition frequencies $\omega_{\text{q}}^{0\rightarrow 1}$ and $\omega_{\text{q}}^{1\rightarrow 2}$ be sufficiently different to be individually adressable. In general, the larger the anharmonicity the better. In practise, the amount of anharmonicity sets a limit on how short the pulses used to drive the qubit can be. This is discussed in detail in Sec. \ref{sec:DRAG}.

To introduce the nonlinearity required to modify the harmonic potential, we use the Josephson junction -- a nonlinear, dissipationless circuit element that forms the backbone in superconducting circuits\cite{Josephson1962,Josephson1964}. By replacing the linear inductor of the QHO with a Josephson junction, playing the role of a nonlinear inductor, we can modify the functional form of the potential energy. The potential energy of the Josephson junction can be derived from Eq. (\ref{Eq:ElementEnergy}) and the two Josephson relations

\begin{equation}
I = I_c \sin (\phi), \hspace{0.5cm}V = \frac{\hbar}{2e} \frac{d\phi}{dt},
\label{Eq:JosephsonRelations}
\end{equation}

\noindent resulting in a modified Hamiltonian

\begin{equation}
H = 4E_C n^2 - E_J \cos(\phi),
\label{Eq:HTransmon}
\end{equation}

\noindent where $E_C = e^2/(2C_{\Sigma})$, $C_{\Sigma} = C_s + C_J$ is the total capacitance, including both shunt capacitance $C_s$ and the self-capacitance of the junction $C_J$, and $E_J = I_c \Phi_0/2\pi$ is the Josephson energy, with $I_c$ being the critical current of the junction\footnote{The critical current is the maximum supercurrent that the junction can support before it switches to the resistive state with non-zero voltage.}. After introducing the Josephson junction in the circuit, the potential energy no longer takes a manifestly parabolic form (from which the harmonic spectrum originates), but rather features a cosinusoidal form, see the second term in Eq. (\ref{Eq:HTransmon}), which makes the energy spectrum non-degenerate. Therefore, the Josephson junction is the key ingredient that makes the oscillator anharmonic and thus allows us to identify a uniquely addressable quantum two-level system, see Fig. \ref{Fig:EnergyPotentials}(d).

Once the nonlinearity has been added, the system dynamics is governed by the dominant energy in Eq. (\ref{Eq:HTransmon}), reflected in the $E_J/E_C$ ratio. Over time, the superconducting qubit community has converged towards circuit designs with $E_J \gg E_C$. In the opposite case when $E_J \leq E_C$, the qubit becomes highly sensitive to charge noise, which has proven more challenging to mitigate than flux noise, making it very hard to achieve high coherence. Another motivation is that current technologies allow for more flexibility in engineering the inductive (or potential) part of the Hamiltonian. Therefore, working in the $E_J \leq E_C$ limit, makes the system more sensitive to the change in the potential Hamiltonian. Therefore, we will focus here on the state-of-the-art qubit modalities that fall in the regime $E_J \gg E_C$. For readers who are interested in the physics in the $E_J \leq E_C$ regime, such as the earlier Cooper-pair box charge qubit, we refer to Refs. \onlinecite{Nakamura1999,Vion2002,You2003,Duty2004}.

To access the $E_J \gg E_C$ regime, one preferred approach is to make the charging $E_C$ small by shunting the junction with a large capacitor, $C_s \gg C_J$, effectively making the qubit less sensitive to charge noise -- a circuit commonly known as the transmon qubit\cite{Koch2007}. In this limit, the superconducting phase $\phi$ is a good quantum number, i.e. the spread (or \textit{quantum fluctuation}) of $\phi$ values represented by the quantum wavefunction is small. The low-energy eigenstates are therefore, to a good approximation, localized states in the potential well, see Fig. \ref{Fig:EnergyPotentials}(d). We may gain more insight by expanding the potential term of Eq. (\ref{Eq:HTransmon}) into a power series (since $\phi$ is small), that is

\begin{equation}
E_J \cos(\phi) = \frac{1}{2}E_J \phi^2 - \frac{1}{24}E_J \phi^4 + \mathcal{O}(\phi^6).
\label{Eq:EJexpansion}
\end{equation}

The leading quadratic term in Eq. (\ref{Eq:EJexpansion}) alone will result in a QHO, recall Eq. (\ref{Eq:Hqho}). The second term, however, is quartic which modifies the eigensolution and disrupts the otherwise harmonic energy structure. Note that, the negative coefficient of the quartic term indicates that the anharmonicity $\alpha = \omega_{\text{q}}^{1\rightarrow 2} - \omega_{\text{q}}^{0\rightarrow 1}$ is negative and its limit in magnitude thus cannot be made arbitrarily large. For the case of the transmon, $\alpha = - E_C$ is usually designed to be $100 - 300\,$MHz, as required to maintain a desirable qubit frequency $\freq = (\sqrt{8 E_J E_C} - E_C)/\hbar = 3-6\,$GHz, while keeping an energy ratio sufficiently large ($E_J/E_C \geq 50$) to suppress charge sensitivity\cite{Koch2007}. Fortunately, the charge sensitivity is exponentially suppressed for increased $E_J/E_C$, while the reduction in anharmonicity only scales as a weak power law, leading to a workable device.

Including terms up to fourth order and using the QHO eigenbases, the system Hamiltonian resembles that of a Duffing oscillator

\begin{equation}
H = \freq a^{\dagger}a + \frac{\alpha}{2}a^{\dagger}a^{\dagger}aa.
\label{Eq:HDuffing}
\end{equation}

Since $|\alpha| \ll \freq$, we can see that the transmon qubit is basically a weakly anharmonic oscillator (AHO). If excitation to higher non-computational states is suppressed over any gate operations, either due to a large enough  $|\alpha|$ or due to robust control techniques such as the DRAG pulse, see Sec. \ref{sec:DRAG}, we may effectively treat the AHO as a quantum two-level system, simplifying the Hamiltonian to

\begin{equation}
H = \freq \frac{\sigma_{z}}{2},
\label{Eq:HTwoLevelSystem}
\end{equation}

\noindent where $\sigma_{z}$ is the Pauli-z operator. However, one should always keep in mind that the higher levels physically exist\cite{Peterer2015}. Their influence on system dynamics should be taken into account when designing the system and its control processes. In fact, there are many cases where the higher levels have proven useful to implement more efficient gate operations\cite{DiCarlo2010}.

In addition to reducing the charge dispersion, the use of a large shunt capacitor also enables us to engineer the electric field distribution of the quantum system, and thus the participation of surface loss mechanisms. In the development of the 3D transmon\cite{Paik2011}, e.g. a 2D transmon coupled to a 3D cavity, it was demonstrated that by making the gap between the two lateral capacitor plates large (compared to the film thickness) the coherence time increases since a smaller portion of the electric field interacts with the lossy interfaces, e.g. metal-substrate and substrate-vacuum interfaces, which has been studied extensively\cite{Shnirman2005,Gao2007,Wenner2011,Zeng2015a,Zeng2015b,Calusine2018}.

\begin{figure*}[htp]
\begin{center}
\includegraphics[width=18.2cm]{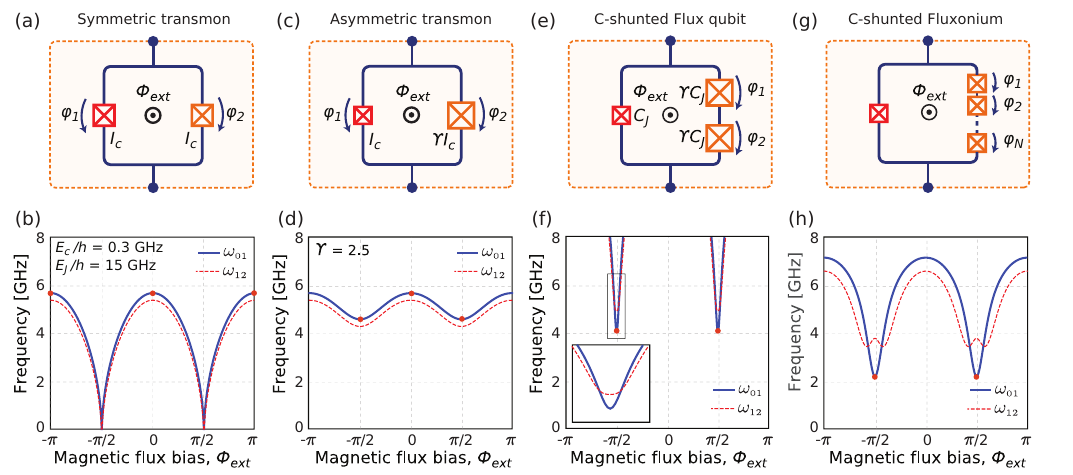}
\caption{Modular qubit circuit representations for capacitively shunted qubit modalities (orange box Fig. \ref{Fig:EnergyPotentials}c) and corresponding qubit transition frequencies for the two lowest energy states as a function of applied magnetic flux in units of $\Phi_0$. \textbf{(a-b)} Symmetric transmon qubit, with Josephson energy $E_J$ are shunted with a capacitor yielding a charging energy $E_C$. \textbf{(c-d)} Asymmetric transmon qubit, with junction asymmetry $\gamma = E_{J2}/E_{J1} = 2.5$. \textbf{(e-f)} Capacitively shunted flux qubit, where a small principle junction (red) is shunted with two larger junctions (orange). Parameters are the same as Yan et al. \cite{Yan2016}. \textbf{(g-h)} C-shunted fluxonium qubit, where the small junction is inductively shunted with a large array of $N$ junctions.}
\label{Fig:QubitModalities}
\end{center}
\end{figure*}

\subsection{\label{sec:HamEng}Qubit Hamiltonian engineering}
\subsubsection{Tunable qubit: split transmon}
To implement fast gate operations with high-fidelity, as needed to implement quantum logic, many (though not all\cite{McKay2016}) of the quantum processor architectures implemented today feature tunable qubit frequencies\cite{DiCarlo2009,Barends2014,Kelly2015,Reagor2018}. For instance, in some cases, we need to bring two qubits into resonance to exchange (swap) energy, while we also need the capability of separating them during idling periods to minimize their interactions. To do this, we need an external parameter which allows us to access one of the degrees of freedom of the system in a controllable fashion.

One widely-used technique is to replace the single Josephson junction with a loop interupted by two identical junctions -- forming a dc superconducting quantum interference device (dc-SQUID)\cite{Tinkham2004}. Due to the interference between the two arms of the SQUID, the effective critical current of the two parallel junctions can be decreased by applying a magnetic flux threading the loop, see Fig. \ref{Fig:QubitModalities}(a). Due to the fluxoid quantization condition, the algebraic sum of branch flux of all of the inductive elements along the loop plus the externally applied flux equal an integer number of superconducting flux quanta, that is

\begin{equation}
\varphi_1 - \varphi_2 + 2\varphi_{e} = 2\pi k,
\label{Eq:FluxoidQuantization}
\end{equation}

\noindent where $\varphi_{e} = \pi \Phi_{\text{ext}}/\Phi_0$. Using this condition, we can eliminate one degree of freedom and treat the SQUID-loop as a single junction, but with the important modification that $E_J$ is tunable (via the SQUID critical current) by means of the external flux $\Phi_{\text{ext}}$. The effective Hamiltonian of the so-called split transmon (ignoring the constant) is

\begin{equation}
H = 4E_C n^2 - \underbrace{2E_J \left| \cos\left(\varphi_e\right)\right|}_{E_{J}'(\varphi_e)}\cos(\phi).
\label{Eq:HsplitTransmon}
\end{equation}

We can see that Eq. (\ref{Eq:HsplitTransmon}) is analogous to Eq. (\ref{Eq:HTransmon}), with $E_J$ replaced by $E_{J}'(\varphi_e) = 2E_J \left|\cos\left(\varphi_e\right)\right|$. The magnitude of the net, effective Josephson energy $E_{J}'$ has a period of $\Phi_0$ in applied flux and spans from $0$ to its maximum value $2E_J$. Therefore, the qubit frequency can be tuned periodically with $\Phi_{\text{ext}}$, see Fig. \ref{Fig:QubitModalities}(b).

While the split transmon enables frequency tunability by the externally applied magnetic field, it also introduces sensitivity to random flux fluctuations, known as flux noise. At any working point, the slope of the qubit spectrum, $\partial \freq/\partial\Phi_{\text{ext}}$, indicates to first order how strongly this flux noise affects the qubit frequency. The sensitivity is generally non-zero, except at multiples of the flux quantum, $\Phi_{\text{ext}} = k\Phi_{\text{0}}$, where $k$ is an integer, where $\partial \freq/\partial\Phi_{\text{ext}} = 0$.

One recent development has focused on reducing the qubit sensitivity to flux noise, while maintaining sufficient tunability to operate our quantum gates. The idea is to make the two junctions in the split transmon asymmetric\cite{Hutchings2017}, see Fig. \ref{Fig:QubitModalities}(c). This yields the following Hamiltonian

\begin{equation}
H = 4E_Cn^2 - \underbrace{{E_{J\Sigma}\sqrt{\cos^2(\varphi_e) + d^2\sin^2(\varphi_e)}}}_{E'_J(\varphi_e)}\cos(\phi),
\label{Eq:AsymmetricTransmon}
\end{equation}

\noindent where $E_{J\Sigma} = E_{J1} + E_{J2}$ and $d = (\gamma - 1)/(\gamma + 1)$ is the junction asymmetry parameter, with $\gamma = E_{J2}/E_{J1}$. Again, we can treat the two junctions as a single-junction transmon, with an effective Josephson energy $E_{J}'(\varphi_e)$. In particular, we can recognize the two special cases; for $d = 0$, the Hamiltonian in Eq. (\ref{Eq:AsymmetricTransmon}) reduces to the symmetric case with $E_{J}'(\varphi_e) = E_{J\Sigma}\left|\cos(\varphi_e)\right|$, as in Eq. (\ref{Eq:HsplitTransmon}) with $E_{J\Sigma} = 2E_J$. In the other limit, when $|d| \rightarrow 1$, $E_{J}'(\varphi_e) \rightarrow E_{J\Sigma}$ and the flux-tunability of the Josephson energy vanishes, which is equivalent to the single junction case, recall Eq. (\ref{Eq:HTransmon}).

From the discussion above we see that going from symmetric to asymmetric transmons does not change the circuit topology. This seemingly trivial modification, however, has profound impact for practical applications. As we can see from the qubit spectra, Fig. \ref{Fig:QubitModalities}(d), the flux sensitivity is suppressed across the entire tunable frequency range. For example, the performance of the cross-resonance gate is optimized with certain frequency detuning between two qubits\cite{Chow2013}. Therefore, by using an asymmetric transmon, a small frequency-tuning range is introduced that is sufficient to compensate for fabrication variations, without introducing unnecessary large susceptibility to flux noise and thus maintaining high coherence. For another example, a surface code scheme based on the adiabatic \textsf{CPHASE}-gate requires specific frequency configuration among qubits in order to avoid frequency crowding issues, and asymmetric transmons fit well with its well-defined frequency range\cite{Versluis2017}. In general, as the quantum processors scale up and fabrication improves, asymmetric transmons are likely to be found in wider applications in the future.

\subsubsection{Towards larger anharmonicity: flux qubit and fluxonium}
We see that split transmon qubits, be it symmetric or not, still share the same topology as the single junction version, yielding a sinusoidal potential. Therefore, the degree to which the properties of these qubits can be engineered has not fundamentally changed. In particular, the limited anharmonicity in transmon-type qubits intrinsically causes significant residual excitation to higher-energy states, undermining performance of gate operations. To go beyond this, it is necessary to introduce additional complexity into the circuit.

One outstanding development in this regard is the invention of the flux qubit\cite{Orlando1999,Mooij1999}, where the qubit loop is interrupted by three (or four) junctions, see Fig. \ref{Fig:QubitModalities}(e). On one branch is one smaller junction; on the other branch are two identical junctions, both a factor $\gamma$ larger in size compared to the small junction. The addition of one more junction as compared to the split transmon is non-trivial, as it changes the circuit topology and reshapes the potential energy profile.

Each junction is associated with a phase variable, and the fluxoid quantization condition again allows us to eliminate one degree of freedom. Consequently, we have a two-dimensional potential landscape, which in comparison to the simpler topology of the transmon, complicates the problem both conceptually and computationally. Fortunately, under the assumed setting that the array junctions are larger in size ($\gamma > 1$), it is usually a good approximation to treat the problem as a particle moving in a quasi-1D potential, which also helps us gain more insight and intuition about the system and draw qualitative conclusions. The Hamiltonian under this \textit{quasi-1D approximation} reads,

\begin{equation}
H \approx 4E_Cn^2 - E_{J}\cos(2\phi + \varphi_e) - 2\gamma E_J\cos(\phi).
\label{Eq:FluxQubit}
\end{equation}

Note that the phase variable in Eq. (\ref{Eq:FluxQubit}) is the sum of the branch phases across the two array junctions, $\phi = (\varphi_1 + \varphi_2)/2$, assuming the same current direction across $\varphi_1$ and $\varphi_2$. The external magnetic flux is denoted $\varphi_e = 2 \pi  \Phi_{ext}/\Phi_0$. The second term in Eq. (\ref{Eq:FluxQubit}) is contributed by the small junction with Josephson energy $E_J$, whereas the third term takes into account the two array junctions, together with Josephson energy $2\gamma E_J$. Clearly, the sum of these two terms no longer has the characteristics of a simple cosinusoid, and the final potential profile as well as the corresponding eigenstates depends on both the external flux $\varphi_e$ and the junction area ratio $\gamma$.

The most common working point for this system is when $\varphi_e = \pi + 2\pi k$, where $k$ is an integer -- that is when half a superconducting flux quantum threads the qubit loop. At this flux bias point, the qubit spectrum reaches its minimum, and the qubit frequency is first-order insensitive to flux noise, see Fig. \ref{Fig:QubitModalities}(f). This point is often referred to as \textit{the flux degeneracy point}, where flux qubits tend to have the optimal coherence time.

At this operation point, the potential energy may assume a single-well ($\gamma \geq 2$) or a double-well ($\gamma < 2$) profile. The single-well case shares some simularities with the transmon qubit, where the quadratic and quartic terms of the Hamiltonian determines the harmonicity and anharmonicity, respectively. The capacitively-shunted flux qubit (CSFQ)\cite{You2006,Yan2016} was explored in this regime, demonstrating long coherence and decently high anharmonicity. Note that as opposed to the transmon qubit, the anharmonicity of the CSFQ is \textit{positive} ($\alpha > 0$). While the improvement in anharmonicity can be associated with reshaping the energy potential, the improved coherence over the first flux qubits can be attributed to the introduction of the capacitive shunt, similar to the modified Cooper-pair box leading to the transmon qubit.

The double-well case obtained for $\gamma < 2$ was demonstrated and investigated much earlier\cite{Orlando1999,Mooij1999}. The intuitive picture based on circulating current states -- so it gets the name persisting-current flux qubit (PCFQ) -- gives a satisfying physical description of the qubit degrees of freedom. However, from the perspective of a quantum engineer, the qubit properties are of more interest, even if sometimes we may lose physical intuition about the system in certain regimes; such as when $\gamma \approx 2$ and there are no clear circulating current states. The most important feature of the PCFQ is that its anharmonicity can be much greater than the transmon and CSFQ and the transition matrix elements $|\langle 1|\hat{n}|0\rangle|, |\langle 1|\hat{\phi}|0\rangle|$ become considerably smaller given equivalent $E_J / E_C$. Therefore, a longer relaxation time can be expected. These features have been demonstrated even more prominently in its close relative, the fluxonium qubit\cite{Pop2014}.

The flux qubit is a striking example that illustrates how one dramatically can engineer the qubit properties through the choice of various circuit parameters. The introduction of array junctions and consequent biharmonic profile generates rich dynamics as well as broad applications. An extention of this idea is the fluxonium qubit, which generated substantial interest recently, due partly to its capability of engineering the transition matrix elements to achieve millisecond $T_1$ time, and due partly to the invention of novel gate schemes applicable to such well-protected qubits\cite{Earnest2018,Lin2018}.

Compared to flux qubits, which usually contain two or three array junctions\cite{Bylander2011}, the number of array junctions in the fluxonium qubit is dramatically increased\cite{Manucharyan2009,Pop2014}, in some cases, to the order of $100$, see Fig. \ref{Fig:QubitModalities}(g). Following the same quasi-1D approximation as for the flux qubit, the last term in Eq. (\ref{Eq:FluxQubit}) becomes $-N\gamma E_J \cos (\phi/N)$, where $N$ denotes the number of array junctions. For large $N$, the argument in the cosine term $\phi/N$ becomes sufficiently small that a second order expansion is a good approximation. This results in the fluxonium Hamiltonian,

\begin{equation}
H \approx 4E_Cn^2 - E_{J}\cos(\phi + \varphi_e) + \frac{1}{2}E_L \phi^2,
\label{Eq:Fluxonium}
\end{equation}

\noindent where $E_L = (\gamma/N)E_J$ is the inductive energy of the effective inductance contributed by the junction array -- often known as superinductance due to its large value~\cite{Manucharyan2009,Masluk2012,ManucharyanPhD2012}. Therefore, we can treat the potential energy as a quadratic term modulated by a sinusoidal term, similar to that of an rf-SQUID type flux qubit\cite{Friedman2002}. However, the kinetic inductance of the Josephson junction array is in general much larger than the geometric inductance of the wire in an rf-SQUID.

Depending on the relative magnitude of $E_J$ and $E_L$, the fluxonium system could involve plasmon states (in the same well) and fluxon states (in different wells). There are a variety of schemes to utilize them for quantum information processing. Generally, the spectrum of the transition between the lowest energy states is similar to that of the flux qubit, see Fig. \ref{Fig:QubitModalities}(h). Both long coherence and high anharmonicity can be expected at the flux sweet spot.

Lastly, we want to point out a further extension -- the $0-\pi$ qubit -- which has even stronger topological protection from noise\cite{Kerman2010,Groszkowski2018}. However, the strongly suppressed sensitivity to external fluctuations also makes it hard to manipulate.

\subsection{\label{sec:interactionHengineering}Interaction Hamiltonian engineering}

To generate entanglement between individual quantum systems -- it is necessary to engineer an interaction Hamiltonian that connects degrees of freedom in those individual systems. In this section, we discuss the physical coupling mechanism and its representation in the qubit eigenbasis. The use of coupling to form 2-qubit gates is discussed in Sec. \ref{sec:QubitControl}.

\subsubsection{Physical coupling: capacitive and inductive}

\begin{figure}[htp]
\begin{center}
\includegraphics[width=8.6cm]{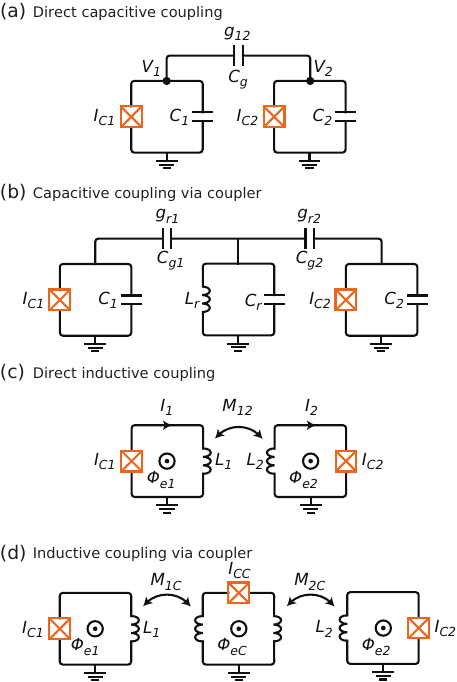}
\caption{Schematic of capacitive and inductive coupling schemes between two superconducting qubits, labeled $1$ and $2$. \textbf{(a)} Direct capacitive coupling, where the voltage nodes of two qubits $V_1$ and $V_2$ are connected by a capacitance $C_g$. \textbf{(b)} Capacitive coupling via a coupler in form of a linear resonator. \textbf{(c)} Direct inductive coupling, where the two qubits are coupled via mutual inductance, $M_{12}$. \textbf{(d)} Inductive coupling via mutual inductances $M_{1C}$ and $M_{2C}$ to a frequency-tunable coupler.}
\label{Fig:QubitInteraction}
\end{center}
\end{figure}

The Hamiltonian of two coupled systems takes a generic form

\begin{equation}
H = H_1 + H_2 + H_{\text{int}},
\label{Eq:Hcoupling}
\end{equation}

\noindent where $H_1$ and $H_2$ denote the Hamiltonians of the individual quantum systems, which could be any combination of the qubit circuits mentioned in Sec. \ref{sec:HOtoTransmon} and \ref{sec:HamEng}. The last term, $H_{\text{int}}$, is the interaction Hamiltonian, which couples variables of both systems. In superconducting circuits, the physical form of the coupling energy is either an electric or magnetic field (or a combination thereof).

To achieve capacitive coupling, a capacitor is placed between the voltage nodes of the two participating circuits, yielding an interaction Hamiltonian $H_{\text{int}}$ of the form

\begin{equation}
H_{\text{int}} = C_g V_1 V_2,
\label{Eq:HintAB}
\end{equation}

\noindent where $C_g$ is the coupling capacitance and $V_1 (V_2)$ is the voltage operator of the corresponding voltage node being connected. Fig. $\ref{Fig:QubitInteraction}$(a) illustrates a realistic example of a direct capacitive coupling between the top nodes of two transmon qubits. Circuit quantization in the limit of $C_g \ll C_1,C_2$ yields

\begin{equation*}
H = \sum_{i = 1,2} \left[ 4E_{C,i}n_{i}^2 - E_{J,i}\cos(\phi_i)\right]
\end{equation*}

\begin{equation}
+ 4e^2\frac{C_g}{C_1 C_2}n_1 n_2,
\label{Eq:HsystCap}
\end{equation}

\noindent where the expressions in brackets are the two Hamiltonians of the individual qubits, [see Eq. (\ref{Eq:HTransmon})], and we take $V_i = (2e/C_i)n_i$ in Eq. (\ref{Eq:HintAB}). From Eq. (\ref{Eq:HsystCap}), we see that the coupling energy depends on the coupling capacitance as well as the matrix elements of the voltage operators. The dependencies are bilinear in the perturbative limit ($C_g \ll C_1, C_2$).

To implement the coupling capacitance, one only need bring the edges of the capacitor pads into close proximity, as has been demonstrated in state-of-the-art planar designs\cite{Barends2013}. The coupling capacitance is determined by the planar capacitor geometry as well as the surrounding environment, such as the dielectric constant of the substrate and the ground plane proximity.

In the case of inductive coupling, a mutual inductance shared by two loops is the coupling mechanism, 
yielding an interaction Hamiltonian that
is of the intuitive form
%
\begin{equation}
H_{\text{int}} = M_{12} I_1 I_2,
\label{Eq:Mint}
\end{equation}

\noindent 
where $M_{12}$ denotes the mutual inductance between the loops and $I_1$ and $I_2$ are the current operators for the currents through the inductors. 
A typical example comprises two closely positioned (rf-SQUID type) flux qubits, as illustrated in Fig.~\ref{Fig:QubitInteraction}(c). The system Hamiltonian can be expressed as (see Refs.~\onlinecite{Burkard2004,Weber2017}):

%



\begin{multline}
    H = \sum_{i = 1,2} \left[ 4E_{C,i}n_{i}^2 - E_{J,i}\cos(\phi_i) + \frac{1}{2}\frac{\Phi_{Li}^2 }{L_i(1-K^2)}\right] 
\\
    - M_{12}(1-K^2) \frac{\Phi_{L1}}{L_1(1-K^2)}\frac{\Phi_{L2}}{L_2(1-K^2)},
    \label{Eq:Mint-Phi}
\end{multline}
%
where the first two terms are the energies associated with the Josephson junctions, the third term captures the inductor energies, and the fourth term is the mutual coupling energy of the form $M I_1 I_2$ (note that in general $\Phi=LI$). $\Phi_{L1}$ and $\Phi_{L2}$ are the magnetic fluxes associated with the currents flowing through the respective inductors, and $K$ is the unitless mutual coupling factor defined by $M_{12} = K \sqrt{L_1 L_2}$.
Importantly, note that in Eq.~(\ref{Eq:Mint-Phi}), $K^2$ renormalizes $L_1$, $L_2$ and $M_{12}$, essentially capturing the loading effect on the circuit due to their presence, and is found by inverting an inductance matrix (see Refs.~\onlinecite{Burkard2004,Weber2017}).

As we did with the charge degree of freedom, we will normalize the inductor and externally applied magnetic fluxes -- in this case, by the reduced quantum unit of flux $\Phi_0/2\pi$ --  to define phases $\phi \equiv \Phi/(\Phi_0/2\pi)$ that are on the same footing as the junction phases $\phi_{1,2}$. Due to fluxoid quantization around the closed loop, these phases must sum to zero or an integer multiple of $2\pi$. For current directions entering the top of the junctions [with $I_1$ counterclockwise and $I_2$ as shown in Fig.~\ref{Fig:QubitInteraction}(c)], $\phi_{i}-\phi_{Li} = \phi_{ei}+ 2\pi n$ with $n=0, \pm1,\ldots$ (see, for example, Ref.~\onlinecite{Orlando1999}). Replacing $\phi_{L1(L2)}$ with $\phi_{1(2)} - \phi_{1(e2)}$  yields:
%
\begin{multline}
= \sum_{i = 1,2} \left[ 4E_{C,i}n_{i}^2 - E_{J,i}\cos(\phi_i) +
                \frac{1}{2} \frac{\left(\frac{\Phi_0}{2\pi}\right)^2 (\phi_{i} - \phi_{ei})^2 }{L_i(1-K^2)}\right]
\\
%
    - M_{12}(1-K^2) \frac{\left(\frac{\Phi_0}{2\pi}\right) (\phi_{1} - \phi_{e1})}{L_1(1-K^2)}\frac{\left(\frac{\Phi_0}{2\pi}\right)(\phi_{2} - \phi_{e2})}{L_2(1-K^2)}.
\label{Eq:Mint-phi}
\end{multline}
%
The coupling is of the form $M_{1,2} I_1 I_2$, with both the mutual coupling and the circulating currents (via the $L_1$ and $L_2$) ``renormalized'' by the factor $(1-K^2)$. To capture the renormalization explicitly, the Hamiltonian is generally simulated using phase operators (Eq.~\ref{Eq:Mint-phi}), rather than current operators.
In the weak-coupling limit $K^2 \ll 1$ (equivalently, $M^2 \ll L_1 L_2 $), the coupling term may be approximated by the Josephson currents with $M_{12} I_1 I_2 \approx M_{12} I_{c1}\sin \phi_1 I_{c2} \sin \phi_2$. Note, however, that this simplifying approximation is only exact in the pathological limit of no coupling. 

%
%
%


To realize a mutual inductance, two looped circuits are brought into close proximity to one another, or, to make it stronger, overlap with each other~\cite{Johnson2011}, and even may share the same wire or Josephson junction inductor~\cite{You2005,Grajcar2006,Niskanen2007,Ashhab2008}.
In the case of a Josephson junction, and for certain metals, the inductance is dominated by \textit{kinetic inductance} contributions, rather than solely geometric inductance~\cite{Yoshihara2016,Niemczyk2010}. Kinetic inductance arises from the mechanical, inertial mass of the charge carriers, but is only practically witnessed in very high-conductance materials like superconductors. A primary feature of kinetic inductance is that its values can vastly exceed those of conventional geometric inductances, which are generally limited by electromagnetic considerations~\cite{Manucharyan2009}.

\subsubsection{\label{sec:couplingAxis}Coupling axis: transverse and longitudinal}

Regardless of its physical realization, the effect of a coupling on system dynamics is determined by its form as represented in the eigenbasis of the individual systems. That is, how $H_{\text{int}}$ appears in the representation spanned by the eigenbasis of $H_1 \otimes H_2$.

Let us start with the previous example of two capacitively coupled transmon qubits [Fig. \ref{Fig:QubitInteraction}(a)]. Using second quantization, the system Hamiltonian in Eq. (\ref{Eq:HsystCap}) can be expressed as

\begin{equation*}
H = \sum_{i \in {1,2}} \left[ \omega_i a_{i}^{\dagger}a_{i} + \frac{\alpha_i}{2}a_{i}^{\dagger}a_{i}^{\dagger}a_{i}a_{i}\right]
\end{equation*}

\begin{equation}
- g \left(a_1 - a_{1}^{\dagger}\right)\left(a_2 - a_{2}^{\dagger}\right),
\label{Eq:HsystDiagC}
\end{equation}

\noindent where the expression within brackets represent the Duffing oscillator Hamiltonian for the qubits and $g$ is the coupling energy. Since we define $V \propto n \propto i(a - a^{\dagger})$, and consequently $I \propto \phi \propto (a + a^{\dagger})$, the original $n_{1}n_{2}$-term becomes what is shown in Eq. (\ref{Eq:HsystDiagC}). Such a coupling is called \textit{transverse}, because the coupling Hamiltonian has non-zero matrix elements only at off-diagonal positions with respect to both oscillators, i.e. $_{i}\langle k|a_{i} - a_{i}^{\dagger}|k\rangle_{i} = 0$ for any integer $k$ and for $i \in {1,2}$ and in this case $_{i}\langle k \pm 1|a_{i} - a_{i}^{\dagger}|k\rangle_{i} \neq 0$.

If we can ignore higher energy levels ($k \geq 2$) either because of sufficient anharmonicity or through careful control protocols that ensure these levels never have influence, we may truncate the Hamiltonian in Eq. (\ref{Eq:HsystDiagC}) to

\begin{equation}
H = \sum_{i \in {1,2}} \frac{1}{2}\omega_i \sigma_{z,i} + g \sigma_{y,1}\sigma_{y,2}.
\label{Eq:HsystDiagCtrunc}
\end{equation}

This is a Hamiltonian of two spins, coupled by an exchange interaction. As we will see in Sec. \ref{sec:capacitivecoupling}, such a Hamiltonian is most commonly used in contemporary implementations and can generate various types of two-qubit entangling gates. Note that, more often, we see that the interaction term is expressed in $\sigma_{x}\sigma_{x}$ instead of $\sigma_{y}\sigma_{y}$. The choice in the context here is arbitrary and does not change the dynamics. However, when both capacitive and inductive couplings are present in the system, both $\sigma_{x}\sigma_{x}$ and $\sigma_{y}\sigma_{y}$ may be needed. In this case, the voltage operator $V \propto i(a - a^{\dagger})$ (reduced to $\sigma_{y}$ after two-level approximation in the lab frame) is transversal to the current operator $I \propto (a + a^{\dagger})$ (reduced to $\sigma_{x}$) and both of them may be transverse to the qubit. A similar example is demonstrated between a qubit and a resonator by Lu et al.\cite{Lu2017}

Transverse coupling can be engineered between a qubit and a harmonic oscillator, see Fig. \ref{Fig:QubitInteraction}(b). In this case, the Hamiltonian becomes

\begin{equation}
H = \frac{1}{2}\freq \sigma_{z} + \omega_{r}a^{\dagger}a + g (\sigma_{+}a + \sigma_{-}a^{\dagger}),
\label{Eq:HsystCircuitQED}
\end{equation}

\noindent where $\freq$ and $\omega_r$ denote the qubit and resonator frequencies, and $\sigma_+ = |0\rangle \langle1|$ and $\sigma_- = |1\rangle\langle 0|$ describes the processes of exciting and de-exciting the qubit, respectively. Here, we have assumed that the coupling is in the dispersive limit, i.e. $g \ll \freq, \omega_r$, hence ignoring the double (de)excitation terms proportional to $\sigma_{+}a^{\dagger}$ and $\sigma_{-}a$, which under typical operation regimes oscillate sufficiently fast to average to zero. The Hamiltonian in Eq. (\ref{Eq:HsystCircuitQED}), is the standard model used for describing how a two-level atom interacts with a resonant cavity that houses it. Such a structure is also known as cavity quantum electrodynamics (cQED), and it is extended to the circuit version here. It has many useful applications in superconducting quantum information architectures, such as high-fidelity readout\cite{Wallraff2004}, see Sec. \ref{sec:readout}, cavity buses\cite{Sandberg2008}, quantum memory\cite{Pierre2014,Yin2013}, quantum computation with cat states\cite{Ofek2016,Wang2016,Axline2016}, etc.

Here, we briefly mention the use of a cavity or resonator to mediate coupling between two qubits, which may be physically well-separated ($\approx 1\,$cm). Since most superconducting resonators are in the GHz frequency range, they can be made much longer than any dimension of a qubit circuit ($\approx 1\,$mm). One can use such a resonator to mediate coupling between two or more otherwise non-interacting qubits. An example is shown in Fig. \ref{Fig:QubitInteraction}(b), where two transmon qubits are both capacitively coupled to the center resonator. The two-level system Hamiltonian is:

\begin{equation*}
H = \sum_{i = 1,2} \left( \omega_i a_{i}^{\dagger}a_{i} + \frac{\alpha_i}{2}a_{i}^{\dagger}a_{i}^{\dagger}a_{i}a_{i}\right) + \omega_r a_{r}^{\dagger}a_{r}
\end{equation*}

\begin{equation}
+ g_{1r} \left(a_{1}^{\dagger}a_{r} + a_{1}a_{r}^{\dagger}\right) + g_{2r} \left(a_{2}^{\dagger}a_{r} + a_{2} a_{r}^{\dagger}\right).
\label{Eq:HsystCircuitQED2}
\end{equation}

It can be shown that in the dispersive limit, i.e. $g_{ir} \ll |\omega_i - \omega_r|$, the resonator can -- after proper transformation and approximation -- be treated as an isolated system, and the composite system simplified to two transversely coupled qubits, see Eq. (\ref{Eq:HsystDiagCtrunc}).

We now turn to the previous example of two inductively coupled flux qubits, see Fig.~\ref{Fig:QubitInteraction}(c). Assume that the double-well potential [Fig. \ref{Fig:QubitModalities}(g)] has a relatively high inter-well barrier, which leads to an exponentially small qubit transition frequency at the energy degeneracy point, ($\Phi_{e} = \pi$). Around this degeneracy point, the off-diagonal matrix element of $\sin(\phi)$ is zero, i.e. the ground and excited states are localized in different wells and $\langle 1 | \sin(\phi) | 1\rangle - \langle 0 | \sin(\phi) | 0\rangle \neq 0$. We can then rewrite the Hamiltonian in Eq. (\ref{Eq:HsystInd}) as

\begin{equation}
H = \sum_{i = 1,2} \frac{1}{2}\omega_i \sigma_{zi} + g \sigma_{z1}\sigma_{z2}.
\label{Eq:HsystDiagL}
\end{equation}

\noindent Now, the coupling axis is the same as the qubit quantization axes and therefore termed \textit{longitudinal coupling}. Note, however, that the physical $\sigma_{x}\sigma_{x}$ and $\sigma_{z}\sigma_{z}$ couplings can change in the qubit frame.

Longitudinal coupling is an important type of interaction, because it can generate entanglement without energy exchange. Moreover, it is found a necessary ingredient in the application of quantum annealing, where certain hard combinatorial optimization problems can be modeled by the Ising Hamiltonian in Eq. (\ref{Eq:HsystDiagL}) and finding its ground state would solve this problem.

An intermediate qubit mode may also be used as a coupler in the longitudinal case. In Fig. \ref{Fig:QubitInteraction}(d), an additional rf-SQUID is used to mediate the coupling. The coupling strength can be tuned by the flux bias of the coupler SQUID\cite{Kounalakis2018}. Note that a tunable coupler may also be realized in a structure with capacitive couplings\cite{McKay2016}. A tunable coupler is useful because it provides a wide range of coupling strengths\cite{Weber2017}, a high on-off ratio\cite{Chen2014} for reducing gate error-rates, and more ways of achieving high-fidelity entangling gates\cite{Roth2017,Didier2018a,Reagor2018,Yan2018}. The trade-off is an additional control line.

In addition to the pure transversal and longitudinal qubit-qubit interactions, there are also examples of mixed types of interaction Hamiltonians\cite{Didier2015}

\begin{equation}
H = \frac{1}{2}\freq \sigma_{z} + \omega_{r}a^{\dagger}a + g \sigma_{z}(a + a^{\dagger}),
\label{Eq:HsystDiagLC}
\end{equation}

\noindent which are longitudinal with respect to a qubit, but transverse with respect to a harmonic oscillator in a qubit-resonator system. Such a model is called longitudinal but one should note that it is only longitudinal to one participating system. It is hard to engineer physically longitudinal coupling with respect to a harmonic oscillator, since either the $E$-field ($V$) or the $B$-field ($I$) is transverse with respect to the eigen field of the harmonic oscillator. Note, however, that a transversal model such as in Eq. (\ref{Eq:HsystCircuitQED}) may be transformed into a longitudinal one in certain operating regimes, see Sec. \ref{sec:readout}.

In some applications, such as for quantum annealing, both longitudinal and transverse couplings are desired ($\sigma_z \sigma_z$ coupling for mapping the problem and $\sigma_x \sigma_x$ coupling for enhancing the annealing performance) and require independent control.

\section{\label{sec:Noise}Noise, decoherence, and error mitigation}

Random, uncontrollable physical processes in the qubit control and measurement equipment or in the local environment surrounding the quantum processor are sources of noise that lead to decoherence and reduce the operational fidelity of the qubits. In this section, we introduce the basics of noise leading to decoherence in superconducting circuits, and we discuss coherent control methods to mitigate certain types of noise.

\subsection{Types of noise}

In a closed system, the dynamical evolution of a qubit state is deterministic. That is, if we know the starting state of the qubit and its Hamiltonian, then we can predict the state of the qubit at any time in the future. However, in open systems, the situation changes. The qubit now interacts with uncontrolled degrees of freedom in its environment, which we refer to as fluctuations or noise. In the presence of noise, as time progresses, the qubit state looks less and less like the state we would have predicted and, eventually, the state is lost. There are many different sources of noise that affect quantum systems, and they can be categorized into two primary types: systematic noise and stochastic noise.

\subsubsection{Systematic noise}

Systematic noise arises from a process that is traceable to a fixed control or readout error. For example, we apply a microwave pulse to the qubit that we believe will impart a 180-degree rotation. However, the control field is not tuned properly and, rather than rotating the qubit 180 degrees, the pulse slightly over-rotates or under-rotates the qubit by a fixed amount. The underlying error is {\em systematic}, and it therefore leads to the same rotation error each time it is applied. However, when such erroneous pulses are used in practice in a variety of control sequences, the observed results may appear to be influenced by random noise. This is because the pulse is generally not applied in the same way for each experiment: it could be applied a different number of times, interspersed with different pulses in different orders, and therefore generally differs from experiment to experiment. However, once systematic errors are identified, they can generally be corrected through proper calibration or the use of improved hardware.

\subsubsection{Stochastic noise}

The second type of noise is stochastic noise, arising from random fluctuations of parameters that are coupled to our qubit~\cite{Ball2016}. For example, thermal noise of a 50$\Omega$ resistor in the control lines leading to the qubit will have voltage and current fluctuations -- Johnson noise -- with a noise power that is proportional to both temperature and bandwidth. Or, the oscillator that provides the carrier for a qubit control pulse may have amplitude or phase fluctuations. Additionally, randomly fluctuating electric and magnetic fields in the local qubit environment -- e.g., on the metal surface, on the substrate surface, at the metal-substrate interface, or inside the substrate -- can couple to the qubit. This creates unknown and uncontrolled fluctuations of one or more qubit parameters, and this leads to qubit decoherence.

\subsubsection{\label{subsubsec:noise-strength}Noise strength and qubit susceptibility}
\begin{figure*}[!t]
\centering
    \includegraphics[width=16cm]{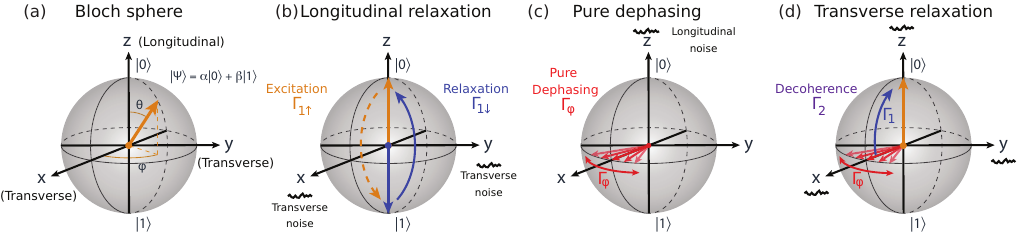}
    \caption{Transverse and longitudinal noise represented on the Bloch sphere.
    \textbf{(a)} Bloch sphere representation of the quantum state $\lvert \psi \rangle = \alpha \lvert 0 \rangle + \beta \lvert 1 \rangle$. The qubit quantization axis -- the $z$ axis -- is \textit{longitudinal} in the qubit frame, corresponding to $\sigma_z$ terms in the qubit Hamiltonian. The $x$-$y$ plane
    is \textit{transverse} in the qubit frame, corresponding to $\sigma_x$ and $\sigma_y$ terms in the qubit Hamiltonian.
    \textbf{(b)} Longitudinal relaxation results from energy exchange between the qubit and its environment, due to transverse noise that couples to the qubit in the $x-y$ plane and drives transitions $\lvert 0 \rangle \leftrightarrow \lvert 1\rangle$. A qubit in state $\lvert 1\rangle$ emits energy to the environment and relaxes to $\lvert 0\rangle$ with a rate $\Gamma_{1\downarrow}$ (blue arched arrow). Similarly, a qubit in state $\lvert 0\rangle$ absorbs energy from the environment, exciting it to $\lvert 1\rangle$ with a rate $\Gamma_{1\uparrow}$ (orange arched arrow). In the typical operating regime $k_{\mathrm{B}} T \ll \hbar \freq$, the up-rate is suppressed, leading to the overall decay rate $\Gamma_1 \approx \Gamma_{1\downarrow}$.
    \textbf{(c)} Pure dephasing in the transverse plane arises from longitudinal noise along the $z$ axis that
    fluctuates the qubit frequency. A Bloch vector along the $x$-axis will diffuse clockwise or counterclockwise around the equator due to the stochastic frequency fluctuations, depolarizing the azimuthal phase with a rate $\Gamma_{\phi}$.
    \textbf{(d)} Transverse relaxation results in a loss of coherence at a rate $\Gamma_2 = \Gamma_1/2 + \Gamma_{\phi}$, due to a combination of energy relaxation and pure dephasing. Pure dephasing leads to decoherence of the quantum state $(1/\sqrt{2})(\lvert 0 \rangle + \lvert 1 \rangle)$, initially pointed along the $x$-axis. Additionally, the excited state component of the superposition state  may relax to the ground state, a phase-breaking process that loses the orientation of the vector in the $x$-$y$ plane.}
\label{fig:Bloch-sphere-noise}
\end{figure*}
The degree to which a qubit is affected by noise is related to the amount of noise impinging on the qubit, and the qubit's susceptibility to that noise. The former is often a question of materials science and fabrication; that is, can we make devices with lower levels of noise. Or, it may be related to the quality of the control electronics and cryogenic engineering to limit the levels of noise on the control lines that necessarily connect to the qubits to control them. The latter -- qubit susceptibility -- is a question of qubit design. Qubits can be designed to trade off sensitivity to one type of noise at the expense of increased sensitivity to other types of noise. Thus, materials science, fabrication engineering, electronics design, cryogenic engineering, and qubit design all play a role in creating devices with high coherence. In general, one should strive to eliminate the sources of noise, and then design qubits that are insensitive to the residual noise.

The qubit response to noise depends on how the noise couples to it -- either through a longitudinal or a transverse coupling as referenced to the qubit quantization axis. This can be visualized using a Bloch Sphere picture of the qubit state, as illustrated in Fig.~\ref{fig:Bloch-sphere-noise} and discussed in detail in Section~\ref{sec:modeling-noise}.


\subsection{\label{sec:modeling-noise}Modeling noise and decoherence}

\subsubsection{Bloch sphere representation}

The \textit{Bloch sphere} is a unit sphere used to represent the quantum state of a two-level system (qubit). Fig.~\ref{fig:Bloch-sphere-noise}(a) shows a Bloch sphere with a \textit{Bloch vector} representing the state $\lvert \psi\rangle = \alpha \lvert 0\rangle + \beta \lvert 1\rangle$. If we visualize the Bloch sphere as the planet Earth, then by convention, the north pole represents state $\lvert 0 \rangle$ and the south pole state $\lvert 1 \rangle$. For pure quantum states such as $\lvert \psi \rangle$, the Bloch vector is of unit length, $\lvert \alpha \rvert^2+\lvert \beta \rvert^2=1$, connecting the center of the sphere to any point on its surface.

The $z$-axis connects the north and south poles. It is called the \textit{longitudinal axis}, since it represents the \textit{qubit quantization axis} for the states $\lvert 0 \rangle$ and $\lvert 1 \rangle$ in the qubit eigenbasis. In turn, the $x$-$y$ plane is the \textit{transverse plane} with \textit{transverse axes} $x$ and $y$. In this Cartesian coordinate system, the unit Bloch vector $\vec{a}=(\sin \theta \cos \phi, \sin \theta \sin \phi, \cos \theta)$ is represented using the polar angle $0\leq \theta \leq \pi$ and the azimuthal angle $0 \leq \phi < 2\pi$, as illustrated in Fig.~\ref{fig:Bloch-sphere-noise} (a). Following our convention, state $\lvert 0 \rangle$ at the north pole is associated with $+1$, and state $\lvert 1 \rangle$ (the south pole) with $-1$. We can similarly represent the quantum state using the angles $\theta$ and $\phi$,
\begin{equation}
    \lvert \psi \rangle = \alpha \lvert 0 \rangle + \beta \lvert 1 \rangle = \cos\frac{\theta}{2} \lvert 0 \rangle + e^{i \phi}\sin \frac{\theta}{2} \lvert 1 \rangle.
    \label{Eq:Bloch-vector-stationary}
\end{equation}
The Bloch vector is stationary on the Bloch sphere in the \textit{rotating frame picture}. If state $\lvert 1 \rangle$ has a higher energy than state $\lvert 0 \rangle$ (as it generally does in superconducting qubits), then in a stationary frame, the Bloch vector would precess around the $z$-axis at the qubit frequency $(E_1-E_0)/\hbar$. Without loss of generality (and much easier to visualize), we instead \textit{choose} to view the Bloch sphere in a reference frame where the $x$ and $y$-axes also rotate around the $z$-axis at the qubit frequency. In this \textit{rotating frame}, the Bloch vector appears stationary as written in Eq.~(\ref{Eq:Bloch-vector-stationary}). The rotating frame will be described in detail in Section~\ref{sec:capacitivecoupling} in the context of single-qubit gates.

For completeness, we note that the density matrix $\rho=\lvert \psi \rangle \langle \psi \vert$ for a pure state $\lvert \psi \rangle$ is equivalently
\begin{align}
    \rho = \frac{1}{2}(I + \vec{a} \cdot \vec{\sigma})
    &= \frac{1}{2}
    \left(
        \begin{matrix}
            1+\cos \theta & e^{-i\phi} \sin \theta \\
            e^{i\phi} \sin \theta  & 1+ \sin \theta
        \end{matrix}
    \right) \\
     &= \left(
        \begin{matrix}
            \cos^2\frac{\theta}{2} & e^{-i\phi}\cos\frac{\theta}{2} \sin\frac{\theta}{2} \\
            e^{i\phi}\cos\frac{\theta}{2} \sin\frac{\theta}{2}  & \sin^2\frac{\theta}{2}
        \end{matrix}
    \right) \\
    &=\left(
        \begin{matrix}
            |\alpha|^2 & \alpha \beta^*  \\
            \alpha^* \beta  & |\beta|^2
        \end{matrix}
    \right)
    \label{eq:rho}
\end{align}
where $I$ is the identity matrix, and $\vec{\sigma} = [\sigma_x, \sigma_y, \sigma_z]$ is a vector of Pauli matrices. If the Bloch vector $\vec{a}$ is a unit vector, then $\rho$ represents a pure state $\psi$ and $\mathrm{Tr}(\rho^2)=1$. More generally, the Bloch sphere can be used to represent \textit{mixed states}, for which $\lvert \vec{a} \rvert < 1$; in this case, the Bloch vector terminates at points \textit{inside} the unit sphere, and $0\leq \mathrm{Tr}(\rho^2)<1$. To summarize, the surface of the unit sphere represents pure states, and its interior represents mixed states\cite{Nielsen2011}.

\subsubsection{Bloch-Redfield model of decoherence}

Within the standard Bloch-Redfield~\cite{Wangsness1953,Bloch1957,Redfield1957} picture of two-level system dynamics, noise sources weakly coupled to the qubits have short correlation times with respect to the system dynamics. In this case, the relaxation processes are characterized by two rates (see Fig.~\ref{fig:Bloch-sphere-noise}):
\begin{align}
    &\textrm{longitudinal relaxation rate:} &\Gamma_1 &\equiv \frac{1}{T_1}  \\
    &\textrm{transverse relaxation rate:} &\Gamma_2 &\equiv \frac{1}{T_2} = \frac{\Gamma_1}{2} + \Gamma_{\varphi} \label{Eq:T2}
\end{align}
which contains the pure dephasing rate $\Gamma_{\varphi}$. We note that the definition of $\Gamma_2$ as a sum of rates presumes that the individual decay functions are exponential, which occurs for Lorentzian noise spectra (centered at $\omega=0$) such as white noise (short correlation times) with a high-frequency cutoff.

The impact of noise on the qubit can be visualized on the Bloch sphere in Fig.~\ref{fig:Bloch-sphere-noise}(a). For an initial state ($t=0$)
\begin{align}
    \lvert \psi \rangle = \alpha |0 \rangle + \beta |1\rangle,
\end{align}
the Bloch-Redfield density matrix $\rho_{\textrm{BR}}$ for the qubit is written~\cite{IthierPhD2005,Ithier2005},
\begin{equation}
\label{eq:rho_RB}
    \rho_{\textrm{BR}} =
    \left(
        \begin{matrix}
            1 + (|\alpha|^2-1) e^{- \Gamma_1 t} & \alpha \beta^* e^{i \delta \omega t} e^{- \Gamma_2 t} \\
            \alpha^* \beta e^{-i \delta \omega t} e^{- \Gamma_2 t} & |\beta|^2 e^{- \Gamma_1 t}
        \end{matrix}
        \right).
\end{equation}
There are a few important distinctions between Eq.~(\ref{eq:rho_RB}) and Eq.~(\ref{eq:rho}), which we list here and then describe in more detail in subsequent sections.
\begin{itemize}
    \item First, we have introduced the \textit{longitudinal decay function} $\exp(-\Gamma_1 t)$, which accounts for longitudinal relaxation of the qubit.
    \item Second, we introduced the \textit{transverse decay function} $\exp(-\Gamma_2 t)$, which accounts for transverse decay of the qubit.
    \item Third, we have introduced an explicit phase accrual $\exp(i \delta \omega t)$, where $\delta \omega = \omega_{\textrm{q}} - \omega_{\textrm{d}}$, which generalizes the Bloch sphere picture to account for cases where the qubit frequency $\omega_{\textrm{q}}$ differs from the rotating-frame frequency $\omega_{\textrm{d}}$, as we will see later when discussing measurements of $T_2$ using Ramsey interferometry\cite{Ramsey1950,Hahn1950}, and in Section~\ref{sec:capacitivecoupling} in the context of single-qubit gates.
    \item And, fourth, we have constructed the matrix such that for $t \gg (T_1, \; T_2)$, the upper-left matrix element will approach unit value, indicating that all population relaxes to the ground state, while the other three matrix elements decay to zero. This is related to the assumption that the environmental temperature is low enough that thermal excitations of the qubit from the ground to excited state rarely occur.
\end{itemize}

\begin{figure*}[!t]
\centering
    \includegraphics[width=16cm]{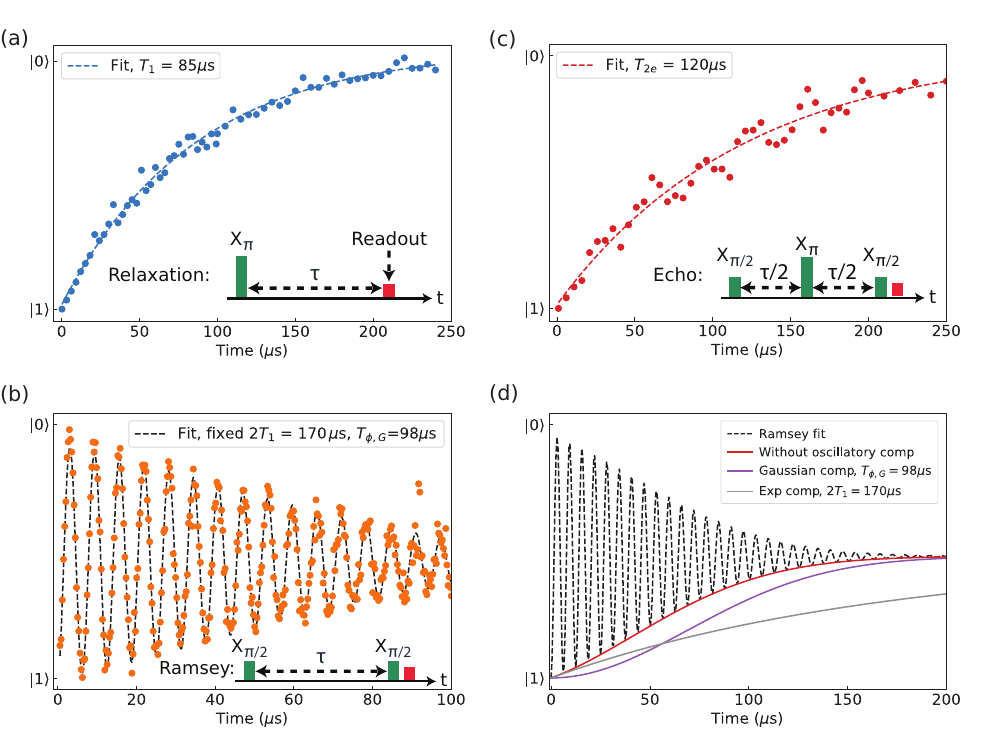}
    \caption{Characterizing longitudinal $(T_1)$ and transverse $(T_2)$ relaxation times of a transmon qubit\cite{Kjaergaard2017}.
    \textbf{(a)} Longitudinal relaxation (energy relaxation) measurement. The qubit is prepared in the excited state using an $\X{\pi}$-pulse and measured after a waiting time $\tau$. For each value $\tau$, this procedure is repeated to obtain an ensemble average of the qubit polarization: $+1$ corresponding to $\lvert 0 \rangle$, and $-1$ corresponding to $\lvert 1 \rangle$.
    The resulting exponential decay function has a characteristic time $T_1 = 85$ $\mu\mathrm{s}$.
    \textbf{(b)} Transverse relaxation (decoherence) measurement via Ramsey interferometry. The qubit is prepared on the equator using an $\X{\pi/2}$-pulse, intentionally detuned from the qubit frequency by $\delta \omega$, causing the Bloch vector to precess in the rotating frame at a rate $\delta \omega$ around the $z$-axis. After a time $\tau$, a second $\X{\pi/2}$ pulse then projects the Bloch vector back on to the $z$ axis, effectively mapping its former position on the equator to a position on the $z$ axis.
    The oscillations decay with an approximately (but not exactly) exponential decay function, with a characteristic time $T_2^* = 95$ $\mu\mathrm{s}$.
    \textbf{(c)} Transverse relaxation (decoherence) measurement via a Hahn echo experiment~\cite{Hahn1950}. The qubit is prepared and measured in the same manner as the Ramsey interfometry experiment, except that a single $X{\pi}$ pulse is applied midway through the free-evolution time $\tau$.
    The decay function is approximately exponential, with a characteristic time $T_{2E} = 120$ $\mu\mathrm{s}$.
    The coherence improvement using the Hahn echo over panel (b) 
    indicates that some low-frequency dephasing noise has been mitigated; however, a small amount 
    remains since $T_{2E}$ has not yet reached the $2 T_1$ limit.
    \textbf{(d)} Coherence function incorporating $T_1$ loss and Gaussian dephasing components of the Ramsey interferometry data in panel (b).
    The Gaussian-distributed $1/f$ noise spectrum of magnetic flux noise leads to a decay function $\exp(-t/2T_1)\exp(-\chi_N) = \exp (-t/2T_1) \exp(-t^2/T_{\varphi,\mathrm{G}}^2)$ in Eq.~(\ref{eq:rho_1_f}). These two decay functions together match well the Ramsey data in panel (b).}
    %
\label{fig:T1-T2}
\end{figure*}

\noindent \textbf{Longitudinal relaxation}

The longitudinal relaxation rate $\Gamma_1$ describes depolarization along the qubit quantization axis, often referred to as ``energy decay'' or ``energy relaxation.'' In this language, a qubit with polarization $p=1$ is entirely in the ground state $(\lvert0\rangle)$ at the north pole, $p=-1$ is entirely in the excited state $(\lvert1\rangle)$ at the south pole, and $p=0$ is a completely depolarized mixed state at the center of the Bloch sphere.

As illustrated in Fig.~\ref{fig:Bloch-sphere-noise}(b), longitudinal relaxation is caused by {\em transverse noise}, via the $x$- or $y$-axis, with the intuition that off-diagonal elements of an interaction Hamiltonian are needed to connect and drive transitions between states $|0\rangle$ and $|1 \rangle$.

Depolarization occurs due to energy exchange with an environment, generally leading to both an ``up transition rate'' $\Gamma_{1\uparrow}$ (excitation from $|0 \rangle$ to $|1 \rangle$), and a ``down transition rate'' $\Gamma_{1\downarrow}$  (relaxation from $|1 \rangle$ to $|0 \rangle$). Together, these form the longitudinal relaxation rate $\Gamma_1$:
\begin{align}
    \Gamma_1 \equiv \frac{1}{T_1} = \Gamma_{1\downarrow} + \Gamma_{1\uparrow}.
\end{align}
$T_1$ is the $1/e$ decay time in the exponential decay function in Eq.~(\ref{eq:rho_RB}), and it is the characteristic time scale over which qubit population will relax to its steady-state value. For superconducting qubits, this steady-state value is generally the ground state, due to Boltzmann statistics and typical operating conditions. Boltzmann equilibrium statistics lead to the ``detailed balance'' relationship $\Gamma_{1\uparrow} = \exp(-\hbar \freq / k_{\mathrm{B}} T) \Gamma_{1\downarrow}$, where $T$ is the temperature and $k_{\mathrm{B}}$ is the Boltzmann constant, with an equilibrium qubit polarization approaching $p=\tanh (\hbar \freq / 2 k_{\mathrm{B}} T)$. Typical qubits are designed at frequency $\freq/2 \pi \approx 5$ GHz and are operated at dilution refrigerator temperatures $T \approx 20$ mK. In this limit, the up-rate $\Gamma_{1\uparrow}$ is exponentially suppressed by the Boltzmann factor $\exp(-\hbar \freq / k_{\mathrm{B}} T)$, and so only the down-rate $\Gamma_{1\downarrow}$ contributes significantly, relaxing the population to the ground state. Thus, qubits generally spontaneously lose energy to their cold environment, but the environment rarely introduces a qubit excitation. As a result, the equilibrium polarization approaches unity [see Eq.~(\ref{eq:rho_RB})]\cite{Siegman1986,Berns2006}.

Only noise at the qubit frequency mediates qubit transitions, whether absorption or emission, and this noise is generally ``well behaved'' (short correlation time, many modes weakly coupled to qubit, no divergences) around the qubit frequency for superconducting qubits. The intuition is that qubit-transition linewidths are relatively narrow in frequency, and so the noise generally does not vary much over this narrow frequency range. Although there are a few notable exceptions, for example, qubit decay in the presence of hot quasiparticles\cite{Catelani2011,Catelani2012,Gustavsson2016}, which can lead to non-exponential decay functions, longitudinal depolarization measurements generally present exponential decay functions consistent with the Bloch-Redfield picture.

An example of a $T_1$ measurement is shown in Fig.~\ref{fig:T1-T2}(a). The qubit is prepared in its excited state using an $\X{\pi}$-pulse, and then left to spontaneously decay to the ground state for a time $\tau$, after which the qubit is measured. A single measurement will project the quantum state into either state $\lvert0\rangle$ or state $\lvert 1 \rangle$, with probabilities that correspond to the qubit polarization. To make an estimate of this polarization, one needs to identically prepare the qubit and repeat the experiment many times. This is analogous to flipping a coin: any single flip will yield heads or tails, but the probability of obtaining a heads or tails can be estimated by flipping the coin many times and taking the ensemble average. The resulting exponential decay has a characteristic time $T_1 = 85$ $\mu\mathrm{s}$.

\noindent \textbf{Pure dephasing}

The \textit{pure dephasing} rate $\Gamma_{\phi}$ describes depolarization in the $x-y$ plane of the Bloch sphere. It is referred to as ``pure dephasing,'' to distinguish it from other phase-breaking processes such as energy excitation or relaxation.

As illustrated in Fig.~\ref{fig:Bloch-sphere-noise}(c), pure dephasing is caused by {\em longitudinal noise} that couples to the qubit via the $z$-axis. Such longitudinal noise causes the qubit frequency $\freq$ to fluctuate, such that it is no longer equal to the rotating frame frequency $\omega_d$, and causes the Bloch vector to precess forward or backward in the rotating frame. Intuitively, we can imagine identically preparing several instances of the Bloch vector along the $x$-axis. For each instance, the stochastic fluctuations of qubit frequency will result in a different precession frequency, resulting in a net fanout of the Bloch vector in the $x-y$ plane. This eventually leads to a complete depolarization of the azimuthal angle $\phi$. Note that this stochastic effect will be captured in the transverse relaxation rate $\Gamma_2$ (next section); it is \textit{not} the deterministic term $\exp(\pm i \delta \omega t)$ that appears in Eq.~(\ref{eq:rho_RB}), which represents intentional detuning of the qubit reference frame.

There are a few important distinctions between pure dephasing and energy relaxation. First, in contrast to energy relaxation, pure dephasing is \textit{not} a resonant phenomenon; noise at any frequency can modify the qubit frequency and cause dephasing. Thus, qubit dephasing is subject to broadband noise. Second, since pure dephasing is elastic (there is no energy exchange with the environment), it is in principle \textit{reversible}. That is, the dephasing can be ``undone'' -- with quantum information being preserved -- through the application of unitary operations, e.g., dynamical decoupling pulses\cite{Bylander2011}, see Sec. \ref{sec:Tphi-S}.

The degree to which the quantum information can be retained depends on many factors, including the bandwidth of the noise, the rate of dephasing, the rate at which unitary operations can be performed, etc. This should be contrasted with spontaneous energy relaxation, which is an \textit{irreversible} process. Intuitively, once the qubit emits energy to the environment and its myriad uncontrollable modes, the quantum information is essentially lost with no hope for its recovery and reconstitution back into the qubit.

\textbf{Transverse relaxation}

\noindent The transverse relaxation rate $\Gamma_2=\Gamma_1/2 + \Gamma_{\varphi}$ describes the loss of coherence of a superposition state, for example $(1/\sqrt{2})(\lvert 0 \rangle + \lvert 1 \rangle)$, pointed along the $x$-axis on the equator of the Bloch sphere as illustrated in Fig.~\ref{fig:Bloch-sphere-noise}(d). Decoherence is caused in part by longitudinal noise, which fluctuates the qubit frequency and leads to pure dephasing $\Gamma_{\varphi}$ (red). It is also caused by transverse noise, which leads to energy relaxation of the excited-state component of the superposition state at a rate $\Gamma_1$ (blue). Such a relaxation event is also a phase-breaking process, because once it occurs, the Bloch vector points to the north pole, $\lvert 0 \rangle$, and there is no longer any knowledge of which direction the Bloch vector \textit{had} been pointing along the equator; the relative phase of the superposition state is lost.

Transverse relaxation $T_2$ can be measured using Ramsey interferometry, as shown and described in Fig.~\ref{fig:T1-T2}(b). The protocol positions the Bloch vector on the equator using a $X_{\pi/2}$-pulse. Typically, the carrier frequency of this pulse is slightly detuned from the qubit frequency by an amount $\delta \omega$. As a result, the Bloch vector will precess around the $z$-axis at a rate $\delta \omega$. This is done for convenience sake, so that the resulting Ramsey measurement will oscillate, making it easier to analyze. After precessing for a time $\tau$, a second $\X{\pi/2}$-pulse projects the Bloch vector back on to the $z$-axis. Repeated measurements are made to take an ensemble averaged estimate of the qubit polarization, as a function of $\tau$. The resulting oscillations in Fig.~\ref{fig:T1-T2}(b) feature an approximately exponential decay function with time $T_2^* = 98$ $\mu\mathrm{s}$. The ``*'' indicates that the Ramsey experiment is sensitive to \textit{inhomogeneous broadening}. That is, it is highly sensitive to quasi-static, low-frequency fluctuations that are constant within one experimental trial, but vary from trial to trial, e.g., due to $1/f$-type noise. This sensitivity to quasi-static noise is related to the corresponding $N=0$ noise filter function shown in Fig.~\ref{fig:T1-T2}(d) being centered at at zero-frequency, as described in more detail in Section~\ref{sec:Tphi-S}.

The Hahn echo shown in Fig.~\ref{fig:T1-T2}(c) is an experiment that is less sensitive to quasi-static noise. By placing a $Y_{\pi}$ pulse at the center of a Ramsey interferometry experiment, the quasi-static contributions to dephasing can be ``refocused,'' leaving an estimate $T_{2E}$ that is less sensitive to inhomogeneous broadening mechanisms. The pulses are generally chosen to be resonant with the qubit transition for a Hahn echo, since any frequency detuning would be nominally refocused anyway. The resulting decay function in Fig.~\ref{fig:T1-T2}(c) is essentially exponential with time $T_{2E} = 120$ $\mu\mathrm{s}$.

With the known $T_1$ and $T_2$ times, one can infer the pure dephasing time $T_{\varphi}$ from Eq.~(\ref{Eq:T2}), provided the decay functions are exponential. In superconducting qubits, however, the broadband dephasing noise (e.g., flux noise, charge noise, critical-current noise, ...) tends to exhibit a $1/f$-like power spectrum. Such noise is singular near $\omega = 0$, has long correlation times, and generally does not fall within the Bloch-Redfield description. The decay function of the off-diagonal terms in Eq.~(\ref{eq:rho_RB}) are generally non-exponential, and for such cases, the simple expression in Eq.~(\ref{Eq:T2}) is not applicable.

\subsubsection{Modification due to $1/f$-type noise}
\label{sec:mod_1_f_noise}
If we assume that the qubit is coupled to many independent fluctuators, then, regardless of their individual statistics, they will in concert generate noise with a Gaussian distribution due to the central limit theorem. We therefore say that the longitudinal fluctuations exhibit Gaussian-distributed $1/f$ noise~\cite{Falci2005,Paladino2014}. For $1/f$ noise spectra, the phase decay function is itself a Gaussian $\exp \left[ -(t/T_{\varphi,\mathrm{G}})^2 \right]$, where we write $T_{\varphi,\mathrm{G}}$ to distinguish it from $T_{\varphi}$ used in Eq.~(\ref{Eq:T2}). Furthermore, this function is separable from the $T_1$-type exponential decay, because the $T_1$-noise remains regular at the qubit frequency. The density matrix in Eq.~(\ref{eq:rho_RB}) becomes, following Refs.~\onlinecite{IthierPhD2005,Bylander2011},
\begin{equation}
\label{eq:rho_1_f}
    \rho 
    = \left(
        \begin{matrix}
            1 + (|\alpha|^2-1) e^{- \Gamma_1 t} & \alpha \beta^* e^{i \delta \omega t} e^{- \frac{\Gamma_1}{2} t} e^{- \chi_N (t)}  \\
            \alpha^* \beta e^{-i \delta \omega t} e^{- \frac{\Gamma_1}{2} t} e^{-\chi_N (t)}  & |\beta|^2 e^{- \Gamma_1 t}
        \end{matrix}
        \right),
\end{equation}
where the decay function $\langle \exp(- \chi_N(t)) \rangle$ contains the \textit{coherence function} $\chi_N(t)$, which generalizes pure dephasing to include non-exponential decay functions. As we shall see later, the subscript $N$ labeling the decay function
refers to the number of $\pi$-pulses used to refocus the low-frequency noise, which impacts the form of the decay function. Because the function is no longer purely exponential, we cannot formally write the transverse relaxation decay function as $\exp(-t/T_2)$. However, an exponential decay remains a practically reasonable approximation for $T_{\varphi} \gtrsim T_1$. We also note that the energy decay component of the transverse relaxation is $\exp(-t/2T_1)$, and so $T_2$ can never be larger than $2T_1$. In the absence of pure dephasing, the maximum $T_2=2T_1$ is reached.

\begin{figure}[!t]
\centering
    \includegraphics[width=8.6cm]{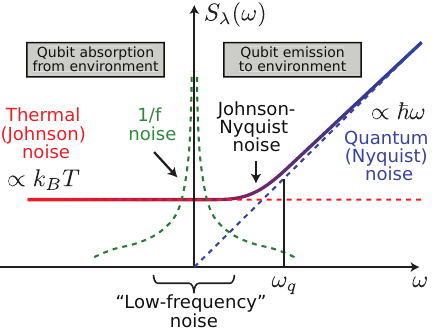}
    \caption{Examples of symmetric and asymmetric noise spectral densities. Noise at positive (negative) frequencies corresponds to the qubit emitting (absorbing) energy to (from) its environment. Thermal noise is proportional to temperature $T$ and carries essentially a white noise spectrum. As it represents a classical fluctuating parameter, such as electric current, the noise power spectral density is symmetric in frequency. When resonant with the qubit, it will drive both stimulated emission and absorption processes. The qubit may also spontaneously emit energy to its environment, represented as Nyquist noise\cite{Nyquist1928}, a quantum mechanical effect that is not symmetric in frequency. At sufficiently low temperatures or high frequencies, $\hbar \omega > 2k_{\mathrm{B}}T$, the Nyquist noise dominates thermal noise. Another common example is $1/f$ noise, which is also a classical noise fluctuation and symmetric in frequency.}
\label{fig:noise-PSD}
\end{figure}

As an example, consider the Ramsey interferometry data in Fig.~\ref{fig:T1-T2}(b). Since the dephasing is relatively weak, the transverse relaxation function as $\exp(-t/T_2)$ is a reasonable fit and yields $T_2 = 95$ $\mu \mathrm{s}$. However, using the value $T_1 = 85$ $\mu \mathrm{s}$ from Fig.~\ref{fig:T1-T2}(a) and dividing out $\exp(-t/2T_1)$ from the data in Fig.~\ref{fig:T1-T2}(b), the remaining pure dephasing decay function is shown in Fig.~\ref{fig:T1-T2}(d) and assumes a Gaussian envelope $\langle \exp(- \chi_N(t)) \rangle = \exp \left[ -(t/T_{\varphi,\mathrm{G}} t)^2 \right]$, with $T_{\varphi,\mathrm{G}} = 98$ $\mu \textrm{s}$. The Hahn echo data in Fig.~\ref{fig:T1-T2}(c) may be treated similarly.

For completeness, in addition to $1/f$ dephasing mechanisms, we note that there are also ``white'' pure dephasing mechanisms, which give rise to an exponential decay function for the dephasing component of $T_2$. One common example is dephasing due to the shot noise of residual photons in the readout resonator coupled to superconducting qubits, as we discuss in Section~\ref{sec:photon-number-fluctuations}.

\subsubsection{Noise power spectral density (PSD)}
The frequency distribution of the noise power for a stationary noise source $\lambda$ is characterized by its PSD $S_{\lambda}(\omega)$
\begin{align}
    \label{eq:PSD_defn}
    S_{\lambda}(\omega) &= \int_{-\infty}^{\infty} d\tau \; \langle \lambda(\tau) \lambda(0) \rangle e^{- i \omega \tau}.
\end{align}
The Wiener-Khintchine theorem states that the PSD is the Fourier transform of the autocorrelation function $c_{\lambda}(\tau) = \langle \lambda(\tau) \lambda(0) \rangle$ of the noise source $\lambda$. Since the integration limits are $(-\infty,\infty)$, this is the bilateral PSD. Symmetrizing the PSD allows one to consider only positive frequencies, which is termed a unilateral PSD. Both unilateral and bilateral PSDs are used, often with the same notation, and so one needs to know how the PSD is defined, keep track of the factors of 2 and $\pi$, and also be aware of the implications for quantum systems.

For classical systems, the noise power spectral density is symmetric. This is because the autocorrelation function of real signals is itself a real function, and the Fourier transform of a real temporal function is symmetric in the frequency domain. Dephasing noise is caused by real, fluctuating fields, and so its PSD is generally symmetric. Examples of such classical noise include thermal (Johnson) noise and $1/f$ noise\cite{Johnson1928} (see Fig.~\ref{fig:noise-PSD}).

In turn, the inverse Fourier transform of the PSD will yield the autocorrelation function:
\begin{align}
      c_{\lambda}(\tau) &= \frac{1}{2\pi} \int_{-\infty}^{\infty} d\omega \; S_{\lambda}(\omega) e^{ i \omega \tau}. 
\end{align}
This implies that integrating the noise power spectral density with $\tau=0$ yields the second moment of the noise, or, for zero-mean fluctuations, the variance.

However, the autocorrelation function for a quantum system may be complex-valued due to the fact that quantum operators generally do not commute at different times. This means that time-ordering of the operators matters, and the PSD need not be symmetric in frequency. This is generally the case for transverse noise causing longitudinal energy relaxation. Noise at a positive frequency $S(\freq)$ corresponds to energy transfer from the qubit to the environment, including both stimulated and spontaneous emission, associated with the down-rate $\Gamma_{1\downarrow}$. Noise at a negative frequency $S(-\freq)$ corresponds to energy transfer to the qubit from the environment, associated with the up-rate $\Gamma_{1\uparrow}$. For a detailed discussion, see Refs.~\onlinecite{Haus2000,Clerk2010}. Spontaneous emission to a cold environment or electromagnetic vacuum, represented by Nyquist noise in Fig.~\ref{fig:noise-PSD}, is an example of an asymmetric noise PSD\cite{Nyquist1928}.

In general, making a connection between $S_{\lambda}(\omega)$ and the measured qubit decay functions is the basis for noise spectroscopy up to second-order statistics~\cite{Bylander2011,Gustavsson2011,Yoshihara2014,Yan2012,Yan2013}. The search for higher-order spectra related to non-Gaussian noise is a current topic of active research\cite{Sung2019}.

\subsection{Common examples of noise}

There are many sources of stochastic noise in superconducting qubits, and we refer the reader to Ref.~\onlinecite{Oliver2013} for a review. Here, we briefly present several of the most common types of noise, their affect on coherence, and refer the reader to the references for a more detailed discussion.

\subsubsection{Charge noise}
\textit{Charge noise} is ubiquitous in solid-state devices. It arises from charged fluctuators present in the defects or charge traps that reside in interfacial dielectrics, the junction tunnel barrier, and in the substrate itself. These are often modeled as an ensemble of fluctuating two-level systems or as bulk dielectric loss~\cite{Wang2015,Dial2016}. For example, in the case of a transmon qubit, the electric field between the capacitor plates traverses and couples to dielectric defects residing on the metal surfaces of the plates (for lateral-plate-type capacitors) or the capacitor dielectric between the plates (for parallel-plate-type capacitors). The electric field variable is transverse with respect to the quantization axis of the transmon qubit, which means that this noise is mainly responsible for energy relaxation ($T_1$). Additionally, if the $E_J/E_C$ ratio of the transmon is not made sufficiently large (smaller than around 60), the qubit frequency itself will also be sensitive to broadband charge fluctuations. In this case, low-frequency charge noise couples longitudinally to the transmon and causes pure dephasing ($T_{\varphi}$).

Charge noise is modeled primarily as a combination of inverse-frequency noise and Nyquist noise, also referred to as \textit{ohmic} noise. At lower frequencies, the spectral density takes the form
\begin{equation}\label{eqn:charge-1_f}
  S_Q(\omega) = A_Q^2 \left( \frac{2 \pi \times 1\mathrm{Hz}}{\omega} \right)^{\gamma_Q},
\end{equation}
with quasi-universal values $A_Q^2 = (10^{-3}e)^2/\mathrm{Hz}$ at 1 Hz, and $\gamma_Q \approx 1$. In addition to large $1/f$ fluctuations, early charge qubits often witnessed discrete, charge offsets reminiscent of random telegraph noise. Together, these two mechanisms severely limited the utility of charge qubits, and served as a strong motivation to move to capacitively shunted charge qubits (transmons), which greatly reduced the qubit longitudinal sensitivity to charge noise. At higher frequencies, the power spectrum takes the form $S_Q(\omega) = B_Q^2 [\omega/ (2 \pi \times 1 \mathrm{Hz})]$,
where the noise strength $B_Q^2$ at 1 Hz can assume a range of values depending on the level of dissipation in the system. Likewise, the cross-over from $1/f$-like behavior to $f$-like behavior generally occurs at around 1 GHz, but will vary higher or lower between samples depending on the degree of dissipation\cite{Astafiev2004,Yan2016}.

\subsubsection{Magnetic flux noise}
Another commonly observed noise in solid-state devices is magnetic \textit{flux noise}.
The origin of this noise is understood to arise from the stochastic flipping of spins (magnetic dipoles) that reside on the surfaces of the superconducting metals comprising the qubit~\cite{KochRoger2007}, resulting in random fluctuations of the effective magnetic field that biases flux-tunable qubits.

For example, in the case of the split transmon, the external magnetic field threading the loop couples longitudinally to the qubit and modulates the transition frequency via the Josephson energy $E_J$ (except at $\varphi_e = 0$, where the qubit is first-order insensitive to magnetic-field fluctuations). Because the flux noise is longitudinal to the transmon, it contributes to pure dephasing ($T_{\varphi}$). However, in the case of the flux qubit, and depending on the flux-bias point, the flux noise may be either longitudinal --  causing dephasing $T_{\varphi}$ -- or it may couple transversely and thus contribute to $T_1$ relaxation~\cite{Bylander2011,Yan2016}.
The noise power spectrum of these fluctuations generally exhibits a ``quasi-universal" dependence,
\begin{equation}\label{eqn:charge-1_f}
  S_{\Phi}(\omega) = A_{\Phi}^2 \left( \frac{2 \pi \times 1\mathrm{Hz}}{\omega} \right)^{\gamma_{\Phi}},
\end{equation}
with $\gamma_{\Phi} \approx 0.8-1.0$ and $A_{\Phi}^2 \approx (1 \; \mu \Phi_0)^2/\mathrm{Hz}$, and has been shown to extend from less than millihertz to beyond gigahertz frequencies~\cite{Wellstood1987,Bylander2011,Yan2012,Yan2013,Slichter2012}.

The large, low-frequency weighting of the $1/f$ power distribution enables the use of engineered error mitigation techniques -- such as dynamical decoupling -- to achieve better coherence~\cite{Falci2004,Bialczak2007,Bylander2011,Anton2012} and for improving single and two-qubit gate fidelity\cite{Pokharel2018}. It was recently demonstrated that $1/f$ flux noise is also a $T_1$-mechanism when extended out to the qubit frequency~\cite{Yan2016}, and one similarly expects a crossover to ohmic flux noise at high enough frequencies\cite{Quintana2017}.

Although much is known about the statistics and number of the defects presumed responsible for flux noise, their precise physical manifestation remains uncertain~\cite{KochRoger2007,Kumar2016}. The fact that the $1/f$ noise is quasi-universal and largely independent of device, strongly suggests a common origin for the noise. Recent studies suggest that adsorbed molecular oxygen may be responsible for flux-noise~\cite{Kumar2016,deGraaf2017}.

\subsubsection{\label{sec:photon-number-fluctuations}Photon number fluctuations}
In the circuit QED architecture, resonator \textit{photon number fluctuation} is another major decoherence source~\cite{Schuster2005}. Residual microwave fields in the cavity have photon-number fluctuations that in the dispersive regime impact the qubit through an interaction term $\chi\sigma_z n$, see Sec. \ref{sec:couplingAxis}, leading to a frequency shift $\Delta_{\mathrm{Stark}}=2\eta \chi \bar{n}$, where $\bar{n}$ is the average photon number, and $\eta=\kappa^2/(\kappa^2 + 4\chi^2)$ effectively scales the photon population seen by the qubit due to the interplay between the qubit-induced dispersive shift of the resonator frequency ($\chi$) and the resonator decay rate ($\kappa$).

In the dispersive limit, the noise is longitudinally coupled to the qubit and leads to pure dephasing at a rate,
\begin{equation}
    \Gamma_{\phi} = \eta \frac{4 \chi^2}{\kappa} \bar{n}.
\end{equation}
The fluctuations originate from residual photons in the resonator, typically due to radiation from higher temperature stages in the dilution refrigerator~\cite{Yeh2017,Yan2018}. The corresponding noise spectral density is of a Lorentzian type,
\begin{equation}
  S(\omega) = 4 \chi^2 \frac{2\eta\bar{n}\kappa}{\omega^2+\kappa^2},
\end{equation}
which exhibits an essentially white noise spectrum up to a $3\mathrm{dB}$ cutoff frequency $\omega = \kappa$ set by the resonator decay rate $\kappa$, see Ref.~\onlinecite{Yan2016}.

\subsubsection{Quasiparticles}

\textit{Quasiparticles}, i.e. unpaired electrons, are another important noise source for superconducting devices~\cite{Catelani2012}. The tunneling of quasiparticles through a qubit junction may lead to both $T_1$ relaxation and pure dephasing $T_{\varphi}$, depending on the type of qubit, the bias point, and the junction through which the tunneling event occurs~\cite{Catelani2011,Gustavsson2016}.

Quasiparticles are naturally excited due to thermodynamics, and the quasiparticle density in equilibrium superconductors should be exponentially suppressed as temperature decreases. However, below about 150 mK, the quasiparticle density observed in superconducting devices -- generally in the range $10^{-8}-10^{-6}$ per Cooper pair -- is much higher than BCS theory would predict for a superconductor in equilibrium with its cryogenic environment at 10 mK. The reason for this excess quasiparticle population is unclear, but it is very likely related to the presence of additional, non-thermal mechanisms that increase the generation rates, ``bottleneck effects'' that occur at millikelvin temperatures to reduce recombination rates, or a combination of both.

It has been shown that the observed $T_1$ and excess excited-state population measured in today's state-of-the-art high-coherence transmon are self-consistent with excess ``hot'' nonequilibrium quasiparticles at the quasi-universal density of around $10^{-7}-10^{-6}$ per Cooper pair\cite{Jin2015,Serniak2018}. Although this quasiparticle generation mechanism is not yet well understood, it has been shown that quasiparticles can be transiently pumped away, improving $T_1$ times and reducing $T_1$ temporal variation~\cite{Gustavsson2016}.

\subsection{Operator form of qubit-environment interaction}

Similar to the way that two qubits are coupled, a qubit may couple and interact with uncontrolled degrees of freedom (DOF) in its environment (the noise sources). The interaction Hamiltonian between the qubit DOF ($\hat{O}_q$) and those of the noise source ($\hat{\lambda}$) may be expressed in a general form

\begin{equation}
\hat{H}_{\text{int}} = \nu \hat{O}_q\hat{\lambda} 
\label{Eq:HintNoise}
\end{equation}

\noindent where $\nu$ denotes the coupling strength -- which is related to the sensitivity of the qubit to environmental fluctuations $\partial\hat{H}_q / \partial\lambda$
-- and we assume that $\hat{O}_q$ is a qubit operator 
within the qubit Hamiltonian $\hat{H}_q$. The noisy environment represented by the operator $\hat{\lambda}$ produces fluctuations $\delta \lambda$. Note that we retained the hats in this section to remind us that these are quantum operators.

\subsubsection{Connecting $T_1$ to $S(\omega)$}
If the coupling is transverse to the qubit, e.g. $\hat{O}_q$ is of the type $\sigma_{x}$ or $(a + a^{\dagger})$ -- see the related case of qubit-qubit coupling treated in Sec. \ref{sec:interactionHengineering} -- then noise at the qubit frequency can cause transitions between the qubit eigenstates. Since this is a stochastic process, the ensemble-average manifests itself as a decay (usually exponential) of the qubit population 
towards a certain equilibrium value (usually the qubit ground state $|0\rangle$ for $k_B T \ll \hbar \freq$). Again, this process is equivalently referred to as ``$T_1$ relaxation", ``energy relaxation", or ``longitudinal relaxation". As stated above, $T_1$ is the characteristic time scale of the decay. Its inverse, $\Gamma_1 = 1/T_{1}$ is called the relaxation rate and depends on the power spectral density of the noise $S(\omega)$ at the transition frequency of the qubit $\omega = \freq$:


\begin{equation}
\Gamma_{1} = \frac{1}{\hbar^2} \left| \langle 0 | \frac{\partial \hat{H}_q}{\partial\lambda} | 1 \rangle \right|^2 S_{\lambda} (\freq),
\label{Eq:Gamma1-PSD}
\end{equation}

\noindent where $\partial\hat{H}_q / \partial\lambda$ is the qubit transverse susceptibility to fluctuations $\delta \lambda$, such that $|\delta \lambda|^2$ 
is the ensemble average value of the environmental noise sources as seen by the qubit.
%
%
%
Eq.~(\ref{Eq:Gamma1-PSD}) is equivalent to Fermi's Golden Rule,
in which the qubit's transverse susceptibility to noise 
is driven by the noise power spectral density.
%
%
The qubit transverse susceptibility can be used to calculate the prefactors; for example, for fluctuations $\delta \lambda = \delta n$, the relevant term in the transmon Hamiltonian in Eq. (\ref{Eq:HTransmon}) is $4E_C(\hat{n} - n_g)^2$, where we allow for an offset charge $n_g$, and the susceptibility is given by $8 E_C\hat{n}$.
We refer the reader to Refs. \onlinecite{You2007,Kerman2008,Oliver2013b} for more details.

\begin{figure}[!t]
\centering
    \includegraphics[width=8.6cm]{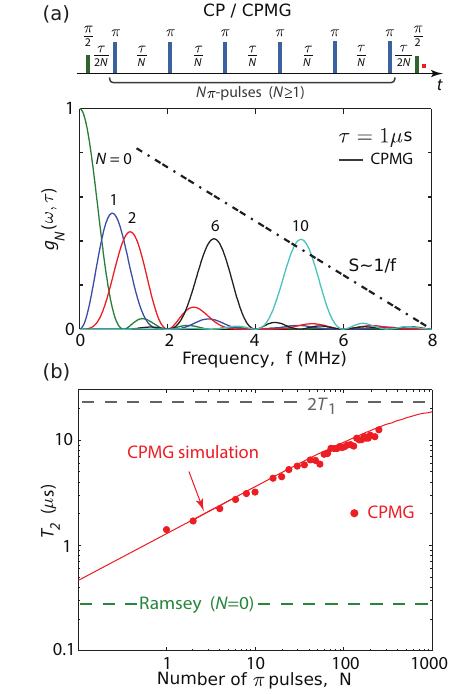}
    \caption{Dynamical error suppression.
    \textbf{(a)} Carr-Purcell-Meiboom-Gill (CPMG) pulse sequence applies $N$ equally spaced $\pi$ pulses within an otherwise free-evolution time $\tau$. Pulses in the time domain correspond to bandpass filters in the frequency domain (lower panel) which serve to shape the noise power spectrum seen by the qubit. The centroid of the bandpass filter shifts to higher frequencies as $N$ is increased. For noise that decreases with frequency, such as $1/f$ noise, larger $N$ corresponds to less integrated noise impinging on the qubit.
    \textbf{(b)} CPMG pulse sequence applied to a flux qubit biased at a point that is highly sensitive to $1/f$ flux noise. The Ramsey $(N=0)$ time is approximately 300 ns, and the Hahn echo $(N=1)$ time is approximately $1.5$ $\mu \mathrm{s}$. Increasing the number of CPMG pulses continues to increase the effective $T_2$ time towards the $2T_1$ limit. Adapted from Ref.~\onlinecite{Bylander2011}.
    }
\label{fig:CPMG}
\end{figure}

\subsubsection{\label{sec:Tphi-S}Connecting $T_{\varphi}$ to $S(\omega)$}

If the coupling to the qubit is instead longitudinal, e.g. $\hat{H}_q$ is of the type $\sigma_z$ or $a^{\dagger}a$, the noise will stochastically modulate the transition frequency of the qubit and thereby  introduce a stochastic phase evolution of a qubit superposition state. This gradually leads to a loss of phase information, and it is therefore called \textit{pure dephasing} (time constant $T_{\varphi}$). Unlike $T_1$ relaxation, which is generally an irreversible (incoherent) error, pure dephasing $T_{\varphi}$ is in principle reversible (a coherent error). The degree of pure dephasing depends on the control pulse sequence applied while the qubit is subject to the noise process.

Consider the relative phase $\varphi$ of a superposition state undergoing free evolution in the presence of noise.
The superposition state's accumulated phase,
\begin{align}
    \varphi(t) = \int_0^t \omega_{q} dt' = \langle\omega_{\mathrm{q}}\rangle t + \delta\varphi(t)
\end{align}
diffuses due to adiabatic fluctuations of the transition frequency,
\begin{align}
    \label{eq:del_phi_defn}
    \delta\varphi(t) = \frac{\partial \freq}{\partial \lambda} \int_0^t \delta\lambda(t')\mathrm{d}t',
\end{align}
where $\partial \freq / \partial \lambda = (1/\hbar) |\langle \partial\hat{H}_q / \partial\lambda \rangle |$ is the qubit's longitudinal sensitivity to $\lambda$-noise.
For noise generated by a large number of fluctuators that are weakly coupled to the qubit, its statistics are Gaussian.
Ensemble averaging over all realizations of the Gaussian-distributed stochastic process $\delta\lambda(t)$,
the dephasing is
\begin{align}
    \label{eq:ens_avg_phase}
    \langle e^{i\,\delta\varphi(t)} \rangle = e^{-\frac{1}{2}\,\langle \delta\varphi^2(t) \rangle} \equiv e^{-\chi_N(t)},
\end{align}
leading to a coherence decay function,

\begin{equation}
\langle e^{- \chi_N(\tau)} \rangle = \exp\left[-\frac{\tau^2}{2} \frac{\partial \freq}{\partial\lambda} \int_{-\infty}^{\infty}g_N(\omega,\tau)S(\omega)d\omega\right],
\label{Eq:PureDephasing}
\end{equation}

\noindent where $g(\omega,\tau)$ is a dimensionless weighting function.

The function $g_N(\omega,\tau)$ can be viewed as a frequency-domain filter of the noise $S_{\lambda}(\omega)$ [see Fig.~\ref{fig:CPMG}(a)].
In general, its filter properties depend on the number $N$ and distribution of applied pulses. For example, considering sequences of $\pi$-pulses~\cite{Martinis2003,Uhrig2007,Cywinski2008,Biercuk2009a,Biercuk2009b,Bylander2011},
\begin{multline} \label{eq:filter_def}
  g_N(\omega,\tau) =
  \frac{1}{(\omega\tau)^2} \, \Big| 1+(-1)^{1+N}\exp(i\omega\tau) + \\
  2\sum_{j=1}^{N} (-1)^j \exp(i\omega\delta_j\tau) \cos(\omega\tau_\pi/2) \Big|^2 ,
\end{multline}
where $\delta_j \in [0,1]$ is the normalized position of the centre of the $j$th $\pi$-pulse between the two $\pi/2$-pulses, $\tau$ is the total free-induction time, and $\tau_\pi$ is the length of each $\pi$-pulse~\cite{Biercuk2009a,Biercuk2009b}, yielding a total sequence length $\tau + N\tau_\pi$.
As the number of pulses increases for fixed $\tau$, the filter function's peak shifts to higher frequencies, leading to a reduction in the net integrated noise for $1/f^{\alpha}$-type noise spectra with $\alpha > 0$.
Similarly, for a fixed $N$, the filter function will shift in frequency with $\tau$.
Additionally, for a fixed time separation $\tau' = \tau/N$ (valid for $N \ge 1$), the filter sharpens and asymptotically peaks at $\omega'/2\pi = 1/2\tau'$ as more pulses are added. $g_N(\omega,\tau)$ is thus called the ``filter function''~\cite{Cywinski2008,Bylander2011}, and it depends on the pulse sequences being applied.
From Eq. (\ref{Eq:PureDephasing}), the pure dephasing decay arises from a noise spectral density that is ``shaped'' or ``filtered'' by the sequence-specific filter function. By choosing the number of pulses, their rotation axes, and their arrangement in time, we can design filter functions that minimize the net noise power for a given noise spectral density within the experimental constraints of the experiment (e.g., pulse-modulation bandwidth of the electronics used to control the qubits).

To give a standard example, we compare the coherence integral for two cases: a Ramsey pulse sequence and a Hahn echo pulse sequence. Both sequences involve two $\pi/2$ pulses separated by a time $\tau$, during which free evolution of the qubit occurs in the presence of low-frequency dephasing noise. The distinction is that the Hahn echo will place a single $\pi$ pulse ($N=1$) in the middle of the free-evolution period, whereas the Ramsey does not use any additional pulses ($N=0$). The resulting filter functions are:

\begin{align}
g_0(\omega,\tau) &= \text{sinc}^2 \frac{\omega \tau}{2} \\
g_1(\omega,\tau) &= \sin^2 \frac{\omega \tau}{4} \text{sinc}^2 \frac{\omega \tau}{4}
\end{align}

\noindent where the subscript $N=0$ and $N=1$ indicate the number of $\pi$-pulses applied for the Ramsey and Hahn echo experiments, respectively. The filter function $g_0(\omega,\tau)$ for the Ramsey case is a sinc-function centered at $\omega=0$. For noise that decreases with frequency, e.g., $1/f$ flux noise in superconducting qubits, the Ramsey experiment windows through the noise in $S(\omega)$ where it has its highest value. This is the worst choice of filter function for $1/f$ noise. In contrast, the Hahn echo filter function has a centroid that is peaked at a higher frequency, away from $\omega=0$. In fact, it has zero value at $\omega=0$. For noise that decreases with frequency, such as $1/f$ noise, this is advantageous. This concept extends to larger numbers $N$ of $\pi$ pulses, and is called a Carr-Purcell-Meiboom-Gill (CPMG) sequence~\cite{Carr1954,Meiboom1958}. In Fig.~\ref{fig:CPMG}(b), the $T_2$ time of a qubit under the influence of strong dephasing noise is increased toward the $2T_1$ limit using a CPMG dynamical error-suppression pulse sequence with an increasing number of pulses, $N$. We refer the reader to Refs.~\onlinecite{Bylander2011,Gustavsson2012a,Gustavsson2012b}, where these experiments were performed with superconducting qubits.

\subsubsection{\label{sec:noise-spectroscopy}Noise spectroscopy}
The qubit is highly sensitive to its noisy environment, and this feature can be used to map out the noise power spectral density. In general, one can map the noise PSD during \textit{free evolution} -- periods of time for which no control is applied to the qubit, except for very short dynamical decoupling pulses -- and during \textit{driven evolution} -- periods of time during which the control fields are applied to the qubit. Both free-evolution and driven-evolution noise is important to characterize, as the noise PSD may differ for these two types of evolution, and both are utilized in the context of universal quantum computation. We refer the reader to Ref.~\onlinecite{Yan2013} for a summary of noise spectroscopy during both types of evolution.

The Ramsey frequency itself is sensitive to longitudinal noise, and monitoring its fluctuations is one means to map out the noise spectral density over the sub-millihertz to $\sim100$ Hz range\cite{Yan2012,SankPhD2014}.

At higher frequencies, the CPMG dynamical decoupling sequence can be used to create narrow-band filters that ``sample'' the noise at different frequencies as a function of the free-evolution time $\tau$ and the number of pulses $N$. This has been used to map out the noise PSD in the range 0.1 - 300 MHz~\cite{Bylander2011}.
One must be careful of the additional small peaks at higher-frequencies, which all contribute to the dephasing used to perform the noise spectroscopy\cite{Loretz2015}.

In fact, using pulse envelopes such as Slepians\cite{Slepian1961} -- which are designed to have concentrated frequency response -- to perform noise spectroscopy is one means to reduce such errors\cite{Biercuk2009a}.

At even higher frequencies, measurements of $T_1$ can be used in conjunction with Fermi's golden rule to map out the transverse noise spectrum above 1 GHz~\cite{Bylander2011,Yan2016,Schoelkopf2002}.

The aforementioned are all examples of noise spectroscopy during free evolution. Noise spectroscopy during driven evolution was also demonstrated using a ``spin-locking'' technique, where a strong drive along $x$ or $y$ axes defines a new qubit quantization axis, whose Rabi frequency is the new qubit frequency in the spin-locking frame. The spin-locking frame is then used to infer the noise spectrum while the qubit is continually subject to a driving field. For more information, we refer the reader to Ref.~\onlinecite{Yan2013}.

\subsection{Engineering noise mitigation}

Here, we briefly review a few examples of techniques that have been developed to reduce noise or reduce its impact on decoherence (sensitivity). We stress that improving gate fidelity is a comprehensive optimization task, one that is full of trade-offs. It is thus important to identify what the limiting factors are, what price we have to pay to diminish these limiting factors, and what advantage we can achieve until reaching a better trade-off. These all require an accurate understanding the limitations on the gate fidelity, the sources of decoherence, the properties of the noise, and how it affects the system performance.

\subsubsection{Materials and fabrication improvements}
Numerous efforts have been undertaken to reduce noise-induced defects due to materials and fabrication~\cite{Oliver2013,Sage2011}. In the case of charge noise, significant efforts have been made 
to reduce the number of defects, such as substrate cleaning~\cite{Quintana2014,Zeng2015a}, substrate annealing~\cite{Kamal2016}, and trenching~\cite{Gambetta2017,Calusine2018}.
In the case of flux noise, several groups have performed experiments to characterize the behavior and properties of magnetic-flux defects~\cite{KochRoger2007,Anton2013,Sendelbach2008}. More recently, a number of groups have tried optical surface treatments to remove these defects~\cite{Kumar2016}.

In the context of residual quasiparticles, it has been shown that adding quasiparticle traps to the circuit design can reduce the quasiparticle number, particularly in devices that create excess quasiparticles, such as classical digital logic or operation in the presence of thermal radiation~\cite{Riwar2016}

\subsubsection{Design improvements}

Another strategy is to reduce qubit sensitivity to the noise by design. A qubit can only lose energy to defects if it couples to them. It has been demonstrated that altering the capacitor geometry to increase the electric-field mode volume reduces the electric field density in the thin dielectric regions that cause loss. This effectively reduces the ``participation'' of the defects and makes the qubits less senstivie to these noise sources.~\cite{Paik2011,Wang2015,Yan2016}.

In another example, the split transmons built using asymmetric junctions have lower sensitivity to flux noise than their symmetric counterparts at the expense of decreased frequency tunability~\cite{Hutchings2017}. This is a good trade-off to make, because generally one is interested in tuning the qubit frequency over a somewhat restricted range (typically around 1 GHz) about the qubit frequency. When such asymmetric transmons are used in a gate scheme such as the adiabatic \textsf{CPHASE}-gate\cite{Barends2014}, (see Sec.\ref{sec:CZgates}) the qubit is less sensitive to flux noise, has a lower dephasing rate, and this should improve the gate fidelity in general.

\subsubsection{Dynamical error suppression}

As introduced in the previous section, it is advantageous to leverage the $1/\omega$ distribution of flux noise, wherein a considerable amount of the noise power resides at low frequencies, and so the noise is ``quasi-static''.
The spin-echo technique~\cite{Hahn1950}, which disrupts the free evolution by a $\pi$-pulse, is extremely effective in mitigating the pure dephasing by refocusing the coherent phase dispersion due to low-frequency noise.
The more advanced versions, such as the CPMG-sequence, use multiple $\pi$-pulses to interrupt the system more frequently, pushing the filter band to even higher frequencies -- a technique known as \textit{dynamical decoupling}\cite{Bylander2011}.

Returning to excess quasiparticles, it has been shown that quasiparticles can be stochastically pumped away from the qubit region, resulting in longer, and more stable $T_1$ times~\cite{Gustavsson2016}. Although the pumping technique uses a series of $\pi$-pulses, this technique differs from dynamical error suppression of coherent errors in that pulses are stochastically applied, and that it addresses incoherent errors ($T_1$).

\subsubsection{Cryogenic engineering}
In the case of photon shot-noise, in addition to applying dynamical decoupling techniques, there have been several recent works aimed at reducing the thermal photon flux that reaches the device. This include optimizing the attenuation of the cryogenic setup~\cite{Yan2018,Krinner2018,Jin2015}, remaking the cryogenic attenuators with more efficient heat sinking~\cite{Yeh2017}, adding absorptive ``black'' material to absorb stray thermal photons~\cite{Barends2011,Corcoles2011}, and adding additional cavity filters for thermalization~\cite{Wang2019}.

\section{\label{sec:QubitControl}Qubit control}
In this section, we will introduce how superconducting qubits are manipulated to implement quantum algorithms. Since the transmon-like variety of superconducting qubits has so far been the most widely deployed modality for implementing quantum programs, the discussion throughout this section will be focused on modern techniques for transmons. Nonetheless, the techniques introduced here are applicable to all types of superconducting qubits.

We start with a brief review of the gates used in classical computing as well as quantum computing, and the concept of universality. Subsequently we discuss the most common technique of driving single qubit gates via a capacitive coupling of a microwave line, coupled to the qubit. We introduce the notion of ``virtual $\Z{}$ gates" and ``DRAG" pulsing. In the latter part of this section, we review some of the most common implementations of two-qubit gates in both tunable and fixed-frequeny transmon qubits. The single-qubit and two-qubit operations together form the basis of many of the medium-scale superconducting quantum processors that exist today.

Throughout this section, we write everything in the computational basis $\{|0\rangle,|1\rangle\}$ where $|0\rangle$ is the $+1$ eigenstate of $\sigma_z$ and $|1\rangle$ is the $-1$ eigenstate. We use capitalized serif-fonts to indicate the rotation operator of a qubit state, e.g. rotations around the $x$-axis by an angle $\theta$ is written as
\begin{equation}
 \X{\theta} = R_X(\theta) = e^{-i\frac{\theta}{2}\sigma_x} = \cos(\theta/2) \mathds{1}-i\sin(\theta/2) \sigma_x
\end{equation}
and we use the shorthand notation `$\X{}$' for a full $\pi$ rotation about the $x$ axis (and similarly for $\Y{} := \Y{\pi}$ and $\Z{} := \Z{\pi}$).

\subsection{\label{sec:ClassicalGatesInQC}Boolean logic gates used in classical computers}
\begin{figure}[!t]
\centering
\includegraphics[width=8.6cm]{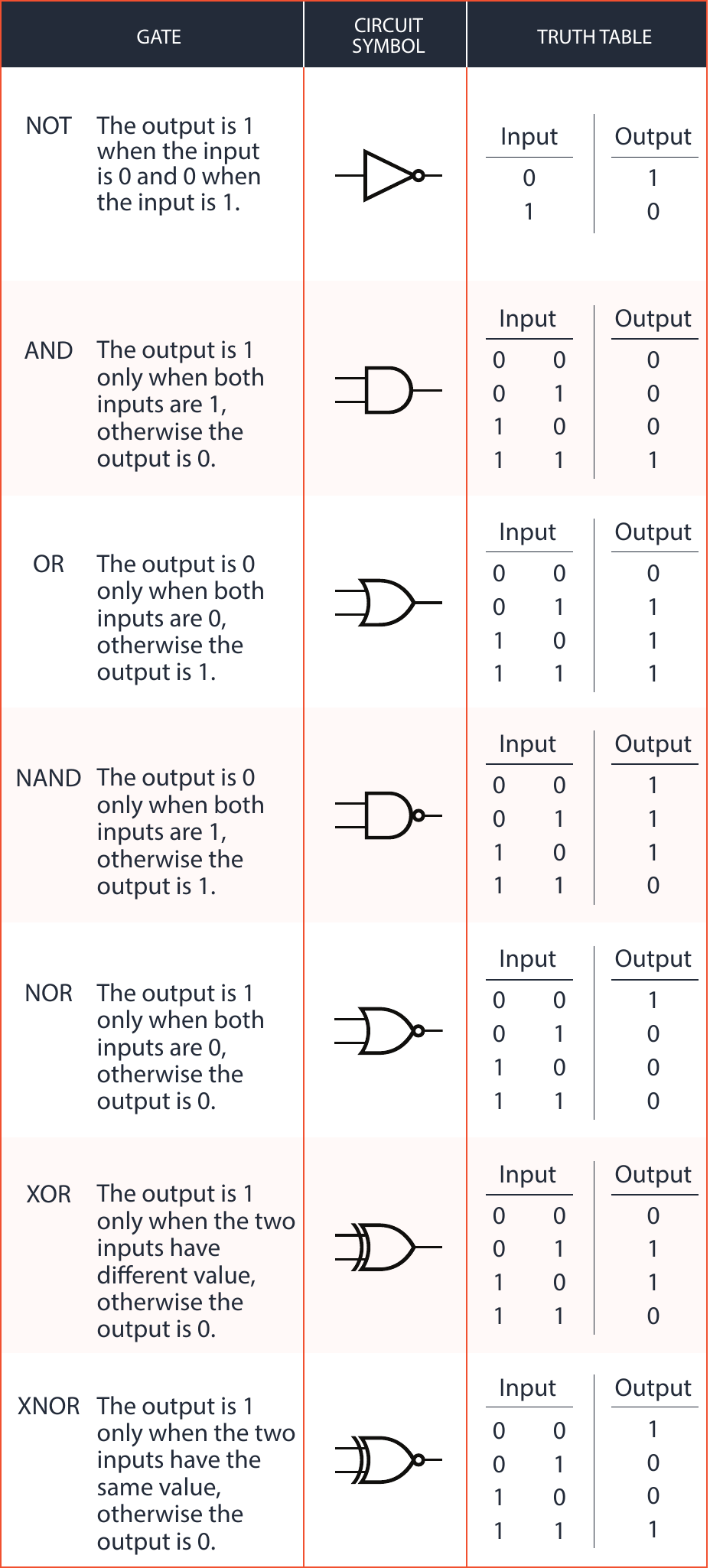}
\caption{Classical single-bit and two-bit boolean logic gates. For each gate, the name, a short description, circuit representation, and input/output truth tables are presented. The numerical values in the truth table correspond to the classical bit values $0$ and $1$. Adapted from Ref.~\onlinecite{MITxPRO}.}
\label{fig:Classical-logic-gates}
\end{figure}

Universal boolean logic can be implemented on classical computers using a small set of single-bit and two-bit gates. Several common classical logic gates are shown in Fig.~\ref{fig:Classical-logic-gates} along with their truth tables. In classical boolean logic, bits can take on one of two values: state $0$ or state $1$. The state $0$ represents logical \textsf{FALSE}, and state $1$ represents logical \textsf{TRUE}.

Beyond the trivial ``identity operation,'' which simply passes a boolean bit unchanged, the only other possible single-bit boolean logic gate is the \textsf{NOT} gate. As shown in Fig.~\ref{fig:Classical-logic-gates}, the \textsf{NOT} gate flips the bit: $0 \rightarrow 1$ and $1 \rightarrow 0$. This gate is \textit{reversible}, because it is trivial to determine the input bit value given the output bit values. As we will see, for two-bit gates, this is not the case.

\begin{figure*}[!t]
\centering
\includegraphics[width=14cm]{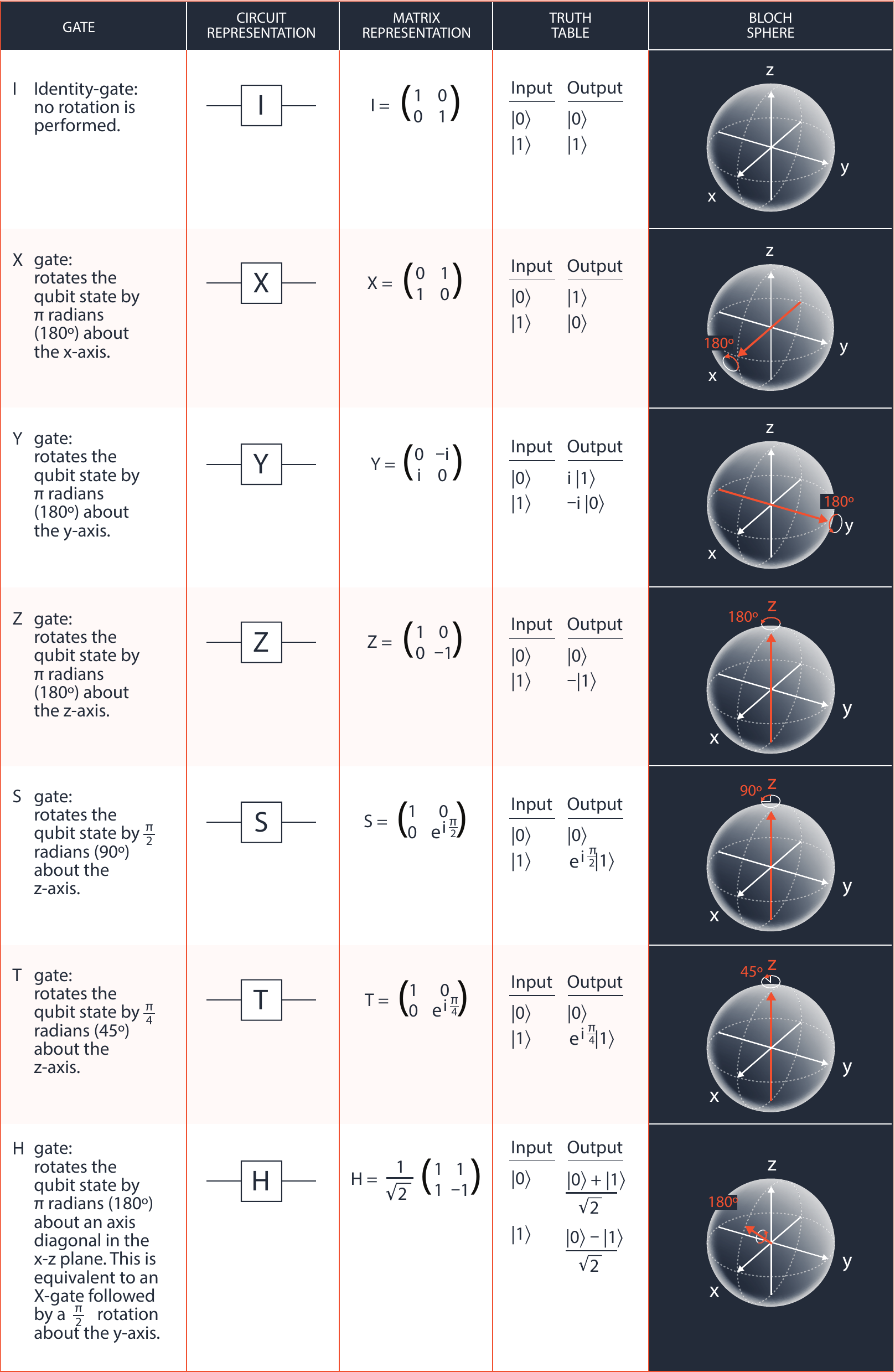}
\caption{Quantum single-qubit gates. For each gate, the name, a short description, circuit representation, matrix representation, input/output truth tables, and Bloch sphere represenation are presented. Matrices are defined in the basis spanned by the state vectors $\vert 0 \rangle \equiv [ 1 \; 0 ]^T$ and $\vert 1 \rangle \equiv [ 0 \; 1 ]^T$. The numerical values in the truth table correspond to the quantum states $\vert 0 \rangle$ and $\vert 1 \rangle$. Adapted from Ref.~\onlinecite{MITxPRO}.}
\label{fig:Quantum-logic-gates-single}
\end{figure*}

\begin{figure*}[!t]
\centering
\includegraphics[width=14.1cm]{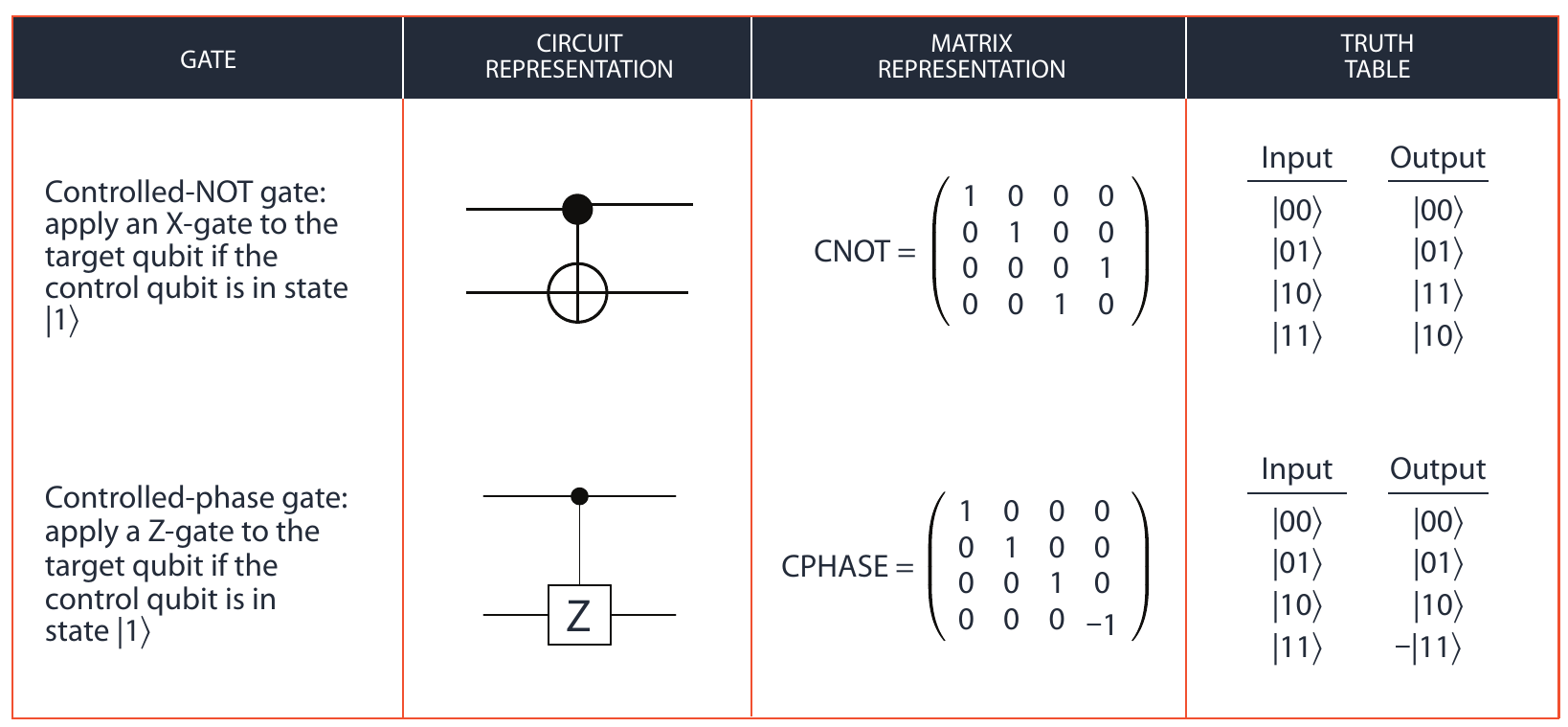}
\caption{Quantum two-qubit gates: the controlled NOT (\CNOT{}) gate and the controlled phase (\CPHASE{} or \textsf{CZ}). For each gate, the name, a short description, circuit representation, matrix representation, and input/output truth tables are presented. Matrices are defined in the basis spanned by the two-qubit state vectors $\vert 00 \rangle \equiv [ 1 \; 0 \; 0 \; 0 ]^T$, $\vert 01 \rangle \equiv [ 0 \; 1 \; 0 \; 0 ]^T$, $\vert 10 \rangle \equiv [ 0 \; 0 \; 1 \; 0 ]^T$, and $\vert 11 \rangle \equiv [ 0 \; 0 \; 0 \; 1 ]^T$, where the first qubit is the \textit{control} qubit, and the second qubit is the \textit{target} qubit. The \CNOT{} gate flips the state of the target qubit conditioned on the control qubit being in state $\vert 1 \rangle$. The \CPHASE{} gate applies a $\Z{}$ gate to the target qubit conditioned on the control qubit being in state $\vert 1 \rangle$. Adapted from Ref.~\onlinecite{MITxPRO}.}
\label{fig:Quantum-logic-gates-coupled}
\end{figure*}

There are several two-bit gates shown in Fig.~\ref{fig:Classical-logic-gates}. A two-bit gate takes two bits as inputs, and it passes as an output the result of a boolean operation. One common example is the \textsf{AND} gate, for which the output is $1$ if and only if both inputs are $1$; otherwise, the output is $0$. The \textsf{AND} gate, and the other two-bit gates shown in Fig.~\ref{fig:Classical-logic-gates}, are all examples of \textit{irreversible} gates; that is, the input bit values cannot be inferred from the output values. For example, for the \textsf{AND} gate, an output of logical $1$ uniquely identifies the input $11$, but an output of $0$ could be associated with $00$, $01$, or $10$. Once the operation is performed, in general, it cannot be ``undone'' and the input information is lost.
\noindent There are several variants of two-bit gates, including,
\begin{itemize}
    \item \textsf{AND} and \textsf{OR};
    \item \textsf{NAND} (a combination of \textsf{NOT} and \textsf{AND}) and \textsf{NOR} (a combination of \textsf{NOT} and \textsf{OR});
    \item \textsf{XOR} (exclusive \textsf{OR}) and \textsf{NXOR} (\textsf{NOT XOR}).
\end{itemize}
The \textsf{XOR} gate is interesting, because it is a \textit{parity} gate. That is, it returns a logical $0$ if the two inputs are the same values (i.e., they have the same parity), and it returns a logical $1$ if the two inputs have different values (i.e., different parity). Still, the \textsf{XOR} and \textsf{NXOR} gates are not reversible, because knowledge of the output does not allow one to uniquely identify the input bit values.

The concept of \textit{universality} refers to the ability to perform any boolean logic algorithm using a small set of single-bit and two-bit gates. A universal gate set can in principle transform any state to any other state in the state space represented by the classical bits. The set of gates which enable universal computation is not unique, and may be represented by a small set of gates. For example, the \textsf{NOT} gate and the \textsf{AND} gate together form a universal gate set. Similarly, the \textsf{NAND} gate itself is universal, as is the \textsf{NOR} gate. The efficiency with which one can implement arbitrary boolean logic, of course, depends on the choice of the gate set.

\subsection{\label{sec:QuantumGatesInQC}Quantum logic gates used in quantum computers}

Quantum logic can similarly be performed by a small set of single-qubit and two-qubit gates. Qubits can of course assume the classical states $\vert 0 \rangle$ and $\vert 1 \rangle$, at the north pole and south pole of the Bloch sphere, but they can also assume arbitrary superpositions $\alpha \vert 0 \rangle + \beta \vert 1 \rangle$, corresponding to any other position on the sphere.

Single-qubit operations translate an arbitrary quantum state from one point on the Bloch sphere to another point by rotating the Bloch vector (spin) a certain angle about a particular axis. As shown in Fig.~\ref{fig:Quantum-logic-gates-single}, there are several single-qubit operations, each represented by a matrix that describes the quantum operation in the computational basis represented by the eigenvectors of the $\sigma_z$ operator, i.e. $\vert 0 \rangle \equiv [ 1 \; 0 ]^T$ and $\vert 1 \rangle \equiv [ 0 \; 1 ]^T$.

For example, the \textit{identity gate} performs no rotation on the state of the qubit. This is represented by a two-by-two identity matrix. The $\X{}$-gate performs a $\pi$ rotation about the $x$ axis. Similarly, the $\Y{}$-gate and $\Z{}$-gate perform a $\pi$ rotation about the $y$ axis and $z$ axis, respectively. The $\textsf{S}$-gate performs a $\pi/2$ rotation about the $z$ axis, and the $\text{T}$-gate performs a rotation of $\pi/4$ about the $z$ axis. The Hadamard gate $\textsf{H}$ is also a common single-qubit gate the performs a $\pi$ rotation about an axis diagonal in the $x$-$z$ plane, see Fig.~\ref{fig:Quantum-logic-gates-single}.

Two-qubit quantum-logic gates are generally \textit{conditional} gates and take two qubits as inputs. Typically, the first qubit is the \textit{control} qubit, and the second is the \textit{target} qubit. A unitary operator is applied to the target qubit, dependent on the state of the control qubit. The two common examples shown in Fig.~\ref{fig:Quantum-logic-gates-coupled} are the controlled NOT (\CNOT{}-gate) and controlled phase (\textsf{CZ} or \CPHASE{} gate). The \CNOT{}-gate flips the state of the target qubit conditioned on the control qubit being in state $\vert 1 \rangle$. The \CPHASE-gate applies a $\Z{}$ gate to the target qubit, conditioned on the control qubit being in state $\vert 1 \rangle$. As we will shown later, the $i\textsf{SWAP}$ gate -- another two-qubit gate -- can be built from the \textsf{CNOT}-gate and single-qubit gates.
\newpage
The unitary operator of the $\CNOT{}$ gate can be written in a useful way, highlighting that it applies an $\X{}$ depending on the state of the control qubit.
\begin{equation}
U_\CNOT{} =
\begin{bmatrix}
1 & 0 & 0 & 0 \\
0 & 1 & 0 & 0 \\
0 & 0 & 0 & 1 \\
0 & 0 & 1 & 0 \\
\end{bmatrix} = |0\rangle\langle  0| \otimes \mathds{1} + |1\rangle\langle 1| \otimes \X{} \label{eq:UCNOT}
\end{equation}
and similarly for the $\CPHASE{}$ gate,
\begin{equation}
U_\CPHASE{} =
\begin{bmatrix}
1 & 0 & 0 & 0 \\
0 & 1 & 0 & 0 \\
0 & 0 & 1 & 0 \\
0 & 0 & 0 & -1 \\
\end{bmatrix} = |0\rangle\langle  0| \otimes \mathds{1} + |1\rangle\langle 1| \otimes \Z{}
\label{eq:UCPHASE}
\end{equation}
Comparing the last equality above with the unitary for the \CNOT{} [\cref{eq:UCNOT}], it is clear that the two gates are closely related. Indeed, a \CNOT{} can be generated from a \CPHASE{} by applying two Hadamard gates,
\begin{equation}
U_\CNOT{} = (\Id \otimes \H{})U_\CPHASE(\Id \otimes \H{}),
\end{equation}
since $\H{}\Z{}\H{} = \X{}$. Due to the form of \cref{eq:UCPHASE}, the \CPHASE{} gate is also denoted the \CZ{} gate, since it applies a controlled $\Z{}$ operator, by analogy with \CNOT{} (a controlled application of $\X{}$ operator). Inspection of the definition of \CPHASE{} in Fig.~\ref{fig:Quantum-logic-gates-coupled} makes no distinction between which qubit acts as the target and which as the control and, consequently, the circuit-diagram is sometimes drawn in a symmetric fashion
\begin{equation}
\CPHASE{}=\begin{minipage}{0.05\textwidth}
\begin{flushleft}
\mbox{\Qcircuit @C=0.5em @R=0.75em{
 & \ctrl{1} &  \qw \\
 & \ctrl{-1} & \qw \
}}
\end{flushleft}
\end{minipage}
\end{equation}
The \CNOT{} in terms of \CPHASE{} can then be realized as
\begin{equation}
\CNOT{}=
\begin{minipage}{0.05\textwidth}
\begin{flushleft}
\mbox{\Qcircuit @C=0.5em @R=0.75em{
 & \qw &\ctrl{1} &  \qw  & \qw \\
 & \gate{\H{}} &\ctrl{-1} & \gate{\H{}} & \qw \
}}
\end{flushleft}
\end{minipage}\label{eq:CNOTfromCPHASE}
\end{equation}

Some two-qubit gates such as \CNOT{} and \CPHASE{} are also called \textit{entangling gates}, because they can take product states as inputs and output entangled states. They are thus an indispensable component of a universal gate set for quantum logic. For example, consider two qubits $A$ and $B$ in the following state:
\begin{equation}
    \vert \psi \rangle = \frac{1}{\sqrt{2}} \left(\vert 0 \rangle + \vert 1 \rangle\right)_A \vert 0 \rangle_B.
\end{equation}
If we perform a \CNOT{} gate, $U_{\CNOT}$, on this state, with qubit A the control qubit, and qubit B the target qubit, the resulting state is (see the truth table in Fig.~\ref{fig:Quantum-logic-gates-coupled}):
\begin{equation}
    U_{\CNOT} \vert \psi \rangle = \frac{1}{\sqrt{2}} \left(\vert 0 \rangle_A \vert 0 \rangle_B + \vert 1 \rangle_A \vert 1 \rangle_B \right) \neq ( \ldots)_A ( \ldots)_B,
\end{equation}
which is a state that cannot be factored into an isolated qubit-A component and a qubit-B component. This is one of the two-qubit entangled \textit{Bell states}, a manifestly quantum mechanical state.

A universal set of single-qubit and two-qubit gates is sufficient to implement arbitrary quantum logic. This means that this gate set can in principle reach \textit{any} state in the multi-qubit state-space. How efficiently this is done depends on the choice of quantum gates that comprise the gate set. We also note that each of the single-qubit and two-qubit gates is \textit{reversible}, that is, given the output state, one can uniquely determine the input state. As we discuss further, this distinction between classical and quantum gates arises, because quantum gates are based on \textit{unitary} operations $U$. If a unitary operation $U$ is a particular gate applied to a qubit, then its hermitian conjugate $U^{\dagger}$ can be applied to recover the original state, since $U^{\dagger}U=I$ resolves an identity operation.

\subsection{\label{sec:GatesInQC}Comparing classical and quantum gates}
\begin{figure}[!t]
\centering
\includegraphics[width=8.6cm]{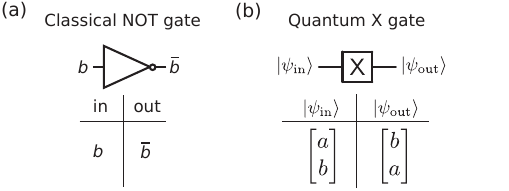}
\caption{Comparison of the classical inverter (\textsf{NOT}) gate and quantum bit flip (\textsf{X}) gate. \textbf{(a)} The classical \textsf{NOT} gate that inverts the state of a classical bit. \textbf{(b)} The quantum \textsf{X} gate, which flips the amplitudes of the two components of a quantum bit.}
\label{fig:ClassicalVQuantum_circuits}
\end{figure}

The gate-sequences used to represent quantum algorithms have certain similarities to those used in classical computing, with a few striking differences. As an example, we consider first the classical \textsf{NOT} gate (discussed previously), and the related quantum circuit version, shown in Fig.~\ref{fig:ClassicalVQuantum_circuits}.

While the classic bit-flip gate inverts any input state, the quantum bit-flip does not in general produce the antipodal state (when viewed on the Bloch sphere), but rather exchange the prefactors of the wavefunction written in the computational basis. The $\mathsf{X}$ operator is sometimes referred to as `the quantum NOT' (or `quantum bit-flip`), but we note that \textsf{X} only acts similar to the classical \textsf{NOT} gate in the case of classical data stored in the quantum bit, i.e. $\textsf{X}|g\rangle = |\bar g\rangle$ for $g\in\{0,1\}$.

As briefly mentioned in Sec.~\ref{sec:QuantumGatesInQC}, \emph{all} quantum gates are \emph{reversible}, due to the underlying unitary nature of the operators implementing the logical operations. Certain other processes used in quantum information processing, however, are irreversible. Namely, measurements (see Sec.~\ref{sec:readout} for detailed discussion) and energy loss to the the environment (if the resulting state of the environment is not known). Here, we will not consider how these processes are modeled, but refer the interested reader to e.g. Ref.~\onlinecite{NielsenChuang}, and will only consider unitary control operations throughout the rest of this section. Finally, we note that quantum circuits are written left-to-right (in order of application), while the calculation of the result of a gate-sequences, e.g the circuit
\begin{equation}
\Qcircuit @C=1em @R=2em{
 \lstick{|\psi_\text{in}\rangle} & \gate{U_0} & \gate{U_1} & \qw & \cdots & & \gate{U_n} & \rstick{|\psi_\text{out}\rangle} \qw\\
}
\label{eq:simplecirc}
\end{equation}
is performed right-to-left, i.e.
\begin{equation}
|\psi_\text{out}\rangle = U_n \cdots U_1U_0 |\psi_\text{in}\rangle.
\end{equation}
As discussed in Sec.~\ref{sec:ClassicalGatesInQC}, the \textsf{NOR} and \textsf{NAND} gates are each individually universal gates for classical computing. Since both of these gates have no direct quantum analogue (because they are not reversible), it is natural to ask which gates \emph{are} needed to build a universal quantum computer. It turns out that the ability to rotate about arbitrary axes on the Bloch-sphere (i.e. a complete single-qubit gate set), supplemented with any entangling 2-qubit operation will suffice for universality \cite{Barenco1995,NielsenChuang}. By using what is known as the `Krauss-Cirac decomposition', any two-qubit gate can be decomposed into a series of \CNOT{} operations \cite{NielsenChuang,Williams2008}.

\subsubsection{\label{sec:gatesetgatesynth}Gate sets and gate synthesis}
A common universal quantum gate set is
\begin{equation}
\mathcal G_0 = \{\X{\theta},\Y{\theta},\Z{\theta},\text{Ph}_\theta,\CNOT{}\}
\end{equation}
where Ph$_\theta = e^{i\theta} \mathds{1}$ applies an overall phase $\theta$ to a single qubit. For completeness we mention another universal gate set which is of particular interest from a theoretical perspective, namely
\begin{equation}
\mathcal G_1 = \{\H, \S, \T, \CNOT{}\},
\end{equation}
As a technical aside, we mention that the restriction to a discrete gate set still gives rise to universality. This fact relies on using the so-called Solovay-Kitaev\cite{Kitaev1997,Dawson2006} theorem, which (roughly) states that any other single-qubit gate can be approximated to an error $\epsilon$ using only $\mathcal O(\log^c (1/\epsilon))$ (where $c>0$) single-qubit gates from $\mathcal G_1$. The gate-set $\mathcal{G}_1$ is typically referred to as the `Clifford + $T$' set, where $\H$, $\S$ and \CNOT{} are all Clifford gates.

Each quantum computing architecture will have certain gates that are simpler to implement at the hardware level than others (sometimes referred to as 'native' gates of the architecture). These are typically the gates for which the Hamiltonian governing the gate-implementation gives rise to a unitary propagator that corresponds to the gate itself. We will show several examples of this in Sections \ref{sec:SWAPgates}, \ref{sec:CZgates}, and \ref{sec:CRgates}. Regardless of which gates are natively available, as long as one has a complete gate set, one can use the Solovay-Kitaev theorem to synthesize any other set efficiently. In general one wants to keep the overall number of time steps in which gates are applied (denoted the \emph{depth} of a circuit) as low as possible, and one wants to use as many of the native gates as possible, to reduce the amount of time spent synthesising. Moreover, running a quantum algorithm also depends on the qubit connectivity of the device. The process of designing a quantum gate sequence that efficiently implements a specific algorithm, while taking into account the considerations outlined above is known as \emph{gate synthesis} and \emph{gate compilation}, respectively. A full discussion of this large research effort is outside the scope of this review, but the interested reader may consult e.g. Refs. \onlinecite{Chong2017,Campbell2017,Alexandru2017} and references therein as a starting point. As a concrete (and trivial) example of how gate identities can be used, in \cref{eq:Hfrompulses} we illustrate how the Hadamard gate from $\mathcal G_1$ can be generated by two single-qubit gates (from $\mathcal G_0$) and an overall phase gate,

\begin{align}
\H &= \text{Ph}_{\frac{\pi}{2}}\Y{\frac{\pi}{2}} \Z{\pi} = i\frac{1}{\sqrt{2}} \begin{bmatrix}
1 & -1\\
1 & 1
\end{bmatrix}
\begin{bmatrix}
-i & 0\\
0 & i
\end{bmatrix}=
\frac{1}{\sqrt{2}}
\begin{bmatrix}
1 & 1\\
1 & -1
\end{bmatrix}
\label{eq:Hfrompulses}
\end{align}
As we show in Sec. \ref{sec:capacitivecoupling}, the gates $\X{\theta}$, $\Y{\theta}$ and $\Z{\theta}$ are all natively available in a superconducting quantum processor.

We now address the question of how single qubit rotations and two-qubit operations are implemented in transmon-based superconducting quantum processors.

\subsubsection{\label{sec:SingleQubitGatesMW}Addressing superconducting qubits}
The modes of addressing transmon-like superconducting qubits can roughly be split into two main categories: $i)$ Capacitive coupling between a resonator (or a feedline) and the superconducting qubit dipole-field allows for microwave control to implement single-qubit rotations (see Sec. \ref{sec:CapCouplingforDriving}) as well as certain two-qubit gates (see Sections \ref{sec:CRgates} and \ref{sec:OtherUWaveGates}). $ii)$ For flux-tunable qubits, local magnetic fields can be used to tune the frequency of individual qubits. This allows the implementation of $z$-axis single-qubit rotation as well as multiple two-qubit gates (see Sections \ref{sec:SWAPgates}, \ref{sec:CZgates} and \ref{sec:TunableCouplingimplementations}).

\subsection{\label{sec:CapCouplingforDriving}Single-qubit gates}
\begin{figure}[!t]
\centering
\includegraphics[width=8.6cm]{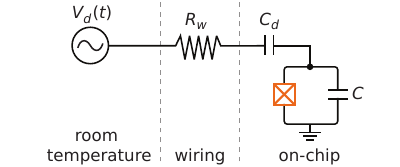}
\caption{Circuit diagram of capacitive coupling of a microwave drive line (characterized by a time-dependent voltage $V_d(t)$) to a generic transmon-like superconducting qubit.}
\label{fig:capacitivecoupling}
\end{figure}

In this section we will review the steps necessary to demonstrate that capacitive coupling of microwaves to a superconducting circuit can be used to drive single-qubit gates. To this end we consider coupling a superconducting qubit to a microwave source (sometimes referred to as a `qubit drive') as shown in \cref{fig:capacitivecoupling}(a).
A full circuit analysis of the circuit in \cref{fig:capacitivecoupling}(a) is beyond the scope of this review, so here we settle for highlighting the steps that elucidate the physics of the qubit/drive coupling. The interested reader may consult a number of lectures notes and pertinent theses (e.g. Refs. \onlinecite{Devoret1997,Vool2017,SankPhD2014,Girvin2014,LangfordCQED}). Here we follow Ref. \onlinecite{SankPhD2014}.

\subsubsection{\label{sec:capacitivecoupling}Capacitive coupling for $\X{}$,$\Y{}$ control}
We start by modeling the qubit as an harmonic oscillator, for which the (classical) circuit Hamiltonian can be calculated by circuit quantization techniques, starting from Kirchoffs laws, and is given by \cite{SankPhD2014}
\begin{equation}
H = \frac{\tilde Q(t)^2}{2C_\Sigma}+\frac{\Phi^2}{2L}+\frac{C_\text{d}}{C_\Sigma}V_\text{d}(t)\tilde Q,
\end{equation}
where $C_\Sigma = C+C_\text{d}$ is the total capacitance to ground and $\tilde Q = C_\Sigma \dot \Phi - C_\text{d}V_\text{d}(t)$ is a renormalized charge variable for the circuit. We can now promote the flux and charge variables to quantum operators and assume weak coupling to the drive-line, so that $\tilde Q \approx \hat Q$, and arrive at
\begin{equation}
H = H_\text{LC} + \frac{C_\text{d}}{C_\Sigma}V_\text{d}(t) \hat Q \label{eq:quantumLCwithDrive},
\end{equation}
where $H_\text{LC} = \hat Q^2/(2C) + \hat \Phi^2/(2L)$ and we have kept only terms that couple to the dynamic variables. Similar to the momentum operator for a harmonic oscillator in $(x,p)$--space, we can express the charge variable in terms of raising and lowering operators, as done in Sec. \ref{sec:circuits}
\begin{equation}
 \hat Q = -iQ_\text{zpf}\left(a-a^\dagger \right)
\end{equation}
where $Q_\text{zpf} = \sqrt{\hbar/2Z}$ is the zero-point charge fluctations and $Z=\sqrt{L/C}$ is the impedance of the circuit to ground. Thus, the $LC$ oscillator capacitively coupled to a drive line can be written as,
\begin{equation}
H = \omega \left(a^\dagger a + \frac{1}{2}\right) - \frac{C_\text{d}}{C_\Sigma}V_d(t) iQ_\text{zpf}\left(a-a^\dagger \right). \label{eq:DriveNlevels}
\end{equation}
Finally, by truncating to the lowest transition of the oscillator we can make the replacement $a \rightarrow \sigma^-$ and $a^\dagger \rightarrow \sigma^+$ throughout and arrive at
\begin{equation}
H = \underbrace{-\frac{\freq}{2}\sigma_z}_{H_0}+\underbrace{\vphantom{\frac{1}{2}}\Omega V_\text{d}(t)\sigma_y}_{H_\d} \label{eq:DriveHamLabFrame}
\end{equation}
where $ \Omega = (C_\text{d}/C_\Sigma)Q_\text{zpf}$ and $\freq = (E_1 - E_0)/\hbar$.\footnote{Starting from a generic qubit Hamiltonian, $H_0 = E_0 |0\rangle\langle 0| + E_1|1\rangle\langle 1|$, we can rewrite in terms of Pauli matrices, and get $H_0 = ((E_0+E_1)/2) \mathds{1} - ((E_1-E_0)/2) \sigma_z$. In the main text we have ignored the constant offset term.}

To elucidate the role of the drive, we move into a frame rotating with the qubit at frequency $\freq$ (also denoted `the rotating frame' or the `the interaction frame'). To see the usefulness of this rotating frame, consider a state $|\psi_0\rangle = ( 1 \quad 1)^T/\sqrt{2}$. By the time-dependent Schr\"odinger equation this state evolves according to
\begin{equation}
|\psi_0(t)\rangle = U_{H_0}|\psi_0\rangle = \frac{1}{\sqrt{2}}\begin{pmatrix} e^{i\freq t/2} \\ e^{-i\freq t/2} \end{pmatrix},
\end{equation}
where $U_{H_0}$ is the propagator corresponding to $H_0$. By calculating e.g. $\langle \psi_0| \sigma_x|\psi_0\rangle = \cos(\freq t)$ it is evident that the phase is winding with a frequency of $\freq$ due to the $\sigma_z$ term. By going into a frame rotating with the qubit at frequency $\freq$, the action of the drive can be more clearly appreciated. To this end we define $U_\rf = e^{iH_0 t} = U_{H_0}^\dagger$ and the new state in the rotating frame is $|\psi_\rf(t)\rangle = U_\rf|\psi_0\rangle$. The time-evolution in this new frame is again found from the Schr\"odinger equation (using the shorthand $\partial_t = \partial / \partial t$),
\begin{align}
i \partial_t |\psi_\rf(t)\rangle &= i(\partial_tU_\rf)|\psi_0\rangle + i U_\rf\left(\partial_t|\psi_0\rangle\right)\\
 &= i\dot U_\rf U_\rf^\dagger|\psi_\rf\rangle + U_\rf H_0 |\psi_0\rangle \\
 &= \underbrace{\left(i\dot U_\rf U_\rf^\dagger  + U_\rf H_0 U^\dagger_\rf\right)}_{\widetilde H_0}|\psi_\rf\rangle \label{eq:RotFrameSchrodinger}
\end{align}
We can think of the term $\widetilde H_0$ in the parentheses in \cref{eq:RotFrameSchrodinger} as the form of $H_0$ in the rotating frame. Simple insertion shows that $\widetilde H_0 = 0$ as expected (the rotating frame should take care of the time-dependence). However, one could also think of the term in brackets in \cref{eq:RotFrameSchrodinger} as a prescription for calculating the form of any Hamiltonian in the rotating frame given by $U_\rf$, by replacing $H_0$ with some other $H$. In general, we will not find $\widetilde H = 0$.

\begin{figure}[!t]
\begin{center}
\includegraphics[width=8.6cm]{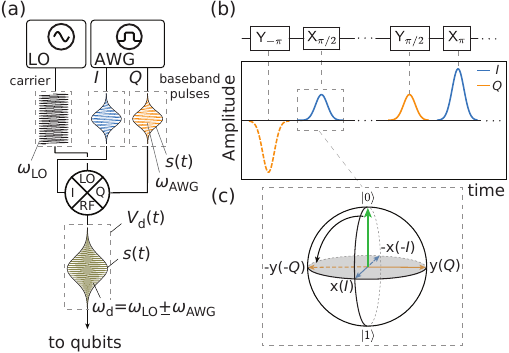}
\caption{\textbf{(a)} Schematic of a typical qubit drive setup. A microwave source supplies a high-frequency signal ($\omega_\textsf{LO}$), while an arbitrary waveform generator (AWG) supplies a pulse-envelope ($s(t)$), sometimes with a low frequency component, $\omega_\textsf{AWG}$, generated by the AWG. The IQ-mixer combines the two signals to generate a shaped waveform $V_d(t)$ with a frequency $\omega_d = \omega_\textsf{LO} \pm \omega_\textsf{AWG}$, typically resonant with the qubit. \textbf{(b)} Example of how a gate sequence is translated into a waveform generated by the AWG. Colors indicate $I$ and $Q$ components. \textbf{(c)} The action of a $\X{\pi/2}$ pulse on a $|0\rangle$ state to produce the $|\!-\!i\rangle = \frac{1}{\sqrt{2}}\left(|0\rangle -i |1\rangle \right)$ state.}
\label{Fig:SignalsPulses}
\end{center}
\end{figure}

Returning to \cref{eq:DriveHamLabFrame}, the form of $H_\text{d}$ in the rotating frame is found to be
\begin{equation}
\widetilde H_\text{d} = \Omega V_\text{d}(t)\left(\cos(\omega_\text{q} t)\sigma_y - \sin(\omega_\text{q} t)\sigma_x\right).
\label{eq:RotatingFrame_drive}
\end{equation}
We can in general assume that the time-dependent part of the voltage ($V_\d(t) = V_0v(t)$) has the generic form
\begin{align}
v(t) & = s(t) \sin(\omega_\d t + \phi) \\
& = s(t)\left(\cos(\phi)\sin(\omega_\d t) + \sin(\phi)\cos(\omega_\d t) \right), \label{eq:IQdef}
\end{align}
where $s(t)$ is a dimensionless envelope function, so that the amplitude of the drive is set by $V_0s(t)$. Adopting the definitions
\begin{align}
I &= \cos(\phi) \text{ (the `in-phase' component)}\\
Q &= \sin(\phi) \text{ (the `out-of-phase' component)}
\end{align}
the driving Hamiltonian in the rotating frame takes the form
\begin{align}
\widetilde H_\text{d} &= \Omega V_0 s(t)\left( I \sin(\omega_\d t) - Q \cos(\omega_\d t)\right)\quad \nonumber\\
& \hphantom{=} \times \left(\cos(\freq t)\sigma_y - \sin(\freq t)\sigma_x\right)
\end{align}
Performing the multiplication and dropping fast rotating terms that will average to zero (i.e. terms with $\freq + \omega_d$), known as the rotating wave approximation (RWA), we are left with
\begin{align}
\widetilde H_\text{d}  & = \frac{1}{2} \Omega V_0 s(t) \left[\vphantom{\frac{2}{2}} \left(-I\cos(\omegad t) + Q\sin(\omegad t)\right)\sigma_x \right. \nonumber \\
& \hphantom{=} + \left. \left(I\sin(\omegad t) - Q\cos(\omegad t)\right)\sigma_y \vphantom{\frac{2}{2}}\right]
\end{align}
where $\omegad = \freq - \omega_\d$. Finally, by re-using the definitions from \cref{eq:IQdef}, the driving Hamiltonian in the rotating frame using the RWA can be written as
\begin{equation}
\widetilde H_\text{d} = -\frac{\Omega}{2}V_0s(t) \begin{pmatrix}
0 & e^{i(\omegad t + \phi)}\\
e^{-i(\omegad t + \phi)} & 0
\end{pmatrix} .\label{eq:finalDriveHam}
\end{equation}
Equation (\ref{eq:finalDriveHam}) is a powerful tool for understanding single-qubit gates in superconducting qubits. As a concrete example, assume that we apply a pulse at the qubit frequency, so that $\omegad = 0$, then
\begin{equation}
\widetilde H_\text{d} = -\frac{\Omega}{2}V_0s(t)\left(I\sigma_x +Q\sigma_y \right),
\end{equation}
showing that an \emph{in-phase} pulse ($\phi =0$, i.e. the $I$-component) corresponds to rotations around the $x$-axis, while an out-of-phase pulse ($\phi=\pi/2$, i.e. the $Q$-component), corresponds to rotations about the $y$-axis. As a concrete example of an in-phase pulse, writing out the unitary operator yields
\begin{equation}
U^{\phi = 0}_\text{rf,d}(t) = \exp\left( \left[\frac{i}{2}\Omega  V_0 \int_0^{t} s(t')\d t'  \right]\sigma_x \right), \label{eq:drivingPropagator}
\end{equation}
which depends only on the macroscopic design parameters of the circuit as well as the envelope of the baseband pulse $s(t)$ and amplitude $V_0$, which can both be controlled using arbitrary waveform generators (AWGs). Equation (\ref{eq:drivingPropagator}) is known as \emph{Rabi driving} and can serve as a useful tool for engineering the circuit parameters needed for efficient gate operation (subject to the available output voltage $V_0$). To see this we define the shorthand
\begin{equation}
\Theta(t) = -\Omega V_0\int_0^{t} s(t')\d t'
\end{equation}
which is the angle by which a state is rotated given the capacitive couplings, the impedance of the circuit, the magnitude $V_0$, and the waveform envelope, $s(t)$. This means that to implement a $\pi$-pulse on the $x$-axis one would solve the equation $\Theta(t) = \pi$ and output the signal in-phase with the qubit drive. In this language, a sequence of pulses (see \cref{Fig:SignalsPulses}(a)) $\Theta_k, \Theta_{k-1},...\Theta_0$ is converted to a sequence of gates operating on a qubit as
\begin{equation}
U_k\cdots U_1U_0 = \mathcal T\prod_{n=0}^k  e^{\left[-\frac{i}{2} \Theta_n(t)\left(I_n \sigma_x + Q_n \sigma_y\right)\right]}, \label{eq:sequencesandwaveforms}
\end{equation}
where $\mathcal T$ is an operator that ensures the pulses are generated in the time-ordered sequence corresponding to $U_k\cdots U_1U_0$.

In \cref{Fig:SignalsPulses} we outline the typical IQ modulation setup used to generate the pulses used in \cref{eq:sequencesandwaveforms}. \cref{Fig:SignalsPulses}(a) shows how a pulse at frequency $\omega_\text{d}$ is generated using a low phase-noise microwave generator (typically denoted `the local oscillator (LO)'), while the pulse is shaped by combining the LO signal in an IQ mixer with pulses generated in an AWG. To allow for frequency multiplexing, the AWG signal will typically be generated with a low-frequency component, $\omega_\text{AWG}$, and the LO signal will be offset, so that $\omega_\text{LO} + \omega_\text{AWG} = \omega_\text{d}$. By mixing in more than one frequency $\omega_\text{AWG1}, \omega_\text{AWG2}, ...$ it is possible to address multiple qubits (or readout resonators) simultaneously, via the superposition of individual drives.

The $I$ ($Q$) input of the $IQ$ mixer will multiply the baseband signal to the in-phase (out-of-phase) component of the LO. In \cref{Fig:SignalsPulses}(b) we schematically show the comparison between $XY$ gates in a quantum circuit and the corresponding waveforms generated in the AWG (omitting for clarity the frequency $\omega_\text{AWG}$ component). The inset in \cref{Fig:SignalsPulses}(b) shows an example of a gate on the Bloch sphere, with indication of $(I,Q)$ axes. More sophisticated and compact approaches exist to reduce the hardware needed for $XY$ qubit control, relative to the setup shown in \cref{Fig:SignalsPulses}, see e.g. \cite{Vesterinen2014,Asaad2016,Ryan2017}.

\subsubsection{\label{sec:EfficientZgateOperation}Virtual $\Z{}$ gate}
As we saw in Sec. \ref{sec:CapCouplingforDriving}, the distinction between $x$-- and $y$--rotations was merely a choice of phase on the microwave signals, and the angle to be rotated is given by $\Theta(t)$, both of which are generated using an AWG. Since the choice of phase $\phi$ has an arbitrary starting point, we could consider $\phi \rightarrow \phi+\pi/2$. This would lead to $I \rightarrow Q$ and $Q \rightarrow -I$. Therefore, changing the phase effectively changes rotations around $x$ to rotations around $y$ (and vice-versa, with a change of sign). This is reminiscent of the result of applying a $\Z{\pi}$ rotation to $x$-- and $y$--rotations, where $\Z{\pi} \X{\pi} = i\Y{\pi}$ and $\Z{\pi} \Y{\pi} = - i\X{\pi}$. This analogy between shifting a phase of an AWG-generated signal and applying $\Z{}$ rotations can be utilized to implement \emph{virtual} $\Z{}$ gates\cite{McKay2017}. As shown by McKay \emph{et al.}, this intuition can be formalized via the following example: consider the case of applying a pulse with an angle $\theta$ on the $I$ channel (i.e. a $\X{\theta}$) followed by another $\theta$ pulse on the $I$ channel, but with a phase $\phi_0$ relative to the first pulse (denoted $\X{\theta}^{(\phi_0)}$, where $\X{}$ indicates we still use the $I$ channel, but the rotation axis is now an angle $\phi_0$ away from the $x$-axis). Using \cref{eq:sequencesandwaveforms} this corresponds to a pulse sequence
\begin{align}
\X{\theta}^{(\phi_0)}\X{\theta} &= e^{-i\frac{\theta}{2}\left(\cos(\phi_0)\sigma_x + \sin(\phi_0) \sigma_y\right)}\X{\theta}\\
& = \Z{-\phi_0}\X{\theta}\Z{\phi_0}\X{\theta}
\end{align}
from which we see that the effect of the offset phase $\phi_0$ is to apply $\Z{\phi_0}$. The equality above can be verified with a little trigonometric footwork. The final $\Z{-\phi_0}$ is due to the rotation being in the frame of reference of the qubit. However, since readout is along $z$-axis (see Sec. \ref{sec:readout}), a final phase rotation about $z$ will not change the measurement outcome. Thus, if one wants to to implement the gate sequence
\begin{equation}
\Qcircuit @C=0.5em @R=0.75em{
\cdots & & & \gate{U_i} & \gate{\Z{\theta_0}} & \gate{U_{i+1}} & \gate{\Z{\theta_1}} & \gate{U_{i+2}}  & \qw & & \cdots}
\end{equation}
where $U_i$'s are arbitrary gates, this can be done by revising the gate sequence (in the control software for the AWG) and changing the phase of subsequent pulses
\begin{equation}
\Qcircuit @C=0.5em @R=0.75em{
\cdots & & &  \gate{U_i \vphantom{U^{(\theta_0)}_{i+1}}}  & \gate{U^{(\theta_0)}_{i+1}} & \gate{U^{(\theta_0+\theta_1)}_{i+2}}  & \qw & & \cdots}
\end{equation}
which reduces the number of overall gates. Moreover, the virtual-$\Z{}$ gates are ``perfect", in the sense that no additional pulses are required, and the gate takes ``zero time", and thus the gate fidelity is nominally unity. As we show in Sections \ref{sec:SWAPgates} and \ref{sec:CZgates}, operation of two-qubit gates can incur additional single-qubit phases. Using the virtual-$\Z{}$ strategy, these phases can be cancelled out, leaving a pure two-qubit interaction.

Finally we mention one more salient feature of the virtual-$\Z{}$ gates. As shown in Ref.\onlinecite{McKay2016}, any single-qubit operation (up to a global phase) can be written as
\begin{equation}
U(\theta,\phi,\lambda) = \Z{\phi-\frac{\pi}{2}}\X{\frac{\pi}{2}}\Z{\pi-\theta} \X{\frac{\pi}{2}}\Z{\lambda-\frac{\pi}{2}}, \label{eq:anySU2}
\end{equation}
for appropriate choice of angles $\theta, \phi, \lambda$. This means that access to a single physical $\X{\frac{\pi}{2}}$ combined with the virtual-$\Z{}$ gives access to a complete single qubit gate set! An explicit example of \cref{eq:anySU2} in action is the Hadamard gate, which can be written as $\H = \Z{\frac{\pi}{2}} \X{\frac{\pi}{2}} \Z{\frac{\pi}{2}}$, but since the $\Z{}$'s can be virtual, it is possible to implement Hadamards as an effective single pulse operation in superconducting qubits.

\subsubsection{\label{sec:DRAG}The DRAG scheme}
\begin{figure}[!t]
\begin{center}
\includegraphics[width=8.6cm]{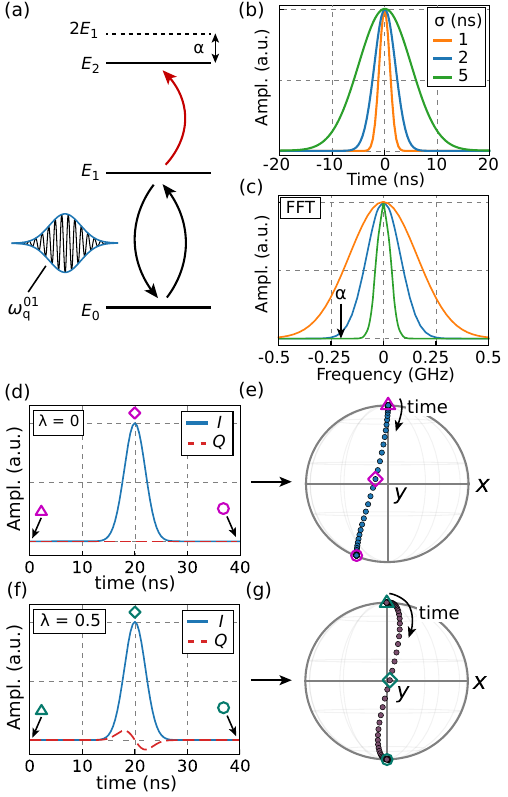}
\caption{\textbf{(a)} Schematic level diagram of a weakly anharmonic transmon qubit subjected to a drive at transition frequency $\omega_\d = \freq$. \textbf{(b)} Gaussian waveform with standard deviation $\sigma$. \textbf{(c)} Fourier transform of (b) showing how the short pulse lengths lead to significant overlap with the $\omega_\text{q}^{1\rightarrow2}$ transition, separated from $\omega_\text{q}$ by the anharmonicity $\alpha$. \textbf{(d)} Waveform of a $\X{\pi}$ pulse without DRAG modulation. \textbf{(e)} Effect of the waveform from (d) on a qubit initialized in the $|0\rangle$ state with $\alpha = -200$~MHz and $\omega_\text{q} = 4$~GHz. The dephasing error is visible as a deviation from the $|1\rangle$ after the pulse. \textbf{(f)} Waveform of a $\X{\pi}$ pulse with DRAG modulation for a qubit with anharmonicity $\alpha = -200$~MHz and DRAG parameter $\lambda = 0.5$ to cancel dephasing errors (see text for details). \textbf{(g)} Effect of the waveform from (f) on the same qubit as (e). Calculated using \textsf{mesolve} in the software package \textsf{QuTiP}\cite{Johansson2013}.}
\label{fig:DRAG}
\end{center}
\end{figure}

In going from \cref{eq:DriveNlevels} to \cref{eq:DriveHamLabFrame} we assumed we could ignore the higher levels of the qubit. However, for weakly anharmonic qubits, such as the transmon (see Sec. \ref{sec:circuits}), this may not be a justified assumption, since $\omega^{1\rightarrow 2}_\text{q}$ only differs from $\omega_\text{q} (\equiv \omega_\text{q}^{0\rightarrow 1})$ by the anharmonicity, $\alpha = \omega^{1\rightarrow 2}_\text{q} - \omega_\text{q}$, which is negative and typically around $200$ to $300$~MHz. This situation is sketched in \cref{fig:DRAG}(a-c), where we illustrate how Gaussian pulses with standard deviations $\sigma = \{1,2,5\}$~ns have spectral content that leads to non-zero overlaps with the $\omega^{1\rightarrow2}_\text{q} = \omega_\text{q}-|\alpha|$ frequency.
This leads to two deleterious effects: (\emph{1}) leakage errors which take the qubit out of the computational subspace, and (\emph{2}) phase errors. Effect 1 can occur because a qubit in the state $|1\rangle$ may be excited to $|2\rangle$ as a $\pi$ pulse is applied, or be excited directly from the $|0\rangle$, since the qubit spends some amount of time in the $|1\rangle$ state during the $\pi$ pulse. Effect 2 occurs because the presence of the drive results in a repulsion between the $|1\rangle$ and $|2\rangle$ levels, in turn changing $\omega_\text{q}^{0\rightarrow 1}$ as the pulse is applied. This leads to the accumulation of a relative phase between $|0\rangle$ and $|1\rangle$ \cite{ChenPhD2018}. The so-called DRAG procedure\cite{Motzoi2009,Gambetta2011,De2015} (Derivative Reduction by Adiabatic Gate) seeks to combat these two effects by applying an extra signal in the out-of-phase component. The trick is to modify the waveform envelope $s(t)$ according to
\begin{equation}
\renewcommand*{\arraystretch}{1.3}
s(t)\rightarrow s'(t) =
\left\{\begin{array}{ccl}
s(t) & \text{ on } & I\\
\displaystyle \lambda\frac{\dot s(t)}{\alpha} & \text{ on } & Q\\
\end{array}\right., \label{eq:DRAG_waveform}
\end{equation}
where $\lambda$ is a dimensionless scaling parameter, and $\lambda = 0$ correspond to no DRAG pulse and $\dot{s}(t)$ is the time derivative of $s(t)$. The theoretically optimal choice for reducing dephasing error is $\lambda = 0.5$ and an optimal choice for reducing leakage error is $\lambda = 1$\cite{Gambetta2011,Motzoi2013}. Interchanging $I$ and $Q$ in \cref{eq:DRAG_waveform} corresponds to DRAG pulsing for the $Q$ component.

In practice there can be a deviation from these two optimal values, often due to pulse distortions in the lines leading to the qubits. Typically, randomized benchmarking experiments combined with single-shot measurements (see Sec. \ref{sec:readout}) of the $|2\rangle$ state is used to determine the optimal value of $\lambda$. The $\lambda=\{0.5,1\}$ tradeoff was demonstrated explicitly in \cite{Chen2016,McKay2017}. However, by extending the original DRAG pulse implementation\cite{Chow2010,Lucero2010}, it is is possible to reduce \emph{both} errors \emph{simultaneously}. By introducing a frequency detuning parameter $\delta f$ to the waveform\cite{Gambetta2011} (defined such that $\delta f = 0$ corresponds to qubit frequency), i.e.
\begin{equation}
s'_{\delta f}(t) = s'(t)e^{i2\pi\delta f t},
\end{equation}
and choosing $\lambda$ to minimize leakage errors, then phase errors can be reduced simultaneously\cite{Chen2016}. Similarly, by a judicious use of the virtual-$\Z{}$ gate, it is also possible to reduce phase errors in combination with DRAG pulsing to reduce leakage \cite{McKay2017}. Modern single-qubit gates using DRAG pulsing now routinely reach fidelities $F_\text{1qb} \gtrsim 0.99$ \cite{Gustavsson2013,Barends2014,Sheldon2016,Sheldon2016a,Chen2016,Rol2017,Reagor2018}. Other techniques also exist for operating single-qubit gates in a spectrally crowded device \cite{Schutjens2013,Theis2016}.

\subsection{\label{sec:SWAPgates}The $i$\textsf{SWAP} two-qubit gate in tunable qubits}
As briefly mentioned in Sec. \ref{sec:GatesInQC}, single-qubit gates supplemented with an entangling two-qubit gate can form the gate set required for universal quantum computation. The two-qubit gates available in the transmon-like superconducting qubit architecture can roughly be split into two broad families as outlined previously: one group requiring local magnetic fields to tune the transition frequency of qubits and one group consisting of all-microwave control. There exist several hybrid schemes that combine various aspects of these two categories and, in particular, the notions of tunable coupling and parametric driving are proving to be important ingredients in modern superconducting qubit processors\cite{Niskanen2007,VanderPloeg2007,Allman2010,Srinivasan2011,Sirois2015,Chen2014,McKay2016,Caldwell2017,Casparis2018,Yan2018,Didier2018,Reagor2018}. In this section, however, we start by introducing the \iSWAP{} gate, and then review the \CPHASE{} (controlled-phase) in Section \ref{sec:CZgates} and the \CR{} (cross-resonance) in Section \ref{sec:CRgates}. We briefly review a few other two-qubit gates and discuss their merits in Sections \ref{sec:OtherUWaveGates} and \ref{sec:TunableCouplingimplementations}.

\subsubsection{Deriving the $i$\textsf{SWAP} unitary}
As we saw in Sec. \ref{sec:circuits}, Eq.\ref{Eq:HsystDiagCtrunc} the interaction term between two capacitively coupled qubits (in the two-level approximation) is given by
\begin{equation}
H_\text{qq} = g \sigma_{y1} \otimes \sigma_{y2}, \label{eq:Hqqsigmay}
\end{equation}
where $g$ is the coupling strength and $\otimes$ is used to emphasize  the tensor product. If the capacitive coupling is mediated through a bus resonator, then \cite{Blais2004,Majer2007}
\begin{equation}
g\rightarrow g_\text{q-r-q} = \frac{g_1g_2(\Delta_1+\Delta_2)}{2\Delta_1\Delta_2}, \label{eq:gqrq}
\end{equation}
where $g_i$ is the resonator coupling to qubit $i$ (dependent on the qubit-resonator coupling capacitance $C_{\text{q}i\text{r}}$) and $\Delta_i = \omega_{\text{q}i} - \omega_\text{r}$ is the detuning of qubit $i$ to the resonator. In the simpler case where the qubits are directly coupled\cite{Wendin2007},
\begin{equation}
g\rightarrow g_\text{q-q} = \frac{1}{2}\sqrt{\omega_{\text{q}1}\omega_{\text{q}2}}\frac{C_\text{q-q}}{\sqrt{C_\text{q-q}+C_1}\sqrt{C_\text{q-q}+C_2}}, \label{eq:gqq}
\end{equation}
where $C_\text{q-q}$ is the qubit-qubit coupling capacitance and $C_i$ is the capacitance of qubit $i$. Throughout this section, we will assume a direct capacitive coupling between qubits of the flux-tunable transmon type, so that $g = g_\text{q-q}$ and $\omega_{\text{q}i} \rightarrow \omega_{\text{q}i}(\Phi_i)$. For simplicity, we suppress the explicit flux dependence of the $\omega_{\text{q}i}$'s and simply refer to the coupling as $g$. Equation (\ref{eq:Hqqsigmay}) can be rewritten as
\begin{equation}
H_\text{qq} = -g\left([\sigma^+ - \sigma^-]\otimes[\sigma^+ - \sigma^-]\right),
\end{equation}
and then using the rotating wave approximation again (\emph{i.e.} dropping fast rotating terms) we arrive at
\begin{equation}
H_\text{qq} = g\left(e^{i\delta\omega_{12}t}\sigma^+\sigma^- + e^{-i\delta\omega_{12}t}\sigma^-\sigma^+\right),
\end{equation}
where we have introduced the notation $\delta \omega_{12} = \omega_{\text{q}1} - \omega_{\text{q}2}$ and suppressed the explicit tensor product between qubit subspaces. If we now change the flux of qubit 1 to bring it into resonance with qubit 2 ($\omega_{\text{q}1} = \omega_{\text{q}2}$), then
\begin{equation}
H_\text{qq} = g\left( \sigma^+\sigma^- + \sigma^-\sigma^+\right) = \frac{g}{2}\left(\sigma_x\sigma_x + \sigma_y\sigma_y \right). \label{eq:Hxxyy}
\end{equation}
The first part of \cref{eq:Hxxyy} shows that a capacitive interaction leads to a swapping of excitations between the two qubits, giving rise to the `swap' in \iSWAP{}. Moreover, due to the last part of \cref{eq:Hxxyy}, this capacitive coupling is also sometimes said to give rise to an `$XY$' interaction\cite{Schuch2003}. The unitary corresponding to a $XY$ (swap) interaction is
\begin{equation}
U_{qq}(t) = e^{-i\frac{g}{2}\left(\sigma_x\sigma_x + \sigma_y\sigma_y \right)t}=
\begin{bmatrix}
1 & 0 & 0 & 0 \\
0 & \cos(gt) & -i\sin(gt) & 0\\
0 & -i\sin(gt) & \cos(gt) & 0\\
0 & 0 & 0 & 1 \\
\end{bmatrix} \label{eq:Uqq}
\end{equation}
Since the qubits are tunable in frequency, we can now consider the effect of tuning the qubits into resonance for a time $t' = \frac{\pi}{2g}$,
\begin{align}
U_{qq}\left(\frac{\pi}{2g}\right) = \begin{bmatrix}
1 & 0 & 0 & 0 \\
0 & 0 & -i & 0\\
0 & -i & 0 & 0\\
0 & 0 & 0 & 1 \\
\end{bmatrix} \equiv i\textsf{SWAP}.
\end{align}
From this result, we see that a capacitive coupling between qubits turned on for a time $t'$ (inversely related to the coupling strength in units of radial frequency) leads to implementing a so called `$i\textsf{SWAP}$' gate\cite{Wendin2007,Steffen2006,Majer2007,Bialczak2010,Neeley2010,Dewes2012}, which acts to swap an excitation between the two qubits, and add a phase of $i = e^{i\pi/2}$. For completeness, we note that for $t'' = \frac{\pi}{4g}$ the resulting unitary,
\begin{align}
U_{qq}\left(\frac{\pi}{4g}\right) = \begin{bmatrix}
1 & 0 & 0 & 0 \\
0 & \sfrac{1}{\sqrt{2}} & -\sfrac{i}{\sqrt{2}} & 0\\
0 & -\sfrac{i}{\sqrt{2}} & \sfrac{1}{\sqrt{2}} & 0\\
0 & 0 & 0 & 1 \\
\end{bmatrix} \equiv \sqrt{i\textsf{SWAP}},
\end{align}
is typically referred to as the `squareroot-$i\textsf{SWAP}$' gate. The $\sqrt{\iSWAP{}}$ gate can be used to generate Bell-like superposition states, e.g. $|01\rangle + i |10\rangle$.

\begin{figure}[!t]
\begin{center}
\includegraphics[width=0.5\textwidth]{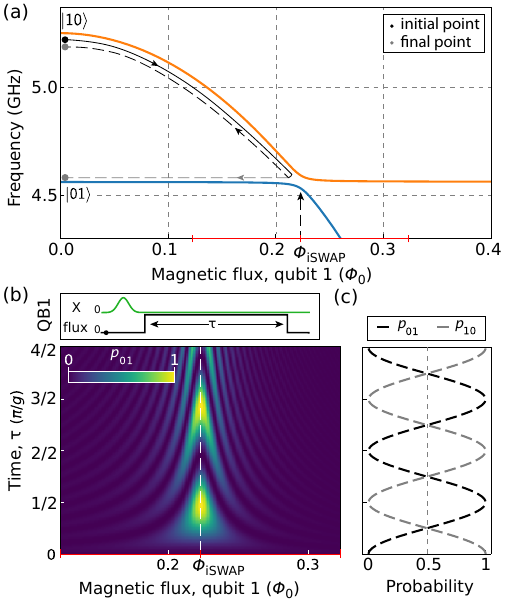}
\caption{\textbf{(a)} Spectrum of two transmon qubits (written in the combined basis as $|\text{QB1},\text{QB2}\rangle$) as the local flux through the loop of qubit 1 is increased. Black/dashed lines with arrows indicate a typical flux trajectory to demonstrate operation of \iSWAP{} gate. \textbf{(b)} Probability of swapping into the $|01\rangle$ state as a function of time and flux. The pulse sequence corresponds to preparing $|10\rangle$ and performing a typical \iSWAP{} operation (for a time $\tau$). \textbf{(c)} Probabilities of $|01\rangle$ (black) and $|10\rangle$ (gray) at $\Phi = \Phi_{\iSWAP{}}$ (white dashed line in (b)) as the time spent at the operating point ($\tau$) is increased. This simulation does not include any decay effects.}
\label{Fig:2qb_iSWAP}
\end{center}
\end{figure}

To elucidate the operating principle behind an $\iSWAP{}$ implementation we show the spectrum of a flux-tunable qubit using typical transmon-like parameters in \cref{Fig:2qb_iSWAP}(a). The \iSWAP{} is performed at the avoided crossing where $\Phi = \Phi_{\iSWAP{}}$. By preparing QB1 in state $|1\rangle$, moving into the avoided crossing, waiting there for a time $\tau$ (see pulse-sequence in inset in \cref{Fig:2qb_iSWAP}(b)), the excitation is swapped back and forth between the two qubits, as shown in \cref{Fig:2qb_iSWAP}(b). In \cref{Fig:2qb_iSWAP}(c), we plot linecuts of (b) at $\Phi_{\iSWAP{}}$, showing the excitation oscillating back and forth between $|01\rangle$ and $|10\rangle$ with the predicted time $t' = \pi/2g$. In turn, the frequency of the oscillation can be used to extract the strength of the coupling, $\frac{2}{t'} = \frac{g}{\pi}$.

So far we have ignored the role of the single-qubit phases acquired by tuning the qubit frequency. Referring to the pulse-sequence shown in the top panel of ~\cref{Fig:2qb_iSWAP}(a), we see that each qubit will acquire a phase given by
\begin{equation}
\theta_z = \int_0^\tau \d t \left( \omega_\text{q} - \omega(t)\right).
\end{equation}
This phase can be conveniently removed either by subsequent application of virtual-$\Z{}$ gates to all following pulses\cite{McKay2017}, or by shaping the waveform of the excursion such that single-qubit phases are exactly cancelled \cite{Salathe2015}.

Equations (\ref{eq:gqq}) and (\ref{eq:Uqq}) together present a useful result from a quantum processor design perspective: The operating regime, frequency and time $\tau$ of the \iSWAP{} can be calculated (typically simulated) to high precision, before any processor fabrication is undertaken. The only `quantum' parts that enter $g_\text{qq}$ (and $g_\text{q-r-q}$) are the qubit frequencies, $\omega_{\text{q}1}(\Phi_1)$ and $\omega_{\text{q}2}(\Phi_2)$. If the Josephson energies of the qubits are known (which they typically are, from fabrication parameters), then by simulating the capacitances in $g_\text{qq}$ or $g_\text{q-r-q}$, the time $\tau$ and the pulseshape needed to implement an \iSWAP{} can be estimated to high precision. Typical values of the coupling strength, $g/(2\pi)$, for architectures using the \iSWAP{} gate is 5-40 MHz and are often very close to expectations from EM simulations \cite{Bialczak2010,Dewes2012,Salathe2015,Casparis2018a}.

\subsubsection{Applications of the $i$\textsf{SWAP} gate}
The $i\textsf{SWAP}$ cannot generate a \CNOT{} gate by itself. Rather, to implement a \CNOT{} gate requires stringing together two $i\textsf{SWAP}$s and several single qubit gates\cite{Schuch2003},
\begin{equation}
\begin{tabular}{ccc}
\Qcircuit @C=0.5em @R=2.2em{
 & \ctrl{1} &  \qw \\
 & \targ & \qw \
} & \raisebox{-1.5em}{=} &
\Qcircuit @C=0.5em @R=1.1em{
 & \qw         &  \gate{\Z{\text{-}\frac{\pi}{2}}}     & \multigate{1}{\iSWAP{}} & \gate{\X{\frac{\pi}{2}}}& \multigate{1}{\iSWAP{}} & \qw & \qw & \qw\\
 & \gate{\X{\frac{\pi}{2}}} & \gate{\Z{\frac{\pi}{2}}} & \ghost{\iSWAP{}} & \qw & \ghost{\iSWAP{}} & \qw & \gate{\Z{\frac{\pi}{2}}} & \qw \\
}
\end{tabular}
\label{eq:CNOT_swap}
\end{equation}
As evident from \cref{eq:CNOT_swap}, the \iSWAP{} gate in general needs to be used twice to generate a single \CNOT{}, leading to a significant overhead when compiling \CNOT{}--dense circuits from \iSWAP{} gates. However, depending on the context, the \iSWAP{} can be used efficiently (i.e. without any two-qubit gate overhead) to mimic the behavior of a \CNOT{}. Typically such circuits will not be completely equivalent, but will share certain salient features for specified input states. As an example of this procedure, Neeley \etal\cite{Neeley2010} demonstrated the generation of a 3-qubit Greenberger-Horne-Zeilinger (GHZ) state (which requires two subsequent \CNOT{}s in the simplest construction), by using only two \iSWAP{}s in a circuit that correctly generates the 3-qubit GHZ state on the $|000\rangle$ input. Moreover, the $XY$--interaction is a powerful tool for certain types of quantum simulation algorithms\cite{Heras2014}. If one is interested in digital quantum simulation of spin--like systems, then the $XY$--interaction can natively simulate e.g. a Heisenberg interaction,
\begin{equation}
H_\text{Heisenberg} = J_x\sigma_x\sigma_x + J_y\sigma_y\sigma_y+J_z\sigma_z\sigma_z.
\end{equation}
This approach to the $XY$ interaction was demonstrated by Salath\'e \etal{}\cite{Salathe2015}, where repeated application of the \iSWAP{} gate interspersed with single-qubit rotations was used to generate successive $XY$, $XZ$ and $YZ$ interactions that lead to an aggregate $H_\text{Heisenberg}$ Hamiltonian.  State-of-the-art operation of the \iSWAP{} gate has also been used to demonstrate a ten-qubit GHZ state\cite{Song2017}.


\subsection{\label{sec:CZgates}The \textsf{CPHASE} two-qubit gate in tunable qubits}
In our discussion of the \iSWAP{} gates, we assumed that the higher energy levels of the superconducting qubit do not play a role. As we show below, it turns out that for the case of transmon qubits (with negative anharmonicity), the higher levels can in fact be utilized to generate a the \CPHASE{} gate directly\cite{Strauch2003,DiCarlo2009}.

Recall from Sec.~\ref{sec:GatesInQC} that the \CPHASE{} gate implements the following unitary,
\begin{equation}
U_\CPHASE{} = \begin{bmatrix}
1 & 0 & 0 & 0\\
0 & 1 & 0 & 0\\
0 & 0 & 1 & 0\\
0 & 0 & 0 & -1\\
\end{bmatrix}
\end{equation}
Our goal for the remainder of this section is to show that the unitary operator of the \CPHASE{} gate appears naturally for capacitively coupled transmon superconducting qubits and review a few of the modern applications of this gate. We have chosen to include a considerable amount of details for the implementation of this gate, as a means to review some of the issues one has to resolve, to engineer high quality two-qubit gates.

The structure of the matrix in \cref{eq:UCPHASE} indicates that we need to apply a phase ($-1 = e^{i\pi}$) to the qubits whenever both are in the excited state $|11\rangle$. Considering the nature of the $XY$ interaction, which couples $|01\rangle \leftrightarrow |10\rangle$ and leads to the \iSWAP{} gate (see previous section), we expect avoided level crossings to exist between higher levels, e.g. $|11\rangle \leftrightarrow |20\rangle$ and $|11\rangle \leftrightarrow |02\rangle$. The flux-tunable implementation of the \CPHASE{} gate relies on this higher-level avoided crossing.

\begin{figure}[!t]
\begin{center}
\includegraphics[width=8.6cm]{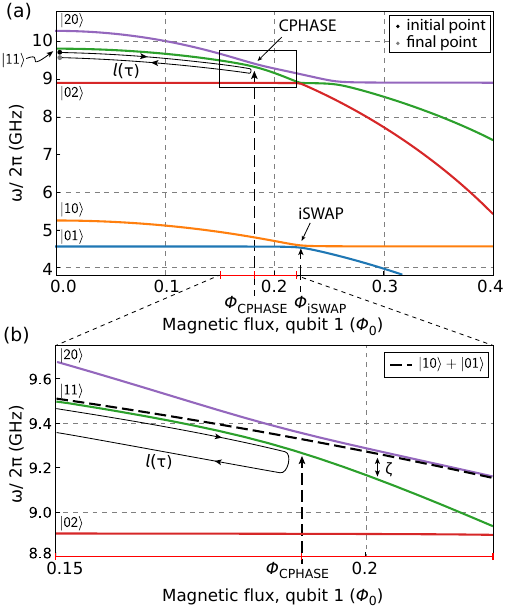}
\caption{\textbf{(a)} Spectrum of two coupled transmon qubits (using typical transmon-like values for Josephson energies and capacitances) as the local magnetic flux for qubit 1 is varied. The two lower branches corresponding to $|01\rangle$ and $|10\rangle$ are involved in the \iSWAP{} gate operation at $\Phi = \Phi_\text{iSWAP}$. The avoided crossing indicated in the black rectangle is used to implement the conditional phase gate (\CPHASE{}), at $\Phi = \Phi_\text{\CPHASE{}}$. Black line with arrows indicate a typical trajectory used to implement a \CPHASE{} gate (starting at the black circle and ending at the gray circle). \textbf{(b)} Zoom in of the $|20\rangle \leftrightarrow |11\rangle$ avoided crossing highlighted in the black box in (a) at $\Phi = \Phi_{\CPHASE{}}$. The parameter $\zeta$ quantifies the difference in energy between $|11\rangle$ and $|01\rangle + |10\rangle$ and $\ell(\tau)$ is the trajectory in ($\Phi,t$)--space.}
\label{fig:2qb_CPHASE}
\end{center}
\end{figure}

To motivate this intuition we plot the spectrum for two coupled transmon qubits, in \cref{fig:2qb_CPHASE}(a), including levels with two excitations, as the local magnetic flux in qubit 1 is being tuned. The Hamiltonian for this spectrum, written in the  $\{|00\rangle, |01\rangle, |10\rangle, |11\rangle, |02\rangle, |20\rangle \}$-basis, is approximately given by,
\begin{equation}
H_\text{2 excitations} = \begin{bmatrix}
E_{00} & 0 & 0 & 0 & 0 & 0 \\
0 & E_{01} & g & 0 & 0 & 0 \\
0 & g & E_{10} & 0 & 0 & 0 \\
0 & 0 & 0 & E_{11} & \sqrt{2}g & \sqrt{2}g \\
0 & 0 & 0 & \sqrt{2}g & E_{02} & 0 \\
0 & 0 & 0 & \sqrt{2}g & 0 & E_{20} \\
\end{bmatrix}, \label{eq:6lvlHam}
\end{equation}
where $E_{nm} = E^\text{q1}_n(\Phi_1) + E^\text{q2}_m(\Phi_2)$ and $E_n(\Phi_i)$ is the flux-dependent energy of the $i$-th level of a transmon \cite{Koch2007}, and the $\{|02\rangle,|20\rangle\} \leftrightarrow |11\rangle$ transitions are scaled by a factor $\sqrt{2}$ due to the higher photon number. In \cref{fig:2qb_CPHASE}, we plot the frequencies $\omega_{nm} = E_{nm} - E_{00}$ calculated from \cref{eq:6lvlHam}, using standard, symmetric, transmon-like parameters, as the local magnetic field of qubit 1 is increased.

The result of the higher levels on the computational basis can be understood by considering a concrete example. By preparing the combined qubit state $|11\rangle$ and moving slowly towards the avoided crossing between $|11\rangle$ and $|20\rangle$ at $\Phi_{\CPHASE{}}$, waiting for some time $\tau$ and moving back (see black line with arrows in \cref{fig:2qb_CPHASE}(b)), the resulting unitary operator in the computational basis is given by
\begin{equation}
U_\text{ad} = \begin{bmatrix}
1 & 0 & 0 & 0 \\
0 & e^{i\theta_{01}(\ell)} & 0 & 0 \\
0 & 0 & e^{i\theta_{10}(\ell)} & 0 \\
0 & 0 & 0 & e^{i\theta_{11}(\ell)} \\
\end{bmatrix},\label{eq:Uadiabatic}
\end{equation}
where
\begin{equation}
\theta_{ij}(\ell(\tau)) = \int_0^{\tau} \d t\, \omega_{ij}[\ell(t)]
\end{equation}
is the phase acquired by the state $|ij\rangle$ along the trajectory $\ell$ in ($\Phi$, $t$)-space during time $\tau$. The movement should be sufficiently slow on a time-scale set by $g$ that the moving state never populates the $|20\rangle$ state, i.e. the movement should be adiabatic. In terms of applied flux, the avoided crossing between the $|11\rangle \leftrightarrow |20\rangle$ state happens before $|10\rangle \leftrightarrow |01\rangle$ (due to the negative anharmonicity of the transmons, $\alpha \approx -E_c$) and consequently $\ell$ does not take the states through the $\Phi_{\iSWAP{}}$ operating point. As shown in \cref{fig:2qb_CPHASE}(b) we can define a parameter (typically denoted $\zeta$) quantifying the difference in phase acquired by the $|11\rangle$ relative to the single excitation states,
\begin{equation}
\zeta = \left(\omega_{11} - (\omega_{01}+\omega_{10})\right).
\end{equation}
The parameter $\zeta$ can be thought of as the result (in the computational space) of the repulsion of $|11\rangle$ due to the $|20\rangle$ state. If we now choose a trajectory $\ell_\pi$, designed so that $\int_0^{\tau}  \zeta(\ell_\pi(t)) dt = \pi$, then
\begin{equation}
\int_0^{\tau}  \zeta(t) dt= \pi  = \theta_{11}(\ell_\pi) - \left(\theta_{01}(\ell_\pi) + \theta_{10}(\ell_\pi)\right).
\end{equation}
Inserting this expression into \cref{eq:Uadiabatic} we see that
\begin{align}
 U_\text{ad} &= \begin{bmatrix}
1 & 0 & 0 & 0 \\
0 & e^{i\theta_{01}(\ell_\pi)} & 0 & 0 \\
0 & 0 & e^{i\theta_{10}(\ell_\pi)} & 0 \\
0 & 0 & 0 & e^{i(\pi + \theta_{01}(\ell_\pi) + \theta_{10}(\ell_\pi))} \\
\end{bmatrix}.
\label{eq:Uadfixedphase}
\end{align}
After the adiabatic excursion, one can now apply single-qubit pulses (or use virtual-\textsf{Z} gates) to exactly cancel the single-qubit phases such that $\theta_{10}(\ell_\pi) = \theta_{01}(\ell_\pi) = 0$. This changes $U_\text{ad}$ to
\begin{equation}
U_\text{ad} = \begin{bmatrix}
1 & 0 & 0 & 0 \\
0 & 1 & 0 & 0 \\
0 & 0 & 1 & 0 \\
0 & 0 & 0 & e^{i\pi} \\
\end{bmatrix} =
\begin{bmatrix}
1 & 0 & 0 & 0 \\
0 & 1 & 0 & 0 \\
0 & 0 & 1 & 0 \\
0 & 0 & 0 & -1 \\
\end{bmatrix} =
 U_\CPHASE{}. \label{eq:howtogetCPHASE}
\end{equation}
From \cref{eq:howtogetCPHASE} it is evident that an adiabatic movement of $|11\rangle$, followed by single-qubit gates (virtual or real) efficiently implements a \CPHASE{} and, through \cref{eq:CNOTfromCPHASE}, also efficiently implements a \CNOT{}. The \CPHASE{} gate is one of the workhorses of modern superconducting qubit processesors with gate fidelities $\gtrsim 0.99$ \cite{Barends2014,Kelly2014}.

One is, of course, free to choose an arbitrary trajectory $\ell_\phi$ that implements the phase $e^{-i\phi}$ on the $|11\rangle$ state. Assuming that the single-qubit phases are properly cancelled, one sees that the arbitrary phase version of the \CPHASE{} gate (typically denoted $\textsf{CZ}_\phi$) can be written as
\begin{align}
\textsf{CZ}_\phi &= \begin{bmatrix}
1 & 0 & 0 & 0 \\
0 & 1 & 0 & 0 \\
0 & 0 & 1 & 0 \\
0 & 0 & 0 & e^{-i\phi} \nonumber
\\
\end{bmatrix} \\
& = \exp\left[-i\frac{\phi}{4}\left(\sigma_z\otimes \sigma_z-\sigma_z\otimes \Id - \Id \otimes \sigma_z\right)\right]. \label{eq:CZtheta}
\end{align}
Because of the form of \cref{eq:CZtheta}, one can think of the avoided crossing with the higher levels outside the computational subspace as giving rise to an effective $\sigma_z\otimes \sigma_z$ coupling within the computational subspace \cite{Strauch2003}.

An alternative to the adiabatic approach outlined above to realize \CPHASE{} is to make a sudden excursion to the $\Phi_{\CPHASE{}}$ operating point, after waiting a time $t=\pi/\sqrt{2}g$, the state will have completed a single Larmor-type rotation from $|11\rangle$ to $|02\rangle$ and back again to $|11\rangle$, but in the process, acquired an overall $\pi$ phase, similar to the \iSWAP{} gate, but in the $\{|11\rangle, |20\rangle\}$ subspace\cite{DiCarlo2010}. In fact, such excursions near or through avoided crossings leading to adiabatic and non-adiabatic transitions have been studied extensively in the context of interferometry, cooling, spectroscopy, and quantum control\cite{Oliver2005,Berns2006,Valenzuela2006,Ashhab2007,Berns2008,Rudner2008,You2008,Grajcar2008,Bylander2009,Oliver2009,Shevchenko2010}.

The remainder of this subsection is devoted to an overview of some of the recent advances and demonstrations using the \CPHASE{} gate since its first demonstration in 2009 where it was used to generate Bell-states and demonstrate two-qubit algorithms \cite{DiCarlo2009}.

\subsubsection{Trajectory design for the \CPHASE{} gate}
The (adiabatic) implemention of $U_\CPHASE$ outlined above assumed that the trajectory $\ell_\pi$ was completely adiabiatic and that the $|11\rangle$ state never left the computational subspace. Since the fidelity of gates is bounded from above by the coherence times of the qubits, short gate times are desirable\cite{Ashhab2012}. This presents a tension for optimally operating the \CPHASE{} gate -- fast operation in conjunction with the need for adiabatic operation. A relevant question is then: what is the \emph{optimal} trajectory $\ell^\star_\pi$ that implements the necessary phase as fast as possible, with as little leakage as possible, for a given size of the avoided crossing between $|11\rangle$ and $|20\rangle$? Given a typical coupling rate $g/2\pi \approx 20$ MHz (as discussed in Sec. \ref{sec:SWAPgates}), one expects a heuristic lower time limit to be $2\pi/g \approx 50$ ns (stronger coupling of course leads to shorter gate times, but will limit the on/off ratio of the gate). Traditional optimal control of adiabatic movement assumes the movement is \emph{through} the avoided crossing (see e.g. Ref. \onlinecite{Bason2012}), but the trajectory $\ell_\pi$ moves close to and then back from the avoided crossing. This modification to the adiabatic movement protocol was addressed by Martinis and Geller \cite{Martinis2014}, specifically in the context of errors for a \CPHASE{} gate implemention. The authors show that non-adiabatic errors can be minimal for gate times only slightly longer than $2\pi/g$ using an optimal waveform (based on a Slepian waveform \cite{Slepian}) to parametrize the trajectory $\ell_\pi^\star(\tau)$.

\subsubsection{The \CPHASE{} gate for quantum error correction}\label{subsubsec:CZforQEC}
Using the approach of Martinis and Geller, Barends \etal{} were able to demonstrate a two-qubit gate fidelity $\mathcal F_\CPHASE{} = 0.9944$ (determined via a technique known as `interleaved randomized benchmarking' \cite{Benhelm2008,Ryan2009,Magesan2011,Corcoles2013}). This implementation had a gate time $\tau = 43$ ns and was implemented with the $\ell^\star_\pi$ waveform\cite{Barends2014}, in an ``xmon" device \cite{Barends2013} -- a transmon with a ``+"-shaped capacitor. A two-qubit gate fidelity $\mathcal F >0.99$ represents a significant milestone, not just from a technical and engineering perspective, but also from a foundational standpoint: The surface code (a quantum error correcting code) has a lenient fault-tolerance threshold of $\sim 1\%$\cite{Fowler2012,Fowler2012,Tomita2014,OBrien2017}. This means, roughly speaking, that if the underlying operations on the qubits have fidelities $\mathcal F>0.99$, then by adding more qubits to the circuit (and correctly implementing the fault-tolerant quantum error correction protocol) the overall error-rate can be reduced, and one can in principle perform arbitrarily long quantum computations, without errors spreading uncontrollably and corrupting the calculation. Because of its relatively lenient threshold under circuit noise (compared to e.g., Steane or Shor codes \cite{NielsenChuang,Nickerson2016,Campbell2017a}) and its use of solely nearest-neighbor coupling, the surface code is one of the most promising quantum error correction codes for medium-to-large scale quantum computing in solid state systems\cite{Fowler2012}. Therefore, surpassing the fault-tolerance threshold using \CPHASE{} represents a significant milestone for the field\cite{Benjamin2015}. Moreover, practical blueprints for implementing scalable subcells of the surface code using the \CPHASE{} as the fundamental two-qubit gate have also been proposed \cite{Versluis2017} as well as \emph{in-situ} calibration protocols for large-scale systems operating with \CPHASE\cite{Kelly2016}. For a full review of the pros and cons of various quantum error correcting codes we refer the interested reader to e.g. an introductory review article Ref. \onlinecite{Devitt2013}, or any of the excellent textbooks and more detailed review articles in Refs. \onlinecite{NielsenChuang,Williams2008,Gottesman2009,Devitt2013,Lidar2013,Terhal2015,Campbell2017a}.

Returning to the \CPHASE{} gate, numerical optimization of $\ell^\star_\pi$ was demonstrated by Kelly \etal{}\cite{Kelly2014} using the interleaved randomized benchmarking sequence fidelity as a cost function to push a native implementation of $\ell^\star_\pi$ with a fidelity $\mathcal F = 0.984$ up to $\mathcal F = 0.993$, surpassing the surface code fault tolerance threshold. In the same work that demonstrated $\mathcal F_\CPHASE{} = 0.9944$, Barends \etal{}\cite{Barends2014} used the \CPHASE{} gate to generate GHZ states, $|\text{GHZ}\rangle = \left(|0\rangle^{\otimes N} + |1\rangle^{\otimes N}\right)/\sqrt{2}$, of up to $N=5$ qubits, with a fidelity for the $N=5$ state of $\mathcal F = \text{Tr}\left(\rho_\text{ideal}\rho_{N=5}\right) = 0.817$. The protocol for generating the GHZ state with $N = 2$ and $N=3$ from \CPHASE{} was originally demonstrated by DiCarlo \etal{} \cite{DiCarlo2009,DiCarlo2010}. The textbook route to generating the  $N=2$ GHZ state, $|\Phi^+\rangle$ (a Bell state) from the all-zero input is
\begin{equation}
\Qcircuit @C=0.5em @R=0.75em{
 \lstick{|0\rangle} & \gate{\H{}} & \ctrl{1} & \qw & \qw \\
 \lstick{|0\rangle} & \qw         & \targ    & \qw & \qw \gategroup{1}{5}{2}{5}{1em}{\}}
}
\raisebox{-1em}{\,\, = \,\,$ |\Phi^+\rangle$}
\end{equation}
An equivalent circuit using \CPHASE{} and native single-qubit gates in superconducting qubits is:
\begin{equation}
\Qcircuit @C=0.75em @R=0.75em{
 \lstick{|0\rangle}  & \gate{\Y{\pi/2}} & \qw                &\ctrl{1} & \qw                  & \qw & \qw \\
 \lstick{|0\rangle}  & \qw               & \gate{\Y{-\pi/2}} &\ctrl{-1}    & \gate{\Y{\pi/2}} & \qw & \qw \gategroup{1}{3}{2}{5}{.7em}{--}
}.
\end{equation}
By repeating the operation inside the dashed box on additional qubits, an $N$-qubit GHZ state can be generated\cite{Barends2014}. Since the demonstration of the $N=5$ GHZ state using the \CPHASE{} gate, the gate has been deployed to demonstrate several important aspects of quantum information processing using superconducting qubits. A nine-qubit implementation of the five-qubit repetition code (five data qubits + four syndrome qubits)\cite{Devitt2013} was demonstrated, and the error suppression factor of a single logical quantum bit was shown to increase as the encoding was changed from three data qubits to five data qubits \cite{Kelly2015}. Similarly, in a five qubit processor the three-qubit repetition code with artificially injected errors was demonstrated\cite{Riste2015}, building on earlier results utilizing a combination of \iSWAP{} and \CPHASE{} gates to perform parallelized stabilizer readout\cite{Saira2014}.

\subsubsection{Quantum simulation and algorithm demonstrations using \CPHASE{}}
As an example of the utility of the \CPHASE{} gate, we briefly discuss a particular demonstration of a digital quantum simulation. In this context, the \CPHASE{} gate has been utilized to simulate a two-site Hubbard model with four fermionic modes, using four qubits\cite{Barends2015}. Using the Jordan-Wigner transformation\cite{Jordan1928a,LasHeras2015}, it is possible to map fermionic operators onto Pauli spin matrices\cite{Jordan1928}. As shown in Ref.\onlinecite{Barends2015} a Hubbard model with two fermionic modes, whose Hamiltonian is given by
\begin{equation}
H_\text{Hubbard, two mode} = -t (b^\dagger_1b_2 + b_2^\dagger b_1) + U b^\dagger_1b^{\vphantom{\dagger}}_1 b^\dagger_2b^{\vphantom{\dagger}}_2
\end{equation}
can be written in terms of Pauli operators as,
\begin{align}
H & = \frac{t}{2}\left(\sigma_x \otimes \sigma_x+\sigma_y\otimes\sigma_y\right)
 \\
& + \frac{U}{4}\left(\sigma_z\otimes\sigma_z + \sigma_z\otimes \Id + \Id \otimes \sigma_z\right),
\end{align}
where $U$ is the repulsion energy and $t$ is the hopping strength. Similar to the Heisenberg interaction discussed briefly in Sec. \ref{sec:SWAPgates}, it is now a question of producing $\alpha \sigma_i\otimes \sigma_i$-type interactions, where the prefactor $\alpha$ can be tuned. Using the $\CZ{}_\phi$ version of \CPHASE{}, a $U_\text{ZZ}(\phi) = \exp\left(-i\frac{\phi}{2} \sigma_z\otimes \sigma_z\right)$ unitary can be generated via
\begin{equation}
\begin{tabular}{ccc}
\Qcircuit @C=0.5em @R=1.3em{
  & \multigate{1}{U_\text{ZZ}(\phi)}  & \qw \\
  & \ghost{U_\text{ZZ}(\phi)}   & \qw
}
& \raisebox{-1.5em}{=} &
\Qcircuit @C=0.5em @R=1em{
  &\gate{\X{\pi}} & \multigate{1}{\CZ{}_\phi}        & \gate{A_\pi} & \multigate{1}{\CZ{}_\phi} & \qw & \qw\\
  & \qw         & \ghost{\CZ{}_\phi}    & \gate{A_\pi} & \ghost{{\CZ{}_\phi}}        & \gate{\X{\pi}} & \qw
},
\end{tabular}\label{eq:ZZfromCZ_v1}
\end{equation}
where $A_\pi \in\{\X{\pi},\Y{\pi}\}$ is used to allow for small and negative angles. Finally, for completeness, we mention an alternative approach to creating $U_\text{ZZ}$, given by\cite{Wendin2017,Havlicek2018}
\begin{equation}
\begin{tabular}{ccc}
\Qcircuit @C=0.5em @R=1.3em{
  & \multigate{1}{U_\text{ZZ}(\phi)}  & \qw \\
  & \ghost{U_\text{ZZ}(\phi)}   & \qw
}
& \raisebox{-1.5em}{=} &
\Qcircuit @C=0.5em @R=1.25em{
  &\ctrl{1} & \qw                  & \ctrl{1} & \qw \\
  & \targ    & \gate{\Z{\phi/2}}    & \targ & \qw
},
\end{tabular}\label{eq:ZZfromCZ_v2}
\end{equation}
which has the benefit of relying on \CPHASE{} (through the \CNOT{}s), and the angle can be controlled using the single-qubit $\Z{}$ gates. We refer the interested reader to two reviews on quantum simulations, see e.g. Refs. \onlinecite{Buluta2009,Georgescu2014}.

The \CPHASE{} gate has also been used in a variety of other contexts, e.g., for calculating the dissociation of diatomic hydrogen ($H_2$) using the variational quantum eigensolver method\cite{OMalley2016}, for feed-forward based teleportation experiments\cite{Baur2012,Steffen2013}, as well as initial steps towards demonstrating quantum supremacy\cite{Neill2018} and a $2\times 2$ implementation of the Harrow-Hassidim-Lloyd algorithm \cite{Harrow2009,Zheng2017}. In the field of hybrid semiconducting nanowire/superconducting qubits (known as the ``gatemon" approach\cite{Larsen2015,DeLange2015,Wang2018b}), where the qubit frequency is modified by electrostatically changing the density of carriers in a semiconducting region with proximity-induced superconductivity, the \CPHASE{} gate was also demonstrated between two nanowire qubits \cite{Casparis2016}.

One may worry that operating a qubit by moving its frequency can lead to overlap with frequencies already used by other qubits, in a system with multiple qubits. This issue is known as \emph{frequency crowding}. While the use of asymmetric transmons (with two sweet spots in the range $[-\Phi_0,+\Phi_0]$, recall \cref{Fig:QubitModalities}(c)) may help alleviate some frequency crowding issues, a more long-term strategy is needed. One way to circumvent the problem is to utilize on/off tunable coupling schemes, in which qubits can exchange energy only if a coupler activates the interaction\cite{Chen2014,McKay2016}. To address this issue in the context of the \CPHASE{} gate, Chen \etal{}\cite{Chen2014} demonstrated a device (named ``the gmon") where the qubit interaction can be tuned with an on/off ratio on the order of 1000, and a \CPHASE{} gate fidelity of $\mathcal F = 0.9907$ was demonstrated.

This concludes the introduction to the physics and operation of the \CPHASE{} gate in its native form. In the remainder of this section we will introduce a few of the microwave-only gates that have been demonstrated in an effort to sidestep the need for local tunability (and the resulting increased sensitivity to noise) as required by the \iSWAP{} and \CPHASE{} gate.


\subsection{\label{sec:CRgates}Two-qubit gates using only microwaves}
One common (potential) drawback for the \iSWAP{} and \CPHASE{} gates is that their operation requires flux-tunable qubits. Introducing a new control knob, such as flux control, in turn also introduces a new noise channel for the system. Furthermore, the need for flux-tunability increases the sensitivity of the devices to flux noise by tuning the qubits from their ``sweet spots", increases the dephasing rate. From this perspective, one could envision using all-microwave-based gates to remedy these issues. To this end, the cross-resonance (``\CR{}") gate was developed for operating fixed-frequency superconducting qubits\cite{Paraoanu2006,Rigetti2010,DeGroot2010}, which typically feature longer lifetimes and reduced sensitivity to flux noise.

\subsubsection{The operational principle of the \textsf{CR} gate}
\begin{figure}[!ht]
\begin{center}
\includegraphics[width=8.6cm]{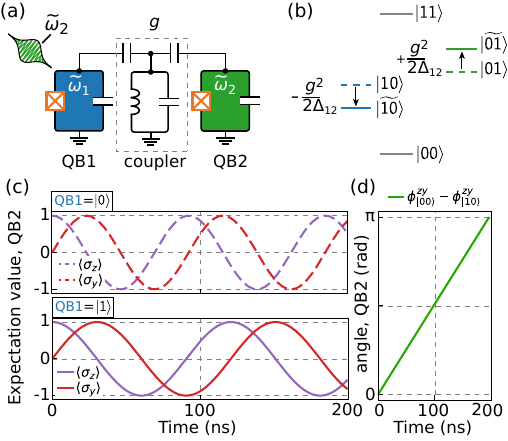}
\caption{\textbf{(a)} Schematic circuit diagram of two fixed frequency transmons coupled through a resonator yielding an overall coupling coefficient $g$. Qubit 1 driven at the frequency of qubit 2 leads to the \CR{} gate. \textbf{(b)} Schematic level diagram of the always-on coupling leading to dressed states $|\widetilde{01}\rangle$ and $|\widetilde{10}\rangle$ with $\Delta_{12} = \omega_1 - \omega_2$. \textbf{(c)} Simulations of the expectation values of $\langle \sigma_z\rangle$ and $\langle \sigma_y \rangle$ for qubit 2 as a drive at the frequency of qubit 2 is applied to qubit 1. Upper panel shows regular Rabi oscillations when qubit 1 is in the $|0\rangle$ state. Bottom panel shows a modified Rabi frequency when qubit 1 is in $|1\rangle$ state, in accordance with \cref{eq:CR_rabi}. (d) Difference in angle in the $(z,y)$ plane as a function of length of the applied drive to qubit 1. At approximately 200~ns $\pi$-phase shift has been acquired.}
\label{fig:2qb_CR}
\end{center}
\end{figure}

To elucidate the operation of the \CR{} gate, we briefly revisit the driving Hamiltonian derived in Sec. \ref{sec:CapCouplingforDriving}. There, we considered only a single qubit. However, if one extends this formalism to two qubits, see Fig. \ref{fig:2qb_CR}(a) denoting the frequency difference by $\Delta_{12} = \omega_{\text{q}1} - \omega_{\text{q}2}$ and the coupling by $g\ll \Delta_{12}$, and performing a Schrieffer-Wolff transformation to go to the dressed state picture, the driving Hamiltonians for qubit 1 and 2 become\cite{Rigetti2010,Julich}
\begin{align}
H_{\d,1} = \Omega V_{\d1}(t)\left(\sigma_x\otimes\Id + \nu_1^-\Id\otimes\sigma_x + \mu_1^{-}\sigma_z\otimes\sigma_x\right)\label{eq:CRGambettaStyle}\\
H_{\d,2} = \Omega V_{\d2}(t)\left(\Id\otimes\sigma_x + \nu_2^+\sigma_x\otimes\Id + \mu_2^{+}\sigma_x\otimes\sigma_z\right)
\end{align}
where
\begin{align}
\mu_i^\pm &= \pm\frac{g}{\Delta_{12}}\frac{\alpha_i}{(\alpha_i \mp \Delta_{12})} \label{eq:muforCR1}\\
\nu_i^\pm &= \pm\frac{g}{\Delta_{12}}\frac{\mp \Delta_{12}}{(\alpha_i \mp \Delta_{12})}
\end{align}
and $\Omega V_{\text{d}i}(t)$ is the driving for qubit $i$. From \cref{eq:CRGambettaStyle}, it is evident that if we drive qubit 1 at the frequency of qubit 2, then to qubit 2, this will look like a combination of $\nu_1^-\Id\otimes\sigma_x$ and $\mu_1^-\sigma_z\otimes\sigma_x$. This means that the Rabi oscillations of qubit 2 will have a frequency given by
\begin{equation}
\Omega^\text{Rabi}_{\text{QB2}} = \Omega V_{\text{d}1}\left(\nu_1^- + z_1 \mu_1^-\right), \label{eq:CR_rabi}
\end{equation}
where $z_1=\langle\sigma_z \Id\rangle$, and $z_1$ depends on the state of qubit 1. This effect is demonstrated in \cref{fig:2qb_CR}(c), where a simulated drive is applied to qubit 1 while the resulting Rabi oscillations in qubit 2 are recorded. We have used typical fixed-frequency transmon parameters from experiments, and we have included a spurious cross-talk term $\eta = 0.03$.\cite{Chow2012,Corcoles2013}. In \cref{fig:2qb_CR}(d), we plot the difference in angle in the $(z,y)$ plane acquired by qubit 2 for different initializations of qubit 1, $\Delta \phi = \phi_{|00\rangle}^{zy} - \phi_{|10\rangle}^{zy}$. For this particular choice of parameters, the cross-resonance gate achieves a $\pi$-phase shift in $\approx 200~$ns.

This strategy was first demonstrated using flux-tunable transmons in Ref.\onlinecite{Chow2011}, where a Bell state with fidelity $\mathcal F_\text{bell} = \langle \Phi^+|\rho|\Phi^+\rangle = 0.90$ was achieved. Using quantum process tomography the gate fidelity was found to be $\mathcal F_\text{QPT} = 0.81$. By moving to fixed-frequency qubits with increased lifetimes, the gate fidelity was increased to $\mathcal F_\text{QPT} = 0.98$ (with subtraction of state initialization and measurement errors)\cite{Chow2012}. For completeness, we note that due to the form of the last term in \cref{eq:CRGambettaStyle}, the \CR{} gate is also sometimes denoted the $\textsf{ZX}_{\theta}$ gate. The unitary matrix representaion of the $\ZX{\theta}$ gate is
\begin{align}
\renewcommand*{\arraystretch}{1.3}
U_{\ZX{\theta}} &= e^{-\frac{i}{2}\theta \sigma_z\otimes\sigma_x} \nonumber \\
&= \begin{bmatrix}
\cos\frac{\theta}{2} & -i\sin\frac{\theta}{2} & 0 & 0\\
-i\sin\frac{\theta}{2} & \cos\frac{\theta}{2} & 0 & 0 \\
0 & 0 & \cos\frac{\theta}{2} & i\sin\frac{\theta}{2} \\
0 & 0 & i\sin\frac{\theta}{2} & \cos\frac{\theta}{2}
\end{bmatrix}
\end{align}
where $\theta = -\mu_1^-\Omega V_{\text{d}1}(t)$, which can be used to generate a \CNOT{} with the addition of only single-qubit gates,
\begin{equation}
\begin{tabular}{ccc}
\Qcircuit @C=0.5em @R=1.9em{
 & \ctrl{1} &  \qw \\
 & \targ & \qw \
} & \raisebox{-1.5em}{=} &
\Qcircuit @C=0.5em @R=0.8em{
 &  \gate{\Z{\frac{\pi}{2}}} &  \gate{\ZX{-\frac{\pi}{2}} }& \qw\\
 &  \gate{\X{\frac{\pi}{2}}} &  \ctrl{-1}    & \qw
},
\end{tabular}
\end{equation}
up to a phase $e^{i\pi/4}$.

\subsubsection{Improvements to the \CR{} gate and quantum error correction experiments using \CR{}}\label{subsubsec:CRforQEC}
Since qubit 1 is being driven off-resonance, an ac-Stark shift will add a term $\propto \sigma_z \Id$ to the driving Hamiltonian of qubit 1. The effect of both the spurious ac Stark shift and the direct $\nu_1^-\Id\sigma_x$ single-qubit rotations was studied in Ref.\onlinecite{Corcoles2013}. By modifying the original $\CR{}$ protocol to effectively ``echo away" the two unwanted contributions from the $\sigma_z\Id$ and $\Id\sigma_x$ terms, the fidelity of the \CR{} gate was improved to $\mathcal F_\text{CR} = 0.8799$\cite{Corcoles2013}, using quantum process tomography. Using interleaved randomized benchmarking of this improved ``echo-CR"-gate (e$\ZX{-\frac{\pi}{2}}$), a gate fidelity of $\mathcal F_{\text{e}\ZX{-\frac{\pi}{2}}} = 0.9347$ was achieved. This gate implementation was used to demonstrate two-qubit parity measurements in a three-qubit device\cite{Chow2014}, as well as detecting bit-flip and phase-flip errors in a Bell state encoded in a four-qubit device\cite{Corcoles2015}, with gate fidelities from interleaved randomized benchmarking in the range 0.94 to 0.96. Using a similar device, but with five qubits, weight-four parity measurement of the forms $ZZZZ$ and $XXXX$ were demonstrated\cite{Takita2016}, where the crosstalk to qubits not involved in the \CR{} gates was studied, leading to the development of a four pulse $\text{e}\CR_{\text{4-pulse}}$ scheme.

Based on improvements in the analysis of the Hamiltonian describing the \CR{} drive, Sheldon \etal{}\cite{Sheldon2016} subsequently demonstrated a version of the \CR{} which reduced the gate time to $\tau = 160$ ns and added an active cancellation tone to the e$\ZX{}$ previously developed. Using this ``active cancellation echo $\ZX{}$" (\text{ace}$\ZX{}$), the fidelity was increased to $\mathcal F_{\text{ace}\ZX{-\frac{\pi}{2}}} = 0.991$, measured with interleaved randomized benchmarking. The same sequence without active cancellation on the same qubits yielded $\mathcal F_{\text{e}\ZX{-\frac{\pi}{2}}} = 0.948$. The interested reader may consult the followup theoretical work \cite{Magesan2018} with more details on the effective Hamiltonian models. Other approaches to fast, high-fidelity cross-resonance gates have also been proposed \cite{Kirchhoff2018}. This series of improvements to the original cross-resonance implementation has increased the gate fidelity to beyond the threshold for fault-tolerance in a surface code, with similar quality to the \CPHASE{} gate. Although improvements should still be made, with the advent of the \CR{} gate, superconducting qubit based quantum computing platforms now offer two entangling two-qubit gates that can be used for implementing surface-code based error correction schemes.

In the initial experiments using \CR{} gates, the gate times were significantly longer than the typical \CPHASE{} gate times ($\tau_{\CPHASE{}}=30$ -- 60 ns and $\tau_{\CR} = 300$ -- 400 ns), which to a large extent accounts for the observed \CR{} gate fidelities. The time scale for \CR{} operation is set by the frequency detuning, the anharmonicity, and the coupling strength, through \cref{eq:muforCR1}. This has the unfortunate drawback that if qubits do not have intended frequencies (due to fabrication variation), it will be immediately manifest as longer gate times, and in turn, reduced gate fidelity. As fabrication techniques are becoming more sophisticated and reliable, this problem may be of reduced importance. However, since the coupling in the \CR{} scheme is always on, there is an inherent tension between well-isolated qubits for high-fidelity single-qubit operations, and coupling qubits, for fast/high-fidelity two qubit gates.

\subsubsection{Quantum simulation and algorithm demonstrations with the \CR{} gate}
Since the form of the \CR{} Hamiltonian ($\sigma_z\otimes\sigma_x$) is not a $(\sigma_x\otimes\sigma_x + \sigma_y\otimes \sigma_y)$-type interaction (leading to \iSWAP{} gate) nor is it an the effective $(\sigma_z\otimes\sigma_z)$-type (leading to \CPHASE{} gate), one could question its applicability to quantum-simulation-type experiments, which often involves terms of the form $\sigma_i\otimes\sigma_i$. However, by developing a variational quantum eigensolver routine that efficiently generates entangled trial states using just the $\ZX{}$ interaction, Kandala \etal \cite{Kandala2017} calculated the ground-state energy for H$_2$, LiH, and BeH$_2$. This experiment was performed on six fixed-frequency qubits, and it employed a technique for compact encoding of the Hamiltonians corresponding to each molecule\cite{Bravyi2017}. As of this writing, this experiment represents the largest molecule for which the ground state has been found using a purely quantum processing approach.

The \CR{} gate is also the native two-qubit gate available on the IBM Quantum Experience quantum processor\cite{IBMqexp}, which is accesible online. Using the IBM Quantum Experience processor, Takita \etal{}\cite{Takita2017} demonstrated an implementation of a two-logical-qubit (four physical qubit) error detection code \cite{Leung1997}. The implementation was inspired by the proposal of Gottesman\cite{Gottesman2016}, which proposed a minimal experiment to claim observation of fault-tolerant encodings\cite{Gottesman2009}, using a four qubit error detection code in a five qubit setup. Due to constraints on the connectivity, the work by Takita \etal{} demonstrated a modified version of the Gottesman encoding, in which two logical qubits are initialized, but only one of them in a fault-tolerant manner. By artificially injecting an error in the state preparation circuit, the authors demonstrate that the probability of correctly preparing a fault tolerant state is greater than the probability of correctly preparing a non-fault-tolerant qubit. This behavior is consistent with expectations for how fault-tolerant encodings work. Simultaneously, Vuillot\cite{Vuillot2017} also used the IBM Quantum Experience machine to study fault-tolerant schemes encoded in that connectivity.

Beyond the applications to error-correction and error-detection, the cross-resonance gate has also been employed in early demonstrations of quantum advantages in machine learning. Rist\'e \etal{}\cite{Riste2017} studied the so-called ``learning parity with noise" problem, in which one attempts to learn a bit-string \textsf{\textbf{k}} by querying an oracle function $f(\textsf{\textbf{D}},\textsf{\textbf{k}}) = \textsf{\textbf{D}}\cdot\textsf{\textbf{k}}\text{ mod }2$ with a user-input bit-string \textsf{\textbf{D}}. In a first implementation of this problem, the authors show that for a specific instance of the bit-string $\textsf{\textbf{k}} = \textsf{11}$, a learner with access to quantum operations needs fewer queries to the function $f$. However, by extending the model of learning parity with noise, the authors demonstrated a consistent advantage of the learner with access to quantum operations\cite{Riste2017}.

The \CR{} gate was also used to demonstrate the implementation of a supervised learning algorithm where the feature space is encoded as quantum data on the Bloch sphere\cite{Havlicek2018}. In typical supervised learning, an algorithm is exposed to a training set of labeled data, and is subsequently asked to classify a new, unlabeled set of data\cite{MacKay2003}. In the support vector machine (SVM) approach to such problems, the data is then mapped non-linearly onto the so-called ``feature space", in which the trained algorithm has constructed a separating hyperplane to classify the data. While a full ``quantum Support Vector Machine" proposal exists, the algorithm assumes that the data are already present in a coherent superposition\cite{Rebentrost2014}. Instead, Havlicek \etal{}\cite{Havlicek2018} proposed, and demonstrated, that mapping the classical data non-linearly onto the Bloch sphere can also be utilized to provide a quantum advantage. For a wider discussion of the important role of quantum data in many quantum machine learning algorithms, the reader is referred to Ref.\onlinecite{Biamonte2017}

\subsubsection{\label{sec:OtherUWaveGates}Other microwave-only gates: \textit{b}\textsf{SWAP}, \textsf{MAP}, and \textsf{RIP}}
\begin{figure}[!t]
\begin{center}
\includegraphics[width=7cm]{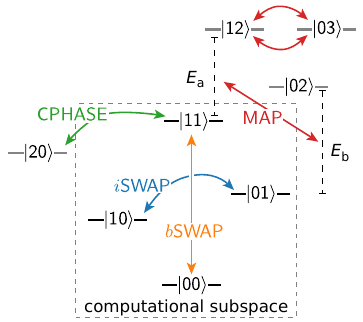}
\caption{Schematic of the level structure of two coupled qubits (including higher levels) with indication of the transitions utilized in the \iSWAP{}, \bSWAP{}, \CPHASE{} and \textsf{MAP} gate. See text for details. Figure inspired from Ref. \onlinecite{Chow2013}.}
\label{fig:2qb_GatesLevelDiagram}
\end{center}
\end{figure}

The \CR{} gate (as outlined above) is not the only all-microwave two-qubit gate available. In particular, the \bSWAP{} gate\cite{Poletto2012} is an interesting alternative. The \bSWAP{} gate directly drives the $|00\rangle\leftrightarrow|11\rangle$ transition, made possible by interactions with the higher levels of the qubit, see Fig. \ref{fig:2qb_GatesLevelDiagram}. Usually, the matrix element for such a transition is small (3rd order in the coupling strength), but if the detuning between the qubits is equal to the anharmonicity, the transition rate is enhanced. Applying a sequence of Schrieffer-Wolff transformations to the coupled-qubit system, and using a carefully chosen drive frequency (close to the midpoint of $\omega_\text{q1}$ and $\omega_\text{q2}$), it can be shown\cite{Julich} that the drive gives rise to a unitary operator
\begin{equation}
U = U_{\bSWAP{}}(\theta,\phi)U_{ZZ}U_{IZ-ZI} \label{eq:Ub}
\end{equation}
with
\begin{equation}
U_{\bSWAP{}}(\theta,\phi) = \begin{bmatrix}
\cos \theta & 0 & 0 & -ie^{-i2\phi}\sin\theta\\
0 & 1 & 0 & 0\\
0 & 0 & 1 & 0\\
-ie^{-i2\phi}\sin\theta & 0 & 0 & \cos \theta\\
\end{bmatrix}, \label{eq:UbSWAP}
\end{equation}
The two unitaries $U_{ZZ}$ and $U_{IZ-ZI}$ only contain terms that commute with $U_{\bSWAP{}}(\theta,\phi)$, and their effects can be offset in post-processing\cite{Julich}. In \cref{eq:UbSWAP}, $\phi$ is the phase of the drive relative to the single-qubit drive pulses, and $\theta = \Omega_B t$ with
\begin{equation}
\Omega_B = \frac{-2g\Omega^2\left(-g \gamma \alpha_\Sigma+\gamma^2\alpha_2(\alpha_1+\Delta_{12}) + \alpha_1(\alpha_2-\Delta_{12})\right)}{(\alpha_1+\Delta_{12})(\alpha_2 -\Delta_{12})\Delta_{12}^2},
\end{equation}
where $\Omega$ is the amplitude of the drive, $\gamma$ is a dimensionless parameter quantifying the coupling coefficient of the drive to qubit 2 in units of coupling strength to qubit 1, and $\alpha_\Sigma = \alpha_1 + \alpha_2$. Explicit derivations leading to \cref{eq:Ub} can be found in the supplement of Ref.\onlinecite{Poletto2012}. By applying $U_{\bSWAP{}}$ for a time that yields $\theta = \pi/2$, and with $\phi = 0$, the resulting gate is denoted \bSWAP{} and can act as the entangling gate (together with single-qubit gates) that forms a universal gate set. Moreover, the power of the \bSWAP{} becomes apparent when one applies it for the time that yields $\theta = \pi/4$, which from the ground state $|00\rangle$ directly produces the entangled Bell state $|00\rangle + e^{i\phi}|11\rangle$. In line with the definition of $\sqrt{\iSWAP{}}$, this gate is denoted the $\sqrt{\bSWAP{}}$. In the work by Poletto \etal{}\cite{Poletto2012}, the fidelity of the \bSWAP{} gate was $\mathcal F_{\bSWAP{}} = 0.9$ (determined from quantum process tomography). The main source of error was the increased dephasing during the relatively long high-power pulse needed to drive the $|00\rangle \leftrightarrow |11\rangle$ transition. The \bSWAP{} gate can be viewed as the superconducting qubit analogue of the M\o{}lmer--S\o{}rensen gate \cite{Molmer1999}. In \cref{fig:2qb_GatesLevelDiagram}, we outline the level diagram of two coupled qubits, along with the higher levels of the qubits. The arrows indicate which coupled states are utilized to implement the corresponding gate. As an application of the \bSWAP{} gate, Colless \etal{}\cite{Colless2018} used this gate to calculate energies of the excited states of a $H_2$ molecule using a two-qubit transmon processor \cite{Colless2018}.

Another all-microwave gate is the so-called "microwave-activated \CPHASE{}" (or `\textsf{MAP}' for short)\cite{Chow2013}. The \textsf{MAP} gate is in spirit similar to the \CPHASE{} gate, where noncomputational states are used to impact a conditional phase inside the computational subspace. In contrast to \CPHASE{}, the \textsf{MAP} gate is implemented without tuning individual qubit frequencies. Rather, the canonical implementation of this gate comprises two fixed-frequency qubits, where the frequencies are carefully designed (and fabricated), such that the $|12\rangle$ and $|03\rangle$ levels are resonant. This leads to a splitting of the otherwise degenerate $|02\rangle\leftrightarrow|01\rangle$, and $|12\rangle \leftrightarrow |11\rangle$ transitions. By driving near resonance with the $|n2\rangle\leftrightarrow |n1\rangle$ transition, an effective $\sigma_z\otimes \sigma_z$ interaction is generated. In a setup comprising two fixed-frequency qubits, the \textsf{MAP} gate was used to implement the unitary
\begin{equation}
U_\textsf{MAP} = \exp\left[-i\frac{\pi}{4}\sigma_z\otimes \sigma_z\right],
\end{equation}
with a gate fidelity $\mathcal F_\textsf{MAP} = 0.87$ (determined via quantum process tomography) in a time $\tau_\textsf{MAP} = 514~$ns\cite{Chow2013}. As the number of qubits in a system increase, one drawback of this gate is the need for a precise matching of higher energy levels across multiple qubits, while simultaneously avoiding spurious couplings to other modes in the system.

The \CR{}, \bSWAP{} and \textsf{MAP} gates all have quite stringent requirements on the spectral landscape of the qubits in order to obtain fast, efficient gate operation. To address this issue, another all-microwave gate was developed, the so-called ``resonator induced phase gate" (``\textsf{RIP}")\cite{Cross2015,Puri2016}. The \textsf{RIP} gate operates by coupling two fixed-frequency qubits to a bus cavity, from which they are far detuned. By adiabatically applying and removing an off-resonant pulse to the cavity, the system undergoes a closed loop in phase space, after which the cavity is left unchanged, but the qubits acquire a state-dependent phase. By a careful choice of the amplitude and detuning of the pulse, and taking into account the dispersive shift of the cavity, a \CPHASE{} gate can be implemented on the two qubits. This effect was experimentally demonstrated by Paik \etal{}\cite{Paik2016} in a 3D transmon system\cite{Paik2011}, where four qubits are coupled to the same bus. In this setup, the \textsf{RIP} gate operation results in unitaries with weight on all four qubits simultaenously. In order to isolate just the desired two-qubit coupling terms, Paik \etal{}  developed a ``refocused" \textsf{RIP} (r\textsf{RIP}) gate that implements
\begin{equation}
U_\textsf{rRIP} = \exp\left[-i \dot \theta \sigma_z \otimes \sigma_z t\right],
\end{equation}
where the coupling rate (for an unmodulated drive) scales as
\begin{equation}
\dot \theta \propto \underbrace{\left(\frac{|\Omega V_\d|}{2\Delta_\text{cd}}\right)^2}_{\bar n} \frac{\chi}{\Delta_\text{cd}},
\end{equation}
where $\bar n$ denotes the average number of photons in the bus, $\chi$ is the dispersive shift, and $\Delta_\text{cd}$ is the detuning of the drive (d) from the cavity (c). By choosing $\dot \theta t = \pi/4$, it is possible to implement the $\CPHASE{}$ gate. The power of the \textsf{RIP} gate lies in its capability to accommodate large differences in qubit frequencies. To demonstrate this, Paik \etal{}\cite{Paik2016} performed two-qubit randomized benchmarking between pairs of qubits in a four-qubit device with frequency differences spanning $0.38$~GHz to $1.8$~GHz, all with fidelities in the range 0.96-0.98 and gate times in the range 285 to 760 ns.

\subsection{\label{sec:TunableCouplingimplementations}Gate implementations with tunable coupling}
Finally, we briefly review tunable coupling architectures, which have recently emerged as a promising alternative. The idea is to engineer an effective qubit-qubit coupling $\widetilde g$ that is tunable (typically by applying a flux), and such gates are referred to as parametric gates. This can be implemented in two different ways: ($i$) The coupling strength between two qubits is tuned by a flux, $g\rightarrow g(\Phi(t))$\cite{Bertet2006,VanderPloeg2007,Chen2016,Wulschner2016,Allman2014}, or ($ii$) the resonant frequency of the coupling element is modified $\omega_\text{coupler} \rightarrow\omega_\text{coupler}(\Phi(t))$\cite{Niskanen2007,Wallquist2006,Harrabi2009,Wang2011,Whittaker2014,Andersen2015,Yan2018}, with a fixed \emph{g}, leading to an effective time-dependent coupling parameter. When the tunable coupling element is driven at frequencies corresponding to the detuning of the qubits from the coupler, an entangling gate can be implemented.

In a setup of type $(ii)$, an implementation of the \iSWAP{} gate was demonstrated by parametrically driving a flux-tunable coupler between two fixed-frequency qubits\cite{McKay2016}, yielding a fidelity $\mathcal F_{\iSWAP{}} = 0.9823$ (using interleaved randomized benchmarking) in a time $\tau = 183$~ns. Similarly, the \bSWAP{} (and \iSWAP{}) gates were recently demonstrated, using a flux-tunable transmon connecting two fixed-frequency transmons. Driving the flux through the tunable qubit at the sum frequency of the fixed-frequency transmons results in the \bSWAP{}\cite{Roth2017} gate. This parametrically driven approach is generally significantly faster than implementations relying solely on fixed-frequency qubits.

A hybrid approach, in which a combination of tunable and fixed-frequency qubits is used, was recently demonstrated for both \iSWAP{} and \CPHASE{} gates \cite{Caldwell2017,Reagor2018,Didier2018a}. This scheme has no added tunable qubits (or resonators) acting as the coupling element, but rather, relies solely on an always-on capacitive coupling between the qubits, and the effective coupling is roughly half that of the always-on coupling. The operational principle here is to modulate the frequency of the tunable qubit (using local flux control) at the transition frequency correponding to $|01\rangle \leftrightarrow |10\rangle$ for \iSWAP{} and $|11\rangle \leftrightarrow |02\rangle$ for \CPHASE{}. Using interleaved randomized benchmarking, the authors demonstrated $\mathcal F_{\iSWAP{}} = 0.94$ ($\tau = 150$~ns), and $\mathcal F^{02}_{\CPHASE{}} = 0.93$ ($\tau = 210$~ns) and $\mathcal F^{20}_{\CPHASE{}} = 0.88$ ($\tau = 290$~ns), showing a slight asymmetry in the direction in which the \CPHASE{} is applied. This hybrid technique was used in Ref.\onlinecite{Reagor2018} to demonstrate a four-qubit GHZ state with fidelity $\mathcal F_\text{4 qubit GHZ} = 0.79$ (using state tomography). Finally, this gate-architecture was used to demonstrate a hybrid quantum/classical implementation of an unsupervised learning task (determining clustering of data), using nineteen qubits and supplemented by a classical computer as part of the minimization loop \cite{Otterbach2017}.


\section{\label{sec:readout}Qubit readout}

\noindent The ability to perform fast and reliable (high fidelity) readout of the qubit states is an important cornerstone of any quantum processor\cite{DiVincenzo2000}.

In this section, we give a brief introduction to how readout is performed on superconducting qubits. We start by reviewing the fundamental theory behind \textit{dispersive readout} -- the most common readout technique used today in the circuit QED architecture -- in which each qubit is coupled to a readout resonator. In the dispersive regime, i.e. when the qubit is detuned from the resonator frequency, the qubit induces a state-dependent frequency shift of the resonator from which the qubit state can be inferred by interrogating the resonator.

Dispersive readout allows us to map the quantum degree of freedom of the qubit onto the classical response of the linear resonator, thus transforming the readout optimization process into obtaining the best signal-to-noise ratio (SNR) of the microwave signal used to probe the resonator.

We then provide guidance on how to optimize system parameters to perform high-fidelity, single-shot readout. After choosing parameters, such as resonant frequencies and coupling rates, we address the filter and amplifier circuitry positioned in-between the qubit plane and the data aquisition hardware outside of the dilution refriderator. On this note, we review the basic principles of Purcell filters as well as parametric amplifiers, both of which are necessary to obtain fast, high-fidelity readout in scaled-up quantum processors.

\subsection{Dispersive readout}

\begin{figure}[t!]
\begin{center}
\includegraphics[width=8.6cm]{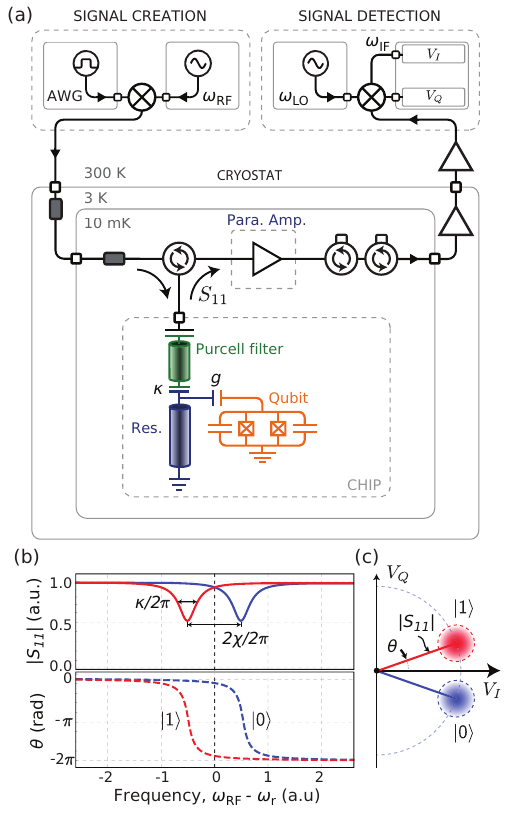}
\caption{\textbf{(a)} Simplified schematic of a representative experimental setup used for dispersive qubit readout. The resonator probe tone is generated, shaped and timed using an arbitrary waveform generator (AWG), and sent down into the cryostat. The reflected signal $S_{11}$ is amplified, first in a parametric amplifier and then in a low-noise HEMT amplifier, before it is downconverted using heterodyne mixing and finally sampled in a digitizer. \textbf{(b)} Reflected magnitude $|S_{11}|$ and phase $\theta$ response of the resonator with linewidth $\kappa$, when the qubit is in its ground state $|0\rangle$ (blue) and excited state $|1\rangle$ (red), separated with a frequency $2\chi/2\pi$. \textbf{(c)} Corresponding complex plane representation, where each point is composed of the in-plane $\mbox{Re}[S_{11}]$ and quadrature $\mbox{Im}[S_{11}]$ components. The highest state discrimination is obtained when probing the resonator just in-between the two resonances, (dashed line in (b)), thus maximizing the distance between the states.}
\label{Fig:DISP}
\end{center}
\end{figure}

A quantum measurement can be described as an entanglement of the qubit degree of freedom with a ``pointer variable" of a measurement probe with a quantum Hamiltonian\cite{Braginsky1996}, followed by classical measurement of the probe. In circuit QED, the qubit (the quantum system) is entangled with an observable of a superconducting resonator (the probe), see Fig. \ref{Fig:DISP}(a), allowing us to gain information about the qubit state by interrogating the resonator -- rather than directly interacting with the qubit. Therefore, the optimization of the readout performance is translated to maximizing the signal-to-noise ratio of a microwave probe tone sent to the resonator, while minimizing the unwanted \textit{back-action} on the qubit.

The qubit-resonator interaction is described by the Jaynes-Cummings Hamiltonian\cite{Jaynes1963,Shore1993,Gerry2005}, previously introduced in Sec. \ref{sec:circuits},

\begin{equation}
H_{\mbox{\tiny{JC}}} = \omega_r \left( a^{\dagger}a + \frac{1}{2}\right) + \frac{\freq}{2}\sigma_z + g \left( \sigma_{+}a + \sigma_{-}a^{\dagger}\right),
\label{Eq:JCH}
\end{equation}

\noindent where $\omega_r$ and $\freq$ denote the resonator and qubit frequencies, respectively, and $g$ is the transverse qubit-resonator coupling rate. The operators $\sigma_{+}$ and $\sigma_{-}$ represent the processes of exciting and de-exciting the qubit, respectively.

In the limit when the detuning between the qubit and the resonator is small compared with their coupling rate, i.e. $\Delta = |\freq - \omega_r| \ll g$, the energy levels of the two systems hybridize and a vacuum Rabi splitting of frequency $\sqrt{n}g/\pi$ opens up, where $n = 1,2,3...$ denotes the resonator mode. In this regime, excitations are coherently swapped between the two systems. Although useful for certain two-qubit gate operations, recall Sec. \ref{sec:SWAPgates}, such transverse interactions change the qubit state (since energy is directly exchanged between the resonator and the qubit) and is therefore not desired in the context of \textit{quantum non-demolition} (QND) readout, in which the outcome of the quantum measurement is not altered in the act of reading out the system.

\begin{figure}[!t]
\begin{center}
\includegraphics[width=8.6cm]{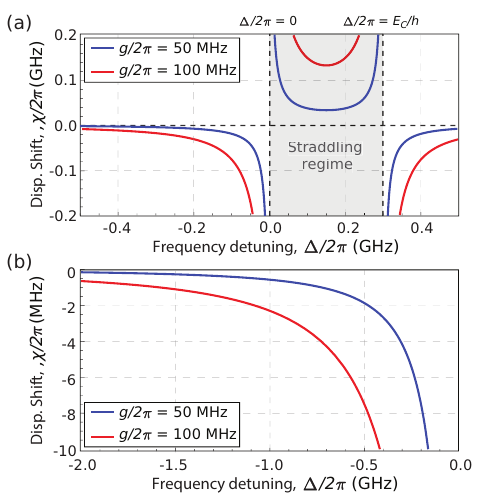}
\caption{\textbf{(a)} Dispersive frequency shift $\chi/2\pi$ as a function of qubit-resonator detuning $\Delta/2\pi$, according to Eq. (\ref{Eq:DispShiftTransmon}), for a transmon qubit with anharmonicity $\alpha/2\pi = -E_C/h = -300\,$MHz, for qubit-resonator coupling rates $g/2\pi = 50\,$MHz (blue) and $g/2\pi = 100\,$MHz (red). The two vertical asymptotes at $\Delta/2\pi = 0$ and $\Delta/2\pi = E_C$ divides the dispersive shift into three regimes; For $\Delta/2\pi < 0$, and $\Delta/2\pi > E_C/h$, the dispersive shift is negative and $\chi/2\pi \rightarrow 0^{-}$ as $\Delta\rightarrow \pm \infty$. For $0<\Delta/2\pi < E_C/2\pi$, the dispersive shift $\chi/2\pi > 0$. This is called the \textit{straddling regime}\cite{Koch2007}. \textbf{(b)} Zoomed-in plot for negative qubit-resonator detuning, the most commonly used operating regime.}
\label{Fig:DispersiveShift}
\end{center}
\end{figure}

In the dispersive limit, i.e., when the qubit is far detuned from the resonator compared with their coupling rate $g$ and the resonator linewidth $\kappa$, $\Delta \gg g, \kappa$, there is no longer a direct exchange of energy between the two systems. Instead, the qubit and resonator push each others' frequencies. To see this, the Hamiltonian can be approximated using second-order perturbation theory\cite{Blais2004,Boissonneault2009} in terms of $g/\Delta$, taken in the limit of few photons in the resonator. This is known as the \textit{dispersive approximation}, after which the Hamiltonian takes the form

\begin{equation}
H_{\text{disp}} = \bigg( \omega_r +\chi\sigma_z\bigg) \left( a^{\dagger}a + \frac{1}{2}\right) + \frac{\widetilde{\omega}_{\text{q}}}{2}\sigma_z,
\label{Eq:JCHdisp}
\end{equation}

\noindent where $\chi = g^2/\Delta$ is the qubit-state dependent frequency shift, a so-called \textit{dispersive shift}, see Fig. \ref{Fig:DISP}(b), allowing us to distinguish the two qubit states. This is an asymptotically longitudinal interaction, yielding a QND measurement. Note that, in addition, the qubit frequency also picks up a \textit{Lamb shift}, $\widetilde{\omega}_{\text{q}} = \freq + g^2/\Delta$, induced by the vacuum fluctuations in the resonator. Also note that the dispersive Hamiltonian in Eq. (\ref{Eq:JCHdisp}) is derived for a two-level atom\footnote{In reality, superconducting qubits, just like natural atoms, have higher energy levels. These higher levels are outside of the computational subspace, but need to be taken into account for most qubit simulations to get accurate predictions of frequency shifts.}. Taking the second excited state into account and introducing the anharmonicity $\alpha = \omega_{\text{q}}^{1\rightarrow 2} - \omega_{\text{q}}^{0\rightarrow 1}$ modifies the expression for the dispersive shift:

\begin{equation}
\chi = \chi_{01} + \frac{\chi_{12}}{2} = -\frac{g_{01}^2}{\Delta}\left(\frac{1}{1 + \Delta/\alpha}\right),
\label{Eq:DispShiftTransmon}
\end{equation}

\noindent which for a transmon qubit with $\alpha < 0$ implies that the dispersive shift will depend on the detuning. This effect is plotted in Fig. \ref{Fig:DispersiveShift}(a), where the second energy level manifests itself as a second vertical asymptote at $\Delta/2\pi = E_C/h$. It is also worth noting that for qubit modalities with positive anharmonicity, e.g. flux qubits, the dispersive shift will also shift the sign\cite{Yan2016}.

In the small photon-number limit, the interaction term of the Hamiltonian in Eq. (\ref{Eq:JCHdisp}) commutes with the qubit observable\footnote{This commutation is approximate and has an asyptotic dependence on the qubit-resonator detuning}, $\sigma_z$, resulting in a QND measurement\cite{Braginsky1996}. This is an important condition for many applications in quantum information processing.

In the case when the resonator photon number $n = a^{\dagger}a$ exceeds a \textit{critical photon number} $n_c \equiv \Delta^2/(4g^2)$, the dispersive Hamiltonian in Eq. (\ref{Eq:JCHdisp}) is no longer a valid approximation\cite{Blais2004,Gambetta2006,Sank2016}. Therefore, the critical photon number sets an upper bound for the power level of the resonator probe signal to maintain (an approximate) QND measurement\footnote{It has been demonstrated that it is still possible to read out the qubit state by applying a very strong resonator drive tone, eventhough this readout scheme is not QND\cite{Reed2010b}}. This limitation could be lifted by implementing a pure (and not only approximate) QND readout using a manifestly longitudinal coupling between a qubit and the resonator. Several groups are currenly pursuing the implementation of \textit{longitudinal readout}, in which QND readout could be performed even with larger number of resonator photons, thus improving the SNR\cite{Kerman2013,Didier2015,Richer2016}.

We can also interpret the dispersive qubit-resonator interaction in another way; by rearranging the terms in Eq. (\ref{Eq:JCHdisp}), we can equivalently write

\begin{equation}
H_{\text{disp}} = \omega_r \left( a^{\dagger}a + \frac{1}{2}\right) + \frac{1}{2}\bigg( \freq + \underbrace{\frac{g^2}{\Delta}}_{\text{Lamb shift}} + \underbrace{\frac{2g^2}{\Delta}a^{\dagger}a}_{\text{ac-Stark shift}}\bigg)\sigma_z,
\label{Eq:JCHdisp2}
\end{equation}

\noindent where the bare qubit frequency is shifted by a fixed amount $g^2/\Delta$, known as the \textit{Lamb shift}\footnote{It is worth mentioning that the observed qubit frequency is always the Lamb-shifted frequency and not the bare qubit frequency.} as well as an amount proportional to the number of photons populating the resonator\cite{Blais2004,Koch2007}. This effect is known as the \textit{ac-Stark shift}. It has the consequence that photon number fluctuations (noise) in the resonator induce small shifts of the qubit frequency, slightly bringing the qubit out of its rotating frame and thus causing dephasing\cite{Schuster2005}. This means that spurious photon occupation and fluctuation in the resonator, be it thermal or coherent photons, shift the qubit frequency and causing dephasing\cite{Gambetta2006,Zhang2017}. For this reason, it is important to make sure that the processor is properly thermalized\cite{Yan2018}, and its control lines well filtered\cite{ReedPhD2013} and attenuated\cite{Yeh2017}, to reduce photon number fluctuation.

\subsection{Measuring the resonator amplitude and phase}

\begin{figure}[!t]
\begin{center}
\includegraphics[width=8.6cm]{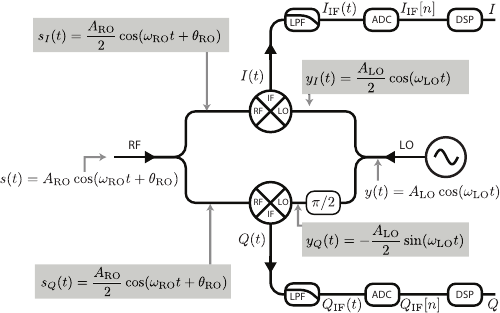}
\caption{Schematic of an I-Q mixer. A readout pulse at frequency $\omega_{\mbox{\tiny{RO}}}$ enters the RF port, where it is equally split into two paths. A local oscillator at frequency $\omega_{\mbox{\tiny{LO}}}$ enters the LO port, where it is equally split into two paths, one of which undergoes a $\pi/2$-radian phase rotation. To perform analog modulation, the two signals in each path are multiplied at a mixer, yielding the outputs $I(t)$ and $Q(t)$, each having frequencies $\omega_{\mbox{\tiny{RO}}} \pm \omega_{\mbox{\tiny{LO}}}$. $I(t)$ and $Q(t)$ are then low-pass-filtered (time averaged) to yield $I_{\mbox{\tiny{IF}}}(t)$ and $Q_{\mbox{\tiny{IF}}}(t)$ at the intermediate frequency $\omega_{\mbox{\tiny{IF}}} = |\omega_{\mbox{\tiny{RO}}} - \omega_{\mbox{\tiny{LO}}}|$, and subsequently digitized using an analog-to-digital (ADC) converter. If $\omega_{\mbox{\tiny{IF}}} \neq 0$, then digital signals $I_{\mbox{\tiny{IF}}}[n]$ and $Q_{\mbox{\tiny{IF}}}[n]$ are further digitally demodulated using digital signal processing (DSP) techniques to extract the amplitude and phase of the readout signal.}
\label{Fig:IQ-mixer}
\end{center}
\end{figure}

In the previous section, we outlined the underlying physics behind the dispersive readout technique, in which we concluded that the qubit induces a state-dependent frequency shift of the resonator. 
We now focus our attention on how to 
probe the resonator to ``read out the qubit,'' that is, best distinguish the two classical resonator signatures corresponding to our qubit states, see Fig. \ref{Fig:DISP}(b)-(c).

The readout circuit can be set up in measuring either reflection or transmission. The best state discrimination is obtained by maximizing the separation between the two states in the $(I,Q)$-plane, i.e. the in-phase and quadrature component of the voltage, see Fig. \ref{Fig:DISP}(c). It can be shown that this separation is maximal when the resonator is probed just in-between the two qubit-state dependent resonance frequencies~\cite{SankPhD2014}, $\omega_{\mbox{\tiny{RF}}} = (\omega_{r}^{|0\rangle} + \omega_{r}^{|1\rangle})/2$. In this case, the reflected magnitude is identical for $|0\rangle$ and $|1\rangle$, and all information about the qubit state is encoded in the phase $\theta$, see dashed line in Fig. \ref{Fig:DISP}(b). In turn, the qubit-resonator detuning should be designed to obey the criterion for maximal state visibility, $\chi = \kappa/2$, which is maximized for phase measurements while constraining qubit dephasing.

Once we have picked the resonator probe frequency, the quantum dynamics of the qubit can be mapped onto the phase of the classical microwave response. In the following, we discuss how we can use heterodyne detection to measure the phase of the resonator response. We assume that the reader is already somewhat familiar with basic mixer operations, such as modulation and de-modulation of signals. For interested readers, we refer to Ref.~\onlinecite{Marki2010}.

\subsubsection{Representation of the readout signal}

A readout event commences with a short microwave tone directed to the resonator at the resonator probe frequency $\omega_{\mbox{\tiny{RO}}}$. After interacting with the resonator, the reflected (or transmitted) microwave signal has the form
\begin{align}\label{Eq:RO-signal}
s(t) &= A_{\mbox{\tiny{RO}}} \cos(\omega_{\mbox{\tiny{RO}}}t + \theta_{\mbox{\tiny{RO}}}), 
\end{align}
\noindent where $\omega_{\mbox{\tiny{RO}}}$ is the \textit{carrier frequency} used to probe the resonator. $A_{\mbox{\tiny{RO}}}$ and $\theta_{\mbox{\tiny{RO}}}$ are, respectively, the qubit-state-dependent amplitude and phase that we want to measure.
%
%
%
One can equivalently use a \textit{complex analytic representation} of the signal,
\begin{align}\label{Eq:analytic-representation}
s(t) &= \operatorname{Re} \left\{ A_{\mbox{\tiny{RO}}} e^{j (\omega_{\mbox{\tiny{RO}}} t + \theta_{\mbox{\tiny{RO}}})} \right\}, \\
     &= \operatorname{Re} \left\{ A_{\mbox{\tiny{RO}}} \cos(\omega_{\mbox{\tiny{RO}}} t + \theta_{\mbox{\tiny{RO}}})
                                + j \sin (\omega_{\mbox{\tiny{RO}}} t + \theta_{\mbox{\tiny{RO}}}) \right\} \nonumber 
\end{align}
where $\operatorname{Re}$ takes the real part of an expression, e.g., $\operatorname{Re} [\exp(jx)] = \operatorname{Re} (\cos x + j\sin x) = \cos x$.

To gain intuition, we can rewrite Eq.~(\ref{Eq:analytic-representation}) in a static ``phasor'' notation that separates out the time dependence $\exp(j \omega_{\mbox{\tiny{RO}}} t)$,
\begin{align}\label{Eq:phasor}
 s(t) &= \operatorname{Re} \left\{ \underbrace{A_{\mbox{\tiny{RO}}} e^{j \theta_{\mbox{\tiny{RO}}}}}_{\mathrm{phasor}} e^{j \omega_{\mbox{\tiny{RO}}} t}  \right\},
\end{align}
where the phasor $A_{\mbox{\tiny{RO}}} \exp(j \theta_{\mbox{\tiny{RO}}}) \equiv A_{\mbox{\tiny{RO}}} \angle \theta_{\mbox{\tiny{RO}}}$ is a shorthand that fully specifies a harmonic signal $s(t)$ at a known frequency $\omega_{\mbox{\tiny{RO}}}$.
To perform qubit readout, we want to measure the ``in-phase'' component $I$ and a ``quadrature'' component $Q$ of the complex number represented by the phasor,
\begin{align}
    A_{\mbox{\tiny{RO}}} e^{j \theta_{\mbox{\tiny{RO}}}} &=
    A_{\mbox{\tiny{RO}}} \cos \theta_{\mbox{\tiny{RO}}} + j  A_{\mbox{\tiny{RO}}} \sin \theta_{\mbox{\tiny{RO}}} \\
    &\equiv I + j Q
\end{align}
to determine the amplitude $A_{\mbox{\tiny{RO}}}$ and the phase $\theta_{\mbox{\tiny{RO}}}$.

\begin{figure*}[htp!]
\begin{center}
\includegraphics[width=18.2cm]{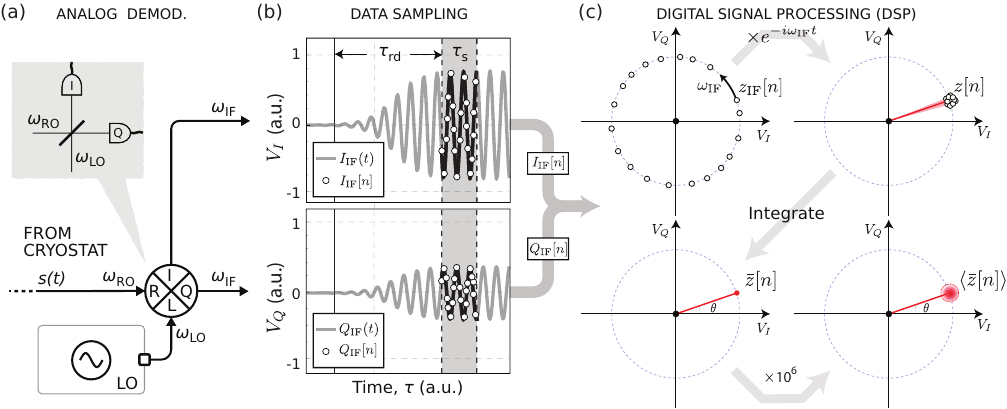}
\caption{Schematic of the heterodyne detection technique. \textbf{(a)} The signal with frequency $\omega_{\mbox{\tiny{RF}}}$ from the cryostat is mixed with a carrier tone with frequency $\omega_{\mbox{\tiny{LO}}}$, yielding two quadratures at a down-converted intermediate frequency $\omega_{\mbox{\tiny{IF}}} = |\omega_{\mbox{\tiny{RO}}} - \omega_{\mbox{\tiny{LO}}}|$, and 90$^{\circ}$ out-of-phase with each other. \textbf{(b)} The two signals are passed into two different analog-to-digital converter (ADC) channels. To avoid sampling the resonator transient, some readout delay ($\tau_{rd}$) corresponding to the resonator linewidth may be added, and the two signals are sampled for a time $\tau_s$. In this case, the white dots represent the sampled points. \textbf{(c)} The sampled traces are post-processed and after some algebra, the sampled data points are averaged into a single point in the $(I,Q)$-plane. To extract statistics of the readout performance, i.e. single-shot readout fidelity, a large number of $(I,Q)$-records are acquired, yielding a 2D-histogram, with a Gaussian distributed spread given by the noise acting on the signal.}
\label{Fig:HeterodyneMixing}
\end{center}
\end{figure*}

\subsubsection{I-Q mixing}
One direct means to extract $I$ and $Q$ is to perform a \textit{homodyne} or \textit{heterodyne} measurement using an analog \textit{I-Q mixer}. Figure ~\ref{Fig:IQ-mixer} shows a basic electrical schematic of an I-Q mixer. The readout signal $s(t)$ and a reference local-oscillator signal $y(t)= A_{\mbox{\tiny{LO}}} \cos \omega_{\mbox{\tiny{LO}}} t$ are fed into the mixer via the RF and LO mixer ports. The mixer then equally splits the signal and local oscillator into two branches and multiplies them in the following way: in the $I$-branch, the signal $s_I(t) = s(t)/2$ is multiplied by the local oscillator $y_I(t)= (A_{\mbox{\tiny{LO}}}/2) \cos \omega_{\mbox{\tiny{LO}}} t$; and in the $Q$-branch, the signal $s_Q(t) = s(t)/2$ is multiplied by a $\pi/2$-phase-shifted version of the local oscillator, $y_Q(t)= - (A_{\mbox{\tiny{LO}}}/2) \sin \omega_{\mbox{\tiny{LO}}} t$. The ``-'' sign arises from the choice of using a $A(\cos \omega t+\phi)$ as the reconstructed real signal. At the $I$ and $Q$ ports, the output signals $I(t)$ and $Q(t)$ contain terms at the sum and difference frequencies, generally referred to as an \textit{intermediate frequency}, $\omega_{\mbox{\tiny{IF}}} = \omega_{\mbox{\tiny{RO}}} \pm \omega_{\mbox{\tiny{LO}}}$. The resulting signals are low-pass filtered, passing only the terms at the difference frequency, $I_{IF}(t)$ and $Q_{IF}(t)$, which are then digitized. After digital signal processing, one obtains the static in-phase ($I$) and quadrature ($Q$) components, from which one calculates the amplitude $A_{\mbox{\tiny{RO}}}$ and the phase $\theta_{\mbox{\tiny{RO}}}$.

Microwave mixers use square-law-type diodes to implement multiplication. The optical analog of a mixer operation is a combination of a balanced (50-50) beamsplitter followed by optical photodetectors, as shown in the inset of Fig.~\ref{Fig:HeterodyneMixing}(a). The signal and local-oscillator optical fields are first combined at the beamsplitter, yielding superpositions of both fields, and then detected at the photodetectors, which act as square-law devices. To build intuition for how this works, tbe square of the sum of two electric fields $(E_1 + E_2)^2 = E_1^2 + E_2^2 + 2 E_1E_2$ has a cross term that is the multiplication of the two fields. We refer the reader to Ref.~\onlinecite{Mandel1995} for further details.

\subsubsection{Homodyne demodulation}
One direct means to extract $I$ and $Q$ is to perform a microwave \textit{homodyne} measurement using an analog I-Q mixer of the type shown in Fig.~\ref{Fig:IQ-mixer}. %
In an analog homodyne measurement, the local oscillator (LO) is chosen to be at the carrier frequency $\omega_{\mbox{\tiny{LO}}}=\omega_{\mbox{\tiny{RO}}}$. Upon mixing, $I(t)$ and $Q(t)$ contain terms at both DC ($\omega_{\mbox{\tiny{IF}}}=0$) and terms at twice the carrier frequency. Time-averaging (filtering) $I(t)$ and $Q(t)$ directly yield the DC terms $I_{IF}(t) =I$ and $Q_{IF}(t) =  Q$:
\begin{align}
 I &= \frac{1}{T} \int_0^T dt \; s_I(t) y_I(t) \nonumber\\
   &= \frac{A_{\mbox{\tiny{RO}}}A_{\mbox{\tiny{LO}}}}{8} \cos (\theta_{\mbox{\tiny{RO}}}),  \\
 Q &= \frac{1}{T} \int_0^T dt \; s_Q(t) y_Q(t) \nonumber \\
   &= \frac{A_{\mbox{\tiny{RO}}}A_{\mbox{\tiny{LO}}}}{8} \sin (\theta_{\mbox{\tiny{RO}}}),
\end{align}
where $T$ is a time interval taken to be an integer number of periods of the readout signal. $I$ and $Q$ are then sampled and used to calculate the amplitude and phase:
\begin{align}
  A_{\mbox{\tiny{RO}}} &\propto \sqrt{I^2 + Q^2}, \\
  \theta_{\mbox{\tiny{RO}}} &= \arctan(Q/I).
\end{align}
Note that the global value of $A_{\mbox{\tiny{RO}}}$ or $\theta_{\mbox{\tiny{RO}}}$ is not what matters; what matters is the \textit{change} in $A_{\mbox{\tiny{RO}}}$ and $\theta_{\mbox{\tiny{RO}}}$ between the qubit being in state $\vline 0 \rangle$ and state $\vline 1 \rangle$. For example, the value of $A$ leaving the resonator and the value $G \times A$ reaching a measurement stage are different, where $G$ represents the net gain in the measurement amplifier chain. However, the gain is the same, independent of the qubit state, whereas $A$ may be different, e.g., $A_{\mbox{\tiny{RO}}}^{(0)}=G \times A_{\lvert 0 \rangle}$ or $A_{\mbox{\tiny{RO}}}^{(1)}=G \times A_{\lvert 1 \rangle}$. Similarly, the propagation phase $\phi$ accumulated while a signal travels between the resonator and the measurement stage is also independent of the qubit state, and simply imparts a phase offset to the qubit-induced phase shift, e.g., $\theta_{\mbox{\tiny{RO}}}^{(0)}=\theta_{\lvert 0 \rangle} + \phi$ or $\theta_{\mbox{\tiny{RO}}}^{(1)}=\theta_{\lvert 1 \rangle} + \phi$.

Homodyning works in principle, but there are two drawbacks. First, signals directly demodulated to DC may be subject to lower signal-to-noise ratios, since they fight against $1/f$ electronics noise, as well as any other noise signals that may have inadvertently been demodulated (e.g., via a square-law detector). The second is that homodyning is not compatible with frequency division multiplexing (FDM), where a single pulse can be used to interrogate $N$ resonators at different frequencies by applying tones at each resonator frequency using the superposition principle, e.g.,
\begin{equation}
s(t) = \sum_{i=1}^N A_{\mbox{\tiny{RO}}}^{(i)} \cos(\omega_{\mbox{\tiny{RO}}}^{(i)}t + \theta^{(i)}).
\end{equation}
Homodyning an FDM signal will put \textit{all} resonator signals at DC, and once downconverted, they cannot be differentiated.
To work around this, it is generally advantageous to use \textit{heterodyning}, which uses a two-step demodulation process via an intermediate frequency $\omega_{\mbox{\tiny{IF}}}$. Such a scheme is easily compatible with the concept of FDM, because a readout signal is first demodulated to unique IF frequencies  $\omega_{\mbox{\tiny{IF}}}^{(i)}$, and then digitally demodulated to extract each $A_{\mbox{\tiny{RO}}}^{(i)}$ and $\theta^{(i)}$.
In the following, we will consider $N=1$ for simplicity, but the process is applicable to larger $N$ provided the frequencies a sufficiently spaced to avoid interference with one another during the demodulation process.

\subsubsection{Heterodyne demodulation}
In a heterodyne scheme, a local oscillator at frequency $\omega_{\mbox{\tiny{LO}}}$ is offset by an intermediate frequency $\omega_{\mbox{\tiny{IF}}}$ to target a unique readout frequency $\omega_{\mbox{\tiny{RO}}}$. Up-conversion techniques such as single-sideband modulation with suppressed carrier (SSB-SC) using balanced I-Q mixers (operated in reverse compared with Fig.~\ref{Fig:IQ-mixer}) are commonly used to create such readout signals. We refer the reader to Ref.~\onlinecite{Marki2010} for more information on how to create such pulses.\\
\indent Here, we want to extract $A_{\mbox{\tiny{RO}}}$ and $\theta_{\mbox{\tiny{RO}}}$ (or their scaled and offset versions) from the reflected/transmitted tone using a heterodyning scheme. The first step is to perform analog I-Q mixing, as illustrated in Fig.~\ref{Fig:HeterodyneMixing}(a).
In contrast to the homodyning case, here, the local oscillator and readout tone are at different frequencies, $\omega_{\mbox{\tiny{IF}}} = |\omega_{\mbox{\tiny{RO}}} - \omega_{\mbox{\tiny{LO}}}| > 0$. Mixing the LO and RO signals yields the signals $I(t)$ and $Q(t)$ with terms at both sum and difference frequencies. Filtering out the sum frequencies using low-pass filtering (time averaging) yields the IF signals:
\begin{align}
  I_{\mbox{\tiny{IF}}}(t) &= \frac{1}{T}\int_0^T (dt) \; s_I(t) y_I(t) \nonumber \\
            &= \frac{A_{\mbox{\tiny{RO}}}A_{\mbox{\tiny{LO}}}}{8} \cos (\omega_{\mbox{\tiny{IF}}}t + \theta_{\mbox{\tiny{RO}}}) \\
  Q_{\mbox{\tiny{IF}}}(t) &= \frac{1}{T}\int_0^T (dt) \; s_Q(t) y_Q(t) \nonumber \\
            &= \frac{A_{\mbox{\tiny{RO}}}A_{\mbox{\tiny{LO}}}}{8} \sin (\omega_{\mbox{\tiny{IF}}}t + \theta_{\mbox{\tiny{RO}}}).
\end{align}
As before, we have omitted any offset phases from the LO or from the wave propagation between the resonator and the measurement. Again, these offset values are not what matters; it is the change in $A_{\mbox{\tiny{RO}}}$ and $\theta_{\mbox{\tiny{RO}}}$ with a change in qubit state that allows state discrimination.

The analog-demodulated $I_{\mbox{\tiny{IF}}}(t)$ and $Q_{\mbox{\tiny{IF}}}(t)$ are now oscillating at a frequency that is generally low enough to be digitized using commonly available analog-to-digital converters (ADCs). The resulting digital signals are now written as $I_{\mbox{\tiny{IF}}}[n]$ and $Q_{\mbox{\tiny{IF}}}[n]$,
\begin{align}
  I_{\mbox{\tiny{IF}}}[n] &= \frac{A_{\mbox{\tiny{RO}}}A_{\mbox{\tiny{LO}}}}{8} \cos (\Omega_{\mbox{\tiny{IF}}}n + \theta_{\mbox{\tiny{RO}}}) \\
  Q_{\mbox{\tiny{IF}}}[n] &= \frac{A_{\mbox{\tiny{RO}}}A_{\mbox{\tiny{LO}}}}{8} \sin (\Omega_{\mbox{\tiny{IF}}}n + \theta_{\mbox{\tiny{RO}}}),
\end{align}
where $n=t/\Delta t$ indexes the sample number of the continuous-time signals $I_{\mbox{\tiny{IF}}}(t)$ and $Q_{\mbox{\tiny{IF}}}(t)$, $\Omega_{\mbox{\tiny{IF}}} = \omega_{\mbox{\tiny{IF}}} \Delta t$ is the digital frequency, and $\Delta t$ is the sampling period (typically around 1 ns). Pulsing the resonator is necessarily accompanied by a ring-up time, related to the quality factor of the resonator, and the first few samples may decrease overall signal-to-noise. Consequently, a delayed window of samples $[n_1:n_2]$ is often used to perform the second digital demodulation of the discrete-time signals $I_{\mbox{\tiny{IF}}}[n_1:n_2]$ and $Q_{\mbox{\tiny{IF}}}[n_1:n_2]$. Note that more complicated windowing functions may also be used to improve state discrimination, but here we use a simple boxcar [see Fig.~\ref{Fig:HeterodyneMixing}(b)].

Digital demodulation comprises the point-by-point multiplication of $I_{\mbox{\tiny{IF}}}[n_1:n_2]$ and $Q_{\mbox{\tiny{IF}}}[n_1:n_2]$ by $\cos \Omega_{\mbox{\tiny{IF}}} n$ and $\sin \Omega_{\mbox{\tiny{IF}}} n$. Averaging the resulting time series eliminates the  $2\Omega_{\mbox{\tiny{IF}}}$ component while retaining the DC component, as in a homodyne measurement, one obtains
\begin{align}
 I &= \frac{1}{M} \sum_{n_1}^{n_2} I_{\mbox{\tiny{IF}}}[n] \cos[\Omega_{\mbox{\tiny{IF}}} n]
        = \frac{A_{\mbox{\tiny{RO}}}A_{\mbox{\tiny{LO}}}}{16} \cos \theta_{\mbox{\tiny{RO}}}, \\
 Q &= \frac{1}{M} \sum_{n_1}^{n_2} Q_{\mbox{\tiny{IF}}}[n] \sin[\Omega_{\mbox{\tiny{IF}}} n]
        = \frac{A_{\mbox{\tiny{RO}}}A_{\mbox{\tiny{LO}}}}{16} \sin \theta_{\mbox{\tiny{RO}}},
\end{align}
where $M=n_2-n_1+1$. As before, $I$ and $Q$ can then be used to find $A_{\mbox{\tiny{RO}}}$ and $\theta_{\mbox{\tiny{RO}}}$.

The same procedure may be view in the complex $I-Q$ plane by the analytic function $z_{\mbox{\tiny{IF}}}[n]$, as illustrated in Fig.~\ref{Fig:HeterodyneMixing}(c-d),
\begin{align}
    z_{\mbox{\tiny{IF}}}[n] &= I_{\mbox{\tiny{IF}}}[n]  + j Q_{\mbox{\tiny{IF}}}[n] \equiv V_{I}[n] + j V_Q[n] \\
         &= \frac{A_{\mbox{\tiny{RO}}}A_{\mbox{\tiny{LO}}}}{8}
            \left[ \cos (\Omega_{\mbox{\tiny{IF}}}n + \theta_{\mbox{\tiny{RO}}}) +  j \sin (\Omega_{\mbox{\tiny{IF}}}n + \theta_{\mbox{\tiny{RO}}}) \right] \\
         &= \frac{A_{\mbox{\tiny{RO}}}A_{\mbox{\tiny{LO}}}}{8} e^{j\theta_{\mbox{\tiny{RO}}}} e^{j\Omega_{\mbox{\tiny{IF}}}n}
\end{align}
where the digital in-phase and quadrature signals are represented here as the voltages $V_{I}[n]$ and $V_Q[n]$ sampled by the ADC, and we have separated the static phasor $(A_{\mbox{\tiny{RO}}}A_{\mbox{\tiny{LO}}}/8) \exp[j\theta_{\mbox{\tiny{RO}}}]$ from the rotating term $\exp[j\Omega_{\mbox{\tiny{IF}}}n]$. One can digitally demodulate the  time series $z_{\mbox{\tiny{IF}}}[n]$ by multiplying by the complex conjugate of the oscillatory exponential,
\begin{align}
    z[n] &= z_{\mbox{\tiny{IF}}}[n].*e^{-j\Omega_{\mbox{\tiny{IF}}}n} 
\end{align}
where $.*$ indicates a point-by-point multiplication, and the result is a vector of length $M$ of nominally identical values of the phasor -- one for each sample point -- with a small amount of additive noise due to noise in measurement chain, digitization errors, etc. A singular phasor value is then estimated by taking average,
\begin{align}
    \bar{z}[n] &= \frac{1}{M} \sum z[n]  \\
         &= \frac{A_{\mbox{\tiny{RO}}}A_{\mbox{\tiny{LO}}}}{8} e^{j\theta_{\mbox{\tiny{RO}}}}.
\end{align}
Such ``single-shot measurements'' may then be repeated a large number of times to obtain an ensemble average $\langle \bar{z}[n] \rangle$.

\subsection{Weak and strong qubit measurements: Impact of noise}

In quantum measurements, noise plays an essential role as it dictates the fidelity of its outcome\cite{Caves1982,Clerk2010}, recall Fig. \ref{Fig:HeterodyneMixing}(c). In the absence of noise, any non-zero dispersive shift (resulting in a resonator field displacement) would suffice to unambigously separate the qubit states, given a properly chosen resonator linewidth. In practice, however, the outcome of the quantum measurement is generally Gaussian distributed in the $(I,Q)$-plane due to presence of classical and quantum noise. In this section, we review the main sources of noise, as well as how it impedes our ability to extract information from the quantum system. For a rigorous discussion of noise and quantum measurements, the interested reader is referred e.g. to the work by Clerk \textit{et al}\cite{Clerk2010} and to the textbook by Haus\cite{Haus2000}.

The total noise added to the signal has multiple origins. One part of the noise is associated with the microwave signal used to probe the resonator, where each photon has an intrinsic quantum noise power of $\hbar\omega/2$ per unit bandwidth. Another contribution comes from the phase-preserving amplifiers, adding both classical noise and at least $\hbar\omega/2$ of noise as required by Heisenberg's uncertainty relation. Finally, any attenuation of the signal prior to the first amplifier will appear as added noise. Combined, these noise sources amount to a \textit{system noise temperature}, which can be characterized using a sensitive thermometer, such as a shot-noise tunnel junction\cite{Simoen2015} or a qubit\cite{Macklin2015} as a sensor.

\begin{figure}[t!]
\begin{center}
\includegraphics[width=8.6cm]{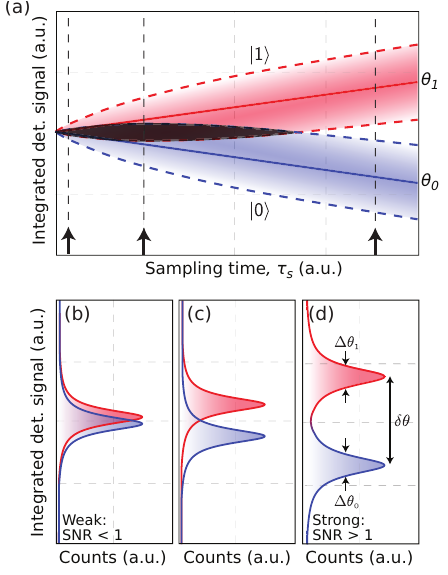}
\caption{\textbf{(a)} Qubit state distribution throughout the course of sampling the readout signal, in the presence of noise. The separation between the peaks (solid lines) increases linearly in time, whereas the peak widths only increase as $\sqrt{t}$. Image inspired by Clerk \textit{et al}\cite{Clerk2010}. The three black arrows represent line cuts for three sampling times: \textbf{(b)} For short sampling time, the states are not separated, resulting in a weak meaurement ($\mbox{SNR} < 1$). \textbf{(c)} After a longer sampling time, the peaks starts to get separated, \textbf{(d)} finally getting fully resolved, resulting in a strong measurement ($\mbox{SNR} > 1$).}
\label{Fig:Noise}
\end{center}
\end{figure}

The noise results in time-dependent fluctuations of the measured signal, which in turn translates into uncertainty in our demodulated signals, see Fig. \ref{Fig:HeterodyneMixing}(c). This can be intuitively understood by considering that our heterodyne detection method requires us to sample for a finite amount of time.

To quantify the impact of the noise on our measurement, we first project the distributed $(I,Q)$ data -- corresponding to $|0\rangle$ and $|1\rangle$ -- onto the axis for which their relative separation in the complex plane is maximized\footnote{When analyzing the readout data, we have the freedom to choose projections.}. The line that is used to separate between $|0\rangle$ and $|1\rangle$ is called a \textit{separatrix}.

The noise can now be quantified by comparing the widths of the Gaussian probability distribution surrounding the mean with the peak separation in the $(I,Q)$-plane, thus defining a signal-to-noise ratio $\mbox{SNR} = \delta\theta/(\Delta\theta_{1} + \Delta\theta_{0})$, see Fig. \ref{Fig:Noise}(a), with $\delta\theta = |\theta_{1} - \theta_{0}|$ representing the signal and $\Delta\theta_0$, $\Delta\theta_1$ represent the noise (2$\sigma$) of each distribution. The SNR allows us to distinguish between a weak and a strong quantum measurement, as illustrated in Fig. \ref{Fig:Noise}(b)-(d).

In a weak measurement, the probabilities are broadly distributed as compared to their relative separation ($\mbox{SNR} < 1$), which means that only partial information of the quantum state is revealed to the observer, see Fig.\ref{Fig:Noise}(b). In a strong measurement, on the other hand, the quantum state is collapsed onto one of the two eigenstates. In this case, the outcome of the measurement can be distinguished unambigously, which is reflected in two fully separated distributions ($\mbox{SNR} > 1$), see Fig. \ref{Fig:Noise}(d).

In many applications of quantum measurements, it is necessary to unambigously (and with high fidelity) tell the outcome without repeating the readout measurement. This is known as \textit{single-shot readout} and it often requires the use of a parametric amplifier -- a preamplifier used to increase system SNR -- which is further discussed in Sec. \ref{sec:TWPA}.

Assuming that the widths of the two distributions are identical, $\Delta\theta_0 = \Delta\theta_1 = \Delta\theta$, the separation error can be calculated by deriving the weight of the overlapping region of the Gaussian distributions as\cite{SankPhD2014}

\begin{equation*}
\epsilon_{\mbox{\tiny{sep}}} = \frac{1}{\sqrt{2\pi(\Delta\theta)^2}} \int_{\theta = \frac{\theta_0 - \theta_1}{2}}^{\infty}\exp\left[{-\frac{(\theta - \theta_1)^2}{2(\Delta\theta)^2}}\right]d\theta
\end{equation*}
\begin{equation}
 = \frac{1}{2}\mbox{erfc}\left[ \frac{\left| \theta_0 - \theta_1\right|}{2\sqrt{2(\Delta\theta)^2}}\right],
\label{Eq:EpsilonError}
\end{equation}

\noindent where erfc(x) denotes the complementary Gaussian error function, defined as

\begin{equation}
\mbox{erfc}(x) = 1 - \frac{2}{\sqrt{\pi}}\int_{x}^{\infty}e^{-t^2}dt.
\label{Eq:EpsilonError2}
\end{equation}

Using the erfc in Eq. (\ref{Eq:EpsilonError2}), the separation error in Eq. (\ref{Eq:EpsilonError}) can be compactly expressed in terms of the signal-to-noise ratio,

\begin{equation}
\epsilon_{\mbox{\tiny{sep}}} = \frac{1}{2}\mbox{erfc}\left[\frac{\mbox{SNR}}{2}\right]
\label{Eq:EpsilonError3}
\end{equation}

Note, however, that the separation error between the two state distributions only tells us the signal-to-noise ratio of our detection scheme. On top of the separation error, fidelity is reduced if the qubit relaxes (or is excited) during the readout. This will result in a count on the ``wrong" side of the threshold. This leads to an additional constraint on the readout; The readout cycle needs to be completed on a timescale much shorter than the qubit relaxation time.

In summary, we see that to optimize the qubit readout fidelity, the readout needs to fulfill the following two requirements:

\begin{itemize}

	\item{\textbf{Fast readout}: The readout cycle needs to be completed within a time that is short compared with the qubit coherence time. The longer the readout time, the more likely the qubit is to relax, thus reducing readout fidelity.}

	\item{\textbf{High signal-to-noise ratio}: The signal-to-noise ratio needs to be sufficiently large to suppress the state separation errors below an acceptable limit where it does not limit the readout fidelity.}

\end{itemize}

In sections \ref{sec:Purcell} and \ref{sec:TWPA}, we review how these two conditions are met by carefully engineering the signal path of the readout circuitry.

\subsection{\label{sec:Purcell} ``Purcell filters" for faster readout}

To ensure high-fidelity readout performance, it is important to perform single-shot readout at a timescale much shorter than the qubit coherence time, $\tau_{ro} \ll T_1$. This motivates us to: \textit{(i)} make the resonator linewidth wide, thus reducing its ring-up time, $\tau_{rd}$, and \textit{(ii)} keep the integration time $\tau_s$ as short as possible, see Fig.\ref{Fig:HeterodyneMixing}(b). The ability to isolate a quantum system from decohering into its environment while, at the same time, being able to read out its state in a short time represents two contradictory criteria, which must be traded-off\cite{Caves1982}.

Even though dispersive readout (in the few-photon limit) has only a small back-action on the qubit state, the qubit will still suffer from $T_1$-relaxation while we are performing a measurement. In fact, this ``decay during the readout" often limits the readout fidelity, reducing it to

\begin{equation}
F(\tau_{ro}) = 1-e^{-\tau_{ro}/T_1},
\label{Eq:FTauRO}
\end{equation}

\noindent where $\tau_{ro} = \tau_{rd} + \tau_{s}/2$ denotes the total time for the readout, consisting of the readout delay $\tau_{rd}$ due to the resonator transient, and half the sampling time $\tau_{s}/2$. The fidelity drop in Eq. (\ref{Eq:FTauRO}) can be interpreted as a manifestation of the competition between the time scales at which our quantum information reaches our detector or the environment first.

The limitation of qubit coherence originates from an enhanced spontaneous emission of photons, induced by its environment. This is known as the Purcell effect\cite{Purcell1946}, and is an important consideration when designing qubit-resonator systems\cite{Houck2008}. The portion of spontaneous emission that is mediated by the resonator describes how qubit relaxation is enhanced by the resonator Q when on-resonance, and suppressed off-resonance. The aim of this section is twofold: first, we develop an intuition for how the Purcell decay limits qubit coherence, and second, how to properly mitigate this limitation by designing a so-called \textit{Purcell filter}, which modifies the impedance seen by the qubit through the readout resonator. This allows us to maintain fast readout, while protecting the qubit from relaxing into its environment.

If we would just choose qubit and resonator operation frequencies guided by the resonator linewidth $\kappa$, qubit-resonator coupling $g$, and the amount of dispersive shift $\chi$, we would reduce the detuning between the qubit and the resonator, thus maximizing the dispersive shift (recall Fig. \ref{Fig:DispersiveShift}). However, this presents a trade-off between two important system parameters; on one hand, we want the qubit to be isolated from the resonator environment off-resonant to avoid Purcell-enhanced decay. On the other hand, looking at the dispersive shift, we want the two rates, $g$ and $\kappa$ to be strong, yielding larger dispersive shift as well as short resonator transient and thus a faster readout.

Fortunately, when operating in the dispersive regime, the qubit and resonator are far detuned from each other $\Delta \gg g,\kappa$, which means that their impedance (environment) can be independently engineered through filter design. In essence, one designs a filter to have strong coupling to the readout port at the resonator frequency (large $\kappa$), but isolates the qubit from its environment at the qubit frequency\cite{Reed2010a,Jeffrey2014}. In other words, an impedance transformation.


Depending on the design of the readout for the quantum processor to which the filter should be coupled, there are different ways to design a Purcell filter; such as quarter-wave stubs\cite{Reed2010a}, low-Q bandpass filters\cite{Jeffrey2014,Kelly2015}, and stepped-impedance filters\cite{Bronn2015}. Which one is optimal depends on system properties such as qubit-resonator detunings, required bandwidth, and allowed insertion loss.

The most promising Purcell filter designs are the ones that allow for frequency multiplexing, such as the low-Q bandpass filter design\cite{Jeffrey2014,Kelly2015}, which in addition to Purcell filtering has the function of a quantum bus, connecting several frequency-multiplexed readout resonators sharing the same amplifier chain.

\begin{figure}[htp!]
\begin{center}
\includegraphics[width=8.6cm]{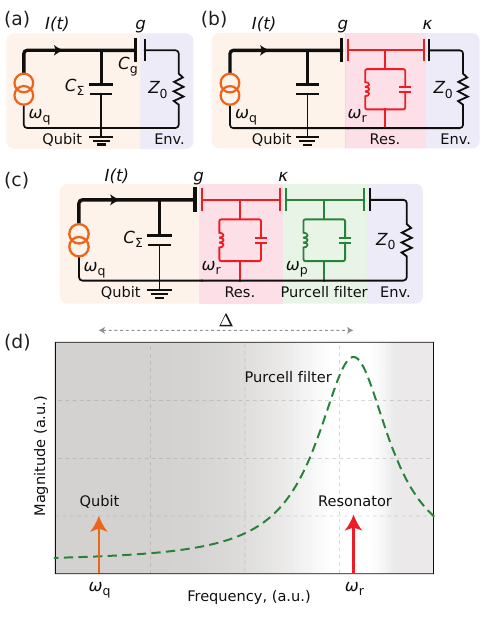}
\caption{\textbf{(a)} Circuit representation of qubit (orange) coupled to an environment (blue) with a load resistor, $Z_0$, via a capacitor $C_g$. To study the decay rate, the Josephson junction has been replaced with a current source, $I(t)$. \textbf{(b)} By adding a resonator (red) with frequency $\omega_r$ in-between the qubit and the $50\,\Omega$ environment, we get the case found in regular dispersive readout. \textbf{(c)} A Purcell-filter (green) is added to the circuit, providing protection for the qubit, while allowing the resonator field to decay fast to the environment. \textbf{(d)} Transmission spectrum of a Purcell filter (dashed green), centered around the resonator frequency (red arrow), whereas the qubit frequency (orange arrow) is far detuned.}
\label{Fig:PurcellFilter}
\end{center}
\end{figure}

The Purcell effect can be framed in terms of Fermi's golden rule, where noise in the environment causes the qubit to decay with some probability. We can gain intuition about the Purcell effect (as well as how the qubit can be protected from it) by replacing the Josephson junction in the qubit circuit with an ac-current source, outputting $I(t) = I_0\sin(\omega t)$, with $I_0 = e\omega$ and study the rate at which power is lost into an environmental load resistor $R = Z_0 = 50\,\Omega$, see Fig. \ref{Fig:PurcellFilter}(a).

Expressing the power lost in the resistor as $P = I_{0}^2(C_g/C_{\Sigma})^2R = (e \omega \beta)^2Z_0$, with $\beta = C_g/C_{\Sigma}$, the qubit Purcell decay rate into the continuum can be written as

\begin{equation}
\gamma^{\text{Purcell}}_{\text{env}} = \frac{1}{T_1} = \frac{P}{\hbar \omega} = \frac{(\beta e \omega)^2 Z_0}{\hbar \omega} = \frac{g^2}{\omega}.
\label{Eq:Decay1}
\end{equation}

To protect the qubit from decaying into the 50$\,\Omega$ environment (as well as for deploying our dispersive readout) we can now add a resonator in parallel with the qubit, see Fig. \ref{Fig:PurcellFilter}(b). The presence of the resonator has the effect of shaping the impedance at the qubit frequency, which in turn modifies the decay rate in Eq. (\ref{Eq:Decay1}) into

\begin{equation}
\gamma^{\text{Purcell}}_{\text{res-env}} = \frac{g^2}{\omega}\frac{\mbox{Re}[Z_r(\omega)]}{Z_0},
\label{Eq:Decay2}
\end{equation}

\noindent where $Z_r(\omega)$ denotes the impedance of the shunted resonator. We can express the real-part of the impedance in terms of the resonator quality factor $Q = \omega_r/\kappa$ and qubit-resonator detuning $\Delta = \omega_{q} - \omega_{r}$,

\begin{equation}
\mbox{Re}[Z_r(\omega)] = \frac{QZ_0}{1 + 2(\Delta/\kappa)^2}.
\label{Eq:Decay3}
\end{equation}

Now, by substituting Eq. (\ref{Eq:Decay3}) into Eq. (\ref{Eq:Decay2}), we see that the Purcell decay rate for the qubit depends on the detuning between the resonator and the qubit. This is intuitive, since the resonator can be thought of as a bandpass filter, with center frequency $\omega_r$ and bandwidth $\kappa$. For resonant condition, i.e. when $\Delta = 0$, the emission rate into the resonator takes the form

\begin{equation}
\gamma^{\text{Purcell}}_{\text{res-env}} = \frac{g^2}{\omega_{r}}\frac{\mbox{Re}[Z_r]}{Z_0} \underset{\Delta=0}{=} \frac{g^2}{\omega_{r}}Q = \frac{g^2}{\kappa}.
\label{Eq:Decay4}
\end{equation}

In the dispersive regime $\Delta \gg g,\kappa$, which is also relevant for us in the context of qubit readout, we can make the approximation $\mbox{Re}[Z_r] \approx QZ_0 (\kappa/\Delta)^2$, yielding the familiar expression for the Purcell decay rate in circuit QED\cite{Houck2008}

\begin{equation}
\gamma^{\text{Purcell}}_{\text{res-env}} = \frac{g^2}{\omega_{r}}\frac{\mbox{Re}[Z_r]}{Z_0} \underset{\Delta\gg g,\kappa}{=} \frac{g^2}{\omega_{r}}Q \left(\frac{\kappa}{\Delta}\right)^2 = \left(\frac{g}{\Delta}\right)^2\kappa.
\label{Eq:Decay5}
\end{equation}

The relation for the Purcell limit in Eq. (\ref{Eq:Decay5}) thus provides us with a useful guide on how to design the coupling rates $g$ and $\kappa$, as well as how large qubit-resonator detuning $\Delta$ is necessary to avoid the Purcell limit.

In recent years, however, the intrinsic coherence times for superconducting qubits have reached above $100\,\mu$s, recall Sec. \ref{sec:circuits}, imposing practical limitations on how to simultaneously optimize $g$ and $\kappa$, to render fast readout without compromising the qubit coherence. Considering the parameters in Eq. (\ref{Eq:Decay5}), it is not possible to just increase the bound on the relaxation time $T_1$, without at the same time trading off the readout speed and contrast.


We can now introduce the Purcell filter [Fig. \ref{Fig:PurcellFilter}(c-d)] in between the readout resonator and the $50\,\Omega$ environment, leading to a reduction of the decay rate according to\cite{Reed2010a}

\begin{equation}
\gamma^{\text{Purcell}}_{\text{res-filter-env}} = \kappa\bigg(\frac{g}{\Delta}\bigg)^2 \bigg( \frac{\freq}{\omega_r}\bigg)\bigg(\frac{\omega_r}{2Q_F\Delta}\bigg),
\label{Eq:Purcell}
\end{equation}

\noindent where $Q_F$ denotes the quality factor of the Purcell filter. This is schematically depicted in Fig. \ref{Fig:PurcellFilter}(d), where the Purcell filter is placed around the resonator frequency, while far detuned from the qubit.

\subsection{Improve signal-to-noise ratio: Parametric amplification}

In light of the aforementioned limited signal-to-noise ratio associated with the low photon number of the dispersive qubit readout, and the short sampling time, the noise temperature of the amplifier chain plays a crucial role in determining the fidelity of the measurement.

A useful benchmark for quantum measurements is the \textit{quantum efficiency}, defined as

\begin{equation}
\eta_{\mbox{\tiny{SQL}}} = \frac{\hbar \omega_{\mbox{\tiny{RF}}}}{k_B T_{\mbox{\tiny{sys}}}}, \hspace{0.5cm} 0< \eta_{\mbox{\tiny{SQL}}} < 1,
\label{Eq:Eta}
\end{equation}

\noindent which quantifies the photon energy to the system noise temperature $T_{\mbox{\tiny{sys}}}$, thus yielding a measure of how close the signal is to the standard quantum limit (SQL), as imposed by Heisenberg's uncertainty relation, adding 1/2 photon of noise when $\eta_{\mbox{\tiny{SQL}}}$ approaches unity. Since the energy of each microwave photon is much smaller than that of optical photons, it is not easy to build a single-photon detector operating in the microwave domain\cite{Hoi2011,Inomata2016}. Instead, for heterodyne detection in circuit QED, a set of cascaded microwave amplifiers are used. The system noise temperature for the amplifier chain can be expressed in terms of the individual gain figures $G_n$ and noise temperatures $T_{N,n}$ of each constituent amplifier\cite{Pozar2004}

\begin{equation}
T_{\text{sys}} = T_{N,1} + \frac{T_{N,2}}{G_1} + \frac{T_{N,3}}{G_1 G_2} + ...
\label{Eq:NoiseTemperature}
\end{equation}

\noindent where $n = 1,2,3,...$ denotes the order of the amplifiers, starting from the qubit chip. From Eq. (\ref{Eq:NoiseTemperature}), we see that the noise temperature $T_{\text{sys}}$ is dominated by the noise contribution from the first amplifier, whereas the gain of the first amplifier has the effect of suppressing the noise from the second amplifier, and so on. If the first amplifier is a low-noise high-electron mobility transistor (HEMT) amplifier ($T_N \approx 2\,$K), the system noise temperature when implemented in a cryostat is around 7-10$\,$K, corresponding to around 10-20 added photons of noise per signal photon around 5 GHz. In practice, this is generally too much noise to perform single-shot readout.

This inherently poor signal-to-noise ratio has revived interest in developing quantum-limited parametric amplifiers (PA) -- tailored for readout of superconducting qubits -- featuring the ability to amplify small microwave signals, and adding only approximately the minimum amount of noise allowed by quantum mechanics\cite{Caves1982,Haus2000,Clerk2010}.

\subsubsection{Quantum-limited amplification processes}

In a linear, phase-insensitive amplifier, an input state $\langle a_{\mbox{\tiny{in}}}\rangle$ is amplified to an output state $\langle a_{\mbox{\tiny{out}}}\rangle$, with an amplitude gain factor $\sqrt{G}$. Microwaves are electromagnetic fields and therefore considered to be coherent light comprising microwave photons. As such, they must obey the commutation relations\cite{Haus1962,Caves1982,Haus2000,FlurinPhD2014}

\begin{equation}
[a_{\mbox{\tiny{in}}},a_{\mbox{\tiny{in}}}^{\dagger}] = [a_{\mbox{\tiny{out}}},a_{\mbox{\tiny{out}}}^{\dagger}] = 1,
\label{Eq:Commutation}
\end{equation}

\noindent from which it can be shown that it is not possible to simultaneously amplify both quadratures of $a_{\mbox{\tiny{in}}}$ without also adding noise. This is known as \textit{Caves theorem} after the work by Caves\cite{Caves1982}, based on earlier work by Haus and Mullen\cite{Haus1962}. This can be seen by considering the scattering relation between the input and output microwave fields

\begin{equation}
a_{\mbox{\tiny{out}}} = \sqrt{G}a_{\mbox{\tiny{in}}}.
\label{Eq:GainRelationNoIdler}
\end{equation}

The gain relation in Eq. (\ref{Eq:GainRelationNoIdler}) constitutes our ideal scenario for an amplifier process. However, the problem is that that this relation does not satisfy the commutation relation in Eq. (\ref{Eq:Commutation}). To satisfy this relation, we need to also take into account the vacuum fluctuations of another mode\cite{Shimoda1957,Clerk2003,Clerk2010,Bultink2018} -- called the \textit{idler} mode $b_{\mbox{\tiny{in}}}$, also satisfying the same communtation relation $[b_{\mbox{\tiny{in}}},b_{\mbox{\tiny{in}}}^{\dagger}] = 1$. To satisfy the commutation relation, the idler mode is amplified by the gain factor $\sqrt{G-1}$. For large gain, it can be shown that a minimum amount of half a photon of noise $\hbar \omega/2$ needs to be added to a signal amplified with gain $\sqrt{G}$.

Finally, taking the idler mode into account, the scattering relation for the coherent output field takes the form

\begin{equation}
a_{\mbox{\tiny{out}}} = \underbrace{\sqrt{G}a_{\mbox{\tiny{in}}}}_{\mbox{\tiny{Amplification}}} + \underbrace{\sqrt{G - 1}b_{\mbox{\tiny{in}}}^{\dagger}}_{\mbox{\tiny{Added idler noise}}}.
\label{Eq:GainRelation}
\end{equation}

\begin{figure*}[htp]
\begin{center}
\includegraphics[width=18.2cm]{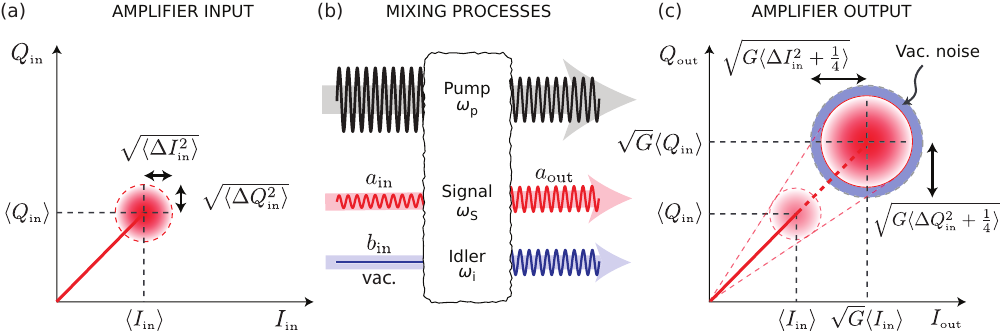}
\caption{Schematic illustration of a quantum-limited, phase-preserving parametric amplification process of a coherent input state, $a_{\mbox{\tiny{in}}} = I_{\mbox{\tiny{in}}} + iQ_{\mbox{\tiny{in}}}$. \textbf{(a)} The state is centered at $(\langle{I_{\mbox{\tiny{in}}}}\rangle,\langle{Q_{\mbox{\tiny{in}}}}\rangle)$ and has a noise represented by the radii of the circles along the real and imaginary axes, respectively. \textbf{(b)} Scattering representation of parametric mixing, where the signal and pump photons are interacting via a purely dispersive nonlinear medium. \textbf{(c)} In the case of phase-preserving amplification, both quadratures get amplified by a factor $\sqrt{G}$, while (in the ideal case) half a photon of noise gets added to the output distribution (blue). Image inspired by Flurin\cite{FlurinPhD2014}.}
\label{Fig:ParaAmpProcess}
\end{center}
\end{figure*}

Generally, this process results in a so-called \textit{phase-insensitive} parametric amplification process, in which both quadratures of the input field gets equally amplified. This is illustrated in Fig. \ref{Fig:ParaAmpProcess}, where the in-phase ($I_{\mbox{\tiny{in}}}$) and quadrature ($Q_{\mbox{\tiny{in}}}$) components of the fields are plotted, before and after the parametric amplifier.

Considering the amplification process in Eq. (\ref{Eq:GainRelation}), we can find a special case for the idler mode, for which noiseless amplification can be accomplished for one of the two quadratures, but at the expense of adding more noise to the other, thus not violating Heisenberg's uncertainty relation for the two field quadratures. This mode of operation is known as \textit{phase-sensitive} amplification, and is obtained when the idler mode oscillates at the same frequency as the signal (or a multiple thereof), but can be shifted with an overall phase $\phi \in [0,2\pi]$. By substituting the idler mode in Eq. (\ref{Eq:GainRelation}) with $b_{\mbox{\tiny{in}}} = e^{i\phi}a_{\mbox{\tiny{in}}}$, the scattering relation becomes

\begin{equation}
a_{\mbox{\tiny{out}}} = \underbrace{\sqrt{G}a_{\mbox{\tiny{in}}}}_{\mbox{\tiny{Amplification}}} + \underbrace{e^{-i\phi}\sqrt{G - 1}a_{\mbox{\tiny{in}}}^{\dagger}}_{\mbox{\tiny{Phase-dep. noise}}}.
\label{Eq:GainRelation2}
\end{equation}

\begin{figure}[htp]
\begin{center}
\includegraphics[width=8.6cm]{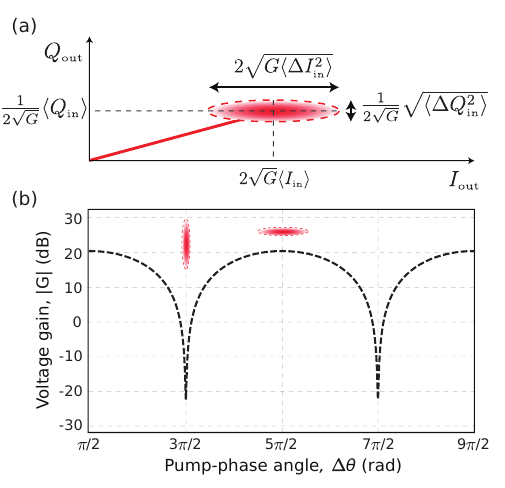}
\caption{Phase-sensitive parametric amplification. \textbf{(a)} In contrast to the phase-insensitive operation, phase-sensitive parametric amplification allows us to suppress the noise along one axis. Consequently, the noise is added to the other quadrature. \textbf{(b)} Voltage gain as a function of pump-phase angle, in which the amplification depends on the phase of the pump, providing either amplification or de-amplification of the quadrature voltage.}
\label{Fig:PhaseSensitiveAmplification}
\end{center}
\end{figure}

The overall phase factor allows us to tune the orientation of the amplification (or de-amplification) by means of the pump phase, thus allowing us to choose a quadrature for which we want to reduce the noise, see Fig. \ref{Fig:PhaseSensitiveAmplification}. Intuitively, this can be understood by considering the interference that occurs when two waves with the same frequency are confined in space, where we obtain constructive or destructive interference, depending on the phase between the two waves. Due to this interference, the noise can be suppressed even below the standard quantum limit (without violating Heisenberg's uncertainty relation). This is known as \textit{single-mode squeezing} and was first observation in superconducting circuits by Yurke \textit{et al.}\cite{Yurke1988}. In particular, after the theoretical prediction by Gardiner\cite{Gardiner1986}, Murch \textit{et al.} showed that the coherence time of a qubit can be enhanced when the qubit is exposed to squeezed vacuum\cite{Murch2013,Toyli2016}. Also \textit{two-mode squeezing} was demonstrated by Eichler \textit{et al.}\cite{Eichler2011}, where the demodulation setup squeezes both quadratures of the acquired signal\cite{Didier2015}.

In the context of qubit readout, however, phase-sensitive amplification tends to be experimentally inconvenient. This is mainly due to its phase-dependent gain, which imposes stringent requirements on continuous phase-calibration of the readout signal.

For a detailed theoretical framework developed for quantum limited amplification, the reader is referred to earlier work by Roy and Devoret\cite{Roy2016}, Clerk \textit{et al.}\cite{Clerk2010}, and Wustmann and Shumeiko\cite{Wustmann2013}.


\subsubsection{Operation of Josephson parametric amplifiers}

In this section, we review the basic operation characteristics of parametric amplifiers, and in particular the Josephson parametric amplifiers (JPAs), that have been exploited for qubit readout. Although many different flavors of parametric systems exist, we here focus on the resonant implementations of the Josephson parametric amplifier (JPA), serving as a good system for reviewing the fundamental concepts around parametric amplification.

All parametric amplifiers operate based on one fundamental principle: the incoming \textit{signal} photons are mixed with an applied \textit{pump} tone via an intrinsic nonlinearity, by which energy from the pump is converted into signal photons and thereby providing gain. As we recall from Sec. \ref{sec:circuits}, such a nonlinearity can be engineered in the microwave domain using Josephson junctions\cite{Dolan1977}, and the resonant parametric amplifiers are built from slightly anharmonic oscillators.

\begin{figure}[htp]
\begin{center}
\includegraphics[width=8.6cm]{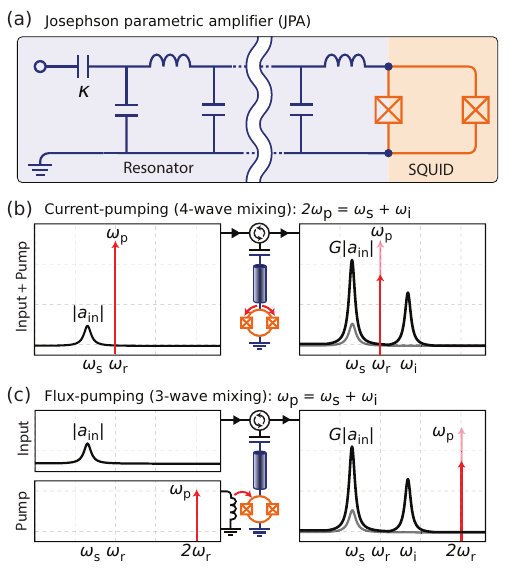}
\caption{Circuit schematics and pump schemes of a Josephson parametric amplifier. \textbf{(a)} The device consists of a quarter-wavelength resonator (blue), represented as lumped elements, shorted to ground via a Kerr-nonlinearity consisting of two parallel Josephson junctions (orange) forming a SQUID. The pump (red) can be applied in two ways; \textbf{(b)} either by modulating the current through the junctions (four-wave mixing) at the resonant frequency, $\omega_p \approx \omega_r$, or \textbf{(c)} by modulating the ac-flux $\Phi_{\mbox{\tiny{ac}}}$ around a static dc-flux point $\Phi_{\mbox{\tiny{dc}}}$ using a separate fast-flux line (three-wave mixing). The flux pump is applied at twice the resonant frequency, $\omega_p \approx 2\omega_r$.}
\label{Fig:JPAoperation}
\end{center}
\end{figure}

The first Josephson parametric amplifiers were built from a coplanar waveguide resonator, made nonlinear by adding a nonlinear Josephson contribution to its total inductance, see Fig. \ref{Fig:JPAoperation}(a). The word \textit{parametric} refers to the process of modulating (or \textit{pumping}) one of the parameters of the system's equation-of-motion (such as frequency or damping) in time\cite{Yurke1989,Wustmann2013,Wustmann2017}. The natural way to perform this parametric pumping is to modulate the nonlinear Josephson inductance, which in turn has the effect of modulating the resonator frequency $\omega_r(t) = 1/\sqrt{L(t)C}$.

Depending on how the pumping is implemented, there are two different mixing processes that can be exploited in Josephson parametric amplifiers, which determines the characteristics of the amplifier. These are illustrated in Fig. \ref{Fig:JPAoperation}(b)-(c) and are referred to as \textit{current-pumping}\cite{Wahlsten1978,Olsson1988,Yurke1989,Siddiqi2004,Tholen2007} and \textit{flux-pumping}\cite{Castellanos-Beltran2007,Yamamoto2008,Castellanos-Beltran2008,Sandberg2008,Wilson2010,Wustmann2013,Sundqvist2013,Sundqvist2014,Wustmann2017}, respectively. The type of mixing process that takes place depends on the leading order of the nonlinearity of the system, as reflected in its Hamiltonian. In the following, we briefly review the difference between these two pump-schemes.

In the current-pumped case, the dynamics of the system has characteristics of a Duffing oscillator\cite{Dykman1998}, with a fourth-order nonlinear term in addition to the harmonic oscillator term in its Hamiltonian

\begin{equation}
H = \omega_{r}c^{\dagger}c + K c^{\dagger}c^{\dagger}cc,
\label{Eq:Hfourwave}
\end{equation}

\noindent where $c$ denotes the resonator field operator and $K$ is the ``Kerr-nonlinearity". This process is a so-called \textit{four-wave mixing} process, since it mixes four photons: one signal ($\omega_s$), one idler ($\omega_i$), and two pump photons ($\omega_p$), obeying the energy conservation relation $\omega_s + \omega_i = 2\omega_p$, see Fig. \ref{Fig:JPAoperation}(b). Pioneered by Yurke\cite{Yurke1989}, this was the first demonstration of microwave amplification using a Josephson parametric amplifier. When the signal and idler modes are at the same frequency, the amplification is said to be \textit{degenerate}. This pumping scheme is the foundation for the Josephson Bifurcation Amplifier (JBA), developed by Siddiqi \textit{et al.}\cite{Siddiqi2004,Vijay2009,Lee2007}, which has been used to perform single-shot qubit readout, by mapping the quantum states onto the high and low resonator field originating from the sharp bifurcation point of the amplifier\cite{Mallet2009}.

In the other case, when the system is flux-pumped, the parametric process is driven by threading a magnetic flux $\Phi_{\mbox{\tiny{ac}}}$ through a SQUID loop, thereby modulating the frequency of the resonator. This results in a \textit{three-wave mixing} process, comprising three photons: one signal, one idler, and one pump photon, with $\omega_s + \omega_i = \omega_p$, see Fig. \ref{Fig:JPAoperation}(c). Therefore, we see that the pump frequency is about twice that of the signal $\omega_p \approx 2\omega_s$ for $\omega_s \approx \omega_i$. For degenerate, flux-pumped systems, the leading nonlinearity is a third-order term, yielding a Hamiltonian

\begin{equation}
H = \omega_{r}c^{\dagger}c + K \left( pc^{\dagger}c^{\dagger} + p^{\dagger}cc\right),
\label{Eq:Hthreewave}
\end{equation}

\noindent where the $p$ operator denotes the flux-pump mode. This approach to building parametric amplifiers was developed by Yamamoto \textit{et al.}\cite{Yamamoto2008}, as well as by Sandberg \textit{et al.}\cite{Sandberg2008}.

The flux-pumping scheme has several practical advantages. First, the large detuning of the pump makes it easier to filter, isolating the readout signal as its passing into the digitizer downstream and preventing the saturation of following amplifier stages. Second, if the resonator is a quarter-wavelength resonator, it has no resonant mode at the pump frequency $\omega_p$, reducing spurious population or saturation of the system as well as backaction on the qubits in the processor. Third, since the flux pump line is a separate on-chip microwave line, no additional directional coupler is needed.

Due to its rich dynamics, flux-pumping has also proven a useful platform to study the quantum dynamics of Josephson parametric oscillators, both in the context of qubit readout\cite{Krantz2016,KrantzPhD2016,Lin2013,Lin2014}, the dynamical Casimir effect\cite{Johansson2009,Johansson2010,Wilson2011}, and to better understand their complex nonlinear dynamics\cite{Dykman1998,Wilson2010,Krantz2013,KrantzLic2013,Svensson2017a,Svensson2018,Bengtsson2018}.

In addition to the degenerate parametric interactions described above, parametric gain can be obtained between different resonant modes; either between different modes of the same resonator\cite{Simoen2015,SimoenPhD2015}, or in-between different resonators\cite{Eichler2014}, as with the Josephson parametric converter (JPC)\cite{Bergeal2010a,Bergeal2010b,Abdo2011,Abdo2013,FlurinPhD2014,Liu2017}. In addition to the possibility of isolating and amplifying certain frequencies, the JPC can implement frequency conversion for which it has some other areas of applications compared with other types of parametric amplifiers.

\subsubsection{\label{sec:TWPA}The traveling wave parametric amplifier}

In the previously described JPA, parametric amplification is realized using resonators that enhance the parametric interaction between the input signal and the Josephson junction nonlinearity. Essentially, the Q-enhancement of the resonator forces each photon to pass through the junction on average Q times before leaving the resonator, thereby enhancing the non-linear interaction. Albeit proven to be able to reach near the standard-quantum limit of noise for readout of a small number of qubits, the future direction of the community is heading towards amplifier technologies which are compatible with multiplexed readout of several qubits coupled to the same amplifier chain\cite{Chen2012,Jerger2012,Barends2014,Kelly2015,Chapman2017}. In this context, resonator-based parametric amplifiers suffer from two major drawbacks: First, the amplifier bandwidth is limited to the resonator linewidth, typically $\approx 10-50\,$MHz, practically limiting the number of multiplexed frequencies that can be amplified. Second, since the Josephson nonlinearity is realized by a small number of junctions, the saturation power is low due to the interplay of higher order nonlinearities, effectively taking the system outside its desired operation regime\cite{Dykman1998,Krantz2013,KrantzLic2013,Liu2017}. In practice, this limits how many readout resonators that can be simultanously read out.

These two bottlenecks can, to a degree, be overcome with microwave engineering. For instance, the linewidth can be made an order of magnitude wider by altering the impedance along the resonator. This is called a stepped-impedance transformer, where the impedance is ramped down from a matched 50$\,\Omega$ at the capacitor down to a small impedance at the SQUID\cite{Roy2015} shorting the device to ground. Also the saturation power can be increased by distributing the nonlinearity across an array consisting of many identical junctions, reducing the Kerr-nonlinearity by a factor $1/N^2$ with $N$ representing the number of junctions in the array. This has been demonstrated by using a an array of SQUIDs in a resonator, rather than a single one\cite{Castellanos-Beltran2008}.

However, despite the above mentioned engineering efforts to improve the resonator-based JPAs, the most prominent approach to date is to get rid of the resonator altogether and, instead, construct a microwave analog to optical parametric amplifiers, where kilometers of weakly nonlinear fibers are used. Such device is called a \textit{traveling wave parametric amplifier} (TWPA) and was developed to surmount the bandwidth and dynamic range limitations of the resonator-based JPAs.

\begin{figure}[t!]
\begin{center}
\includegraphics[width=8.6cm]{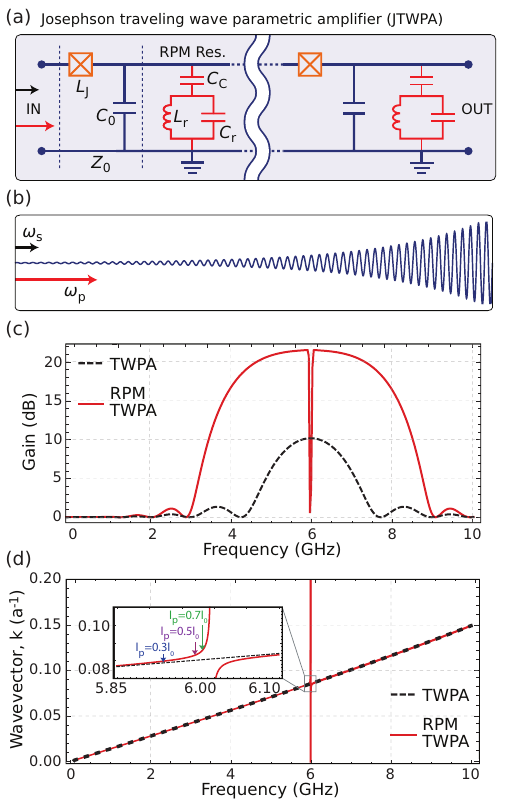}
\caption{\textbf{(a)} Simplified circuit representation of a Josephson traveling wave parametric amplifier (JTWPA). The characteristic impedance for each unit cell is set by the in-line Josephson inductor, $L_J$ (orange) and the shunt capacitor, $C$ (blue). A resonant LC-circuit (red) is used to phase match the four-wave amplification process. \textbf{(b)} Schematic of how the signal gets amplified in each unit cell as it propagates through the device. \textbf{(c)} Gain vs. frequency for a JTWPA, with and without the resonant phase matching (RPM). \textbf{(d)} Dispersion relation of the TWPA, where the LC-resonators collectively open up a stopband at the resonant frequency. By applying the pump close to this frequency, the wave vector of the pump can be set to obtain a phase-matching. The optimal pump frequency depends on the pump power, as indicated in the inset. Image courtesy of Kevin O'Brien\cite{OBrien2014,Macklin2015}}
\label{Fig:TWPA}
\end{center}
\end{figure}

Although operated in similar way, the nonlinearity of TWPAs can be realized in different ways, such as the kinetic inductance of a superconducting film\cite{Eom2012,Bockstiegel2014,Adamyan2016} or using an array of Josephson junctions\cite{OBrien2014,Macklin2015,White2015}, through which the four-wave mixing process is distributed across a nonlinear lumped element transmission line, see Fig. \ref{Fig:TWPA}(a).

The Josephson TWPA consists of a few thousand identical unit cells, each comprising a shunt capacitor to ground and a nonlinear Josephson inductor, together yielding a characteristic impedance of $Z_0 = \sqrt{L_J/C} \approx 50\,\Omega$, see Fig. \ref{Fig:TWPA}(a).

The fact that the nonlinearity is distributed allows for high saturation power, since each Josephson junction is accessed once. However, even though energy conservation is satisfied, the four-wave mixing process in the device, there is a problem with phase (or momentum) conservation. This is associated with the system nonlinearity as well as the large frequency detuning between signal and pump photons, yielding a difference in phase-velocity between the two, which in turn gives rise to a non-flat gain profile, as well as an overall reduction in gain\cite{OBrien2014}.

Again, by taking inspiration from the dispersive engineering developed in quantum optics and photonics, where the refractive index can be periodically altered to engineer the momentum of a transferred signal, the solution to this phase-mismatch problem was introduced by O'Brien \textit{et al.}\cite{OBrien2014}. By introducing resonators at periodic intervals of TWPA unit cells, the pump tone can be given a ``momentum kick", effectively slowing it down and phase-matching the device by means of its wave vector. This technique is called resonant-phase matching (RPM), see Fig. \ref{Fig:TWPA}(d), and requires that the pump frequency is set on the left side of the dispersion feature (where the wave vector diverges), defined by the resonant frequency of the phase-matching resonators. Note, finally, that broadband parametric amplification with high dynamic range has been demonstrated in other Josephson-based circuits, e.g. the superconducting nonlinear asymmetric inductive element (SNAIL) parametric amplifier (SPA)\cite{Frattini2018}.

\section{Summary and outlook}
In this review, we have discussed the phenomenal progress over the last decade in the engineering of superconducting devices, the development of high-fidelity gate-operations, and quantum non-demolition measurements with high signal-to-noise ratio. Putting these advances together, we hope that it is clear that the planar superconducting qubit modality is a promising platform for realizing near-term medium scale quantum processors. While we have focused on highlighting the advances made within the fields of realizing, controlling and reading out planar superconducting qubits specifically used for quantum information processing, there has of course also been tremendous activity in the surrounding fields. In this final section, we briefly mention a few of those fields, and invite the reader to look into the references, for further details.

\underline{Quantum annealing}: Superconducting qubits also form the basis for certain quantum annealing platforms\cite{Farhi2000,Farhi2001}. Quantum annealing operates by finding the ground state of a given Hamiltonian (typically a classical Ising Hamiltonian), and this state will correspond to the solution of an optimization problem. By utilizing a flux-qubit type design (see Sec.\ref{sec:circuits}, the company D-Wave have demonstrated quantum annealing processors\cite{Johnson2011} which have now reached beyond 2000 qubits\cite{DWaveSite}. The benchmarking of quantum annealers and attempts to demonstrate a quantum speedup for a general class of problems is a highly active research field, and we refer the reader, for example, to recent papers Refs. \onlinecite{Denchev2016,Albash2018,King2019} and references therein.

\underline{Cavity based QIP}: A parallel effort to the planar superconducting qubits discussed in this review is the development of 3D cavity-based superconducting qubits. In these systems, quantum information is encoded in superpositions of coherent photonic modes of the cavity\cite{Wang2016}. The cat states can be highly coherent due to the inherently high quality factors associated with 3D cavities\cite{Reagor2013,Axline2016,Pfaff2017}. This approach has a fairly small hardware overhead to encode a logical qubit\cite{Heeres2017}, and lends itself to certain implementations of asymmetric error-correcting codes due to the fact that errors due to single-photon loss in the cavity is a tractable observable to decode. Using this architecture, several important advances were recently demonstrated including extending the lifetime of an error-corrected qubit beyond its constituent parts\cite{Ofek2016}, randomized benchmarking of logical operations\cite{Heeres2017}, a \textsf{CNOT} gate between two logical qubits\cite{Rosenblum2018} as well as Ramsey interference of an encoded quantum error corrected qubit\cite{Hu2019}.

\underline{Cryogenics and software development}: We briefly mentioned the electrical engineering, software development, and cryogenic considerations associated with the control wiring and on-chip layout of medium-scale quantum processors. While dilution refrigerators are now readily available, off-the-shelf commercial products, the details of how to optimally do signal-routing and rapid data processing in a scalable fashion, is also a field in rapid development. However, with the recent demonstrations of enabling technologies such as 3D integration, packages for multi-layered devices and superconducting interconnects\cite{Tolpygo2015,Bejanin2016,Vahidpour2017,Rosenberg2017,Bronn2018,Foxen2018,McConkey2018}, some of the immediate concerns for how to scale the \emph{number} of qubits in the superconducting modality, have been addressed. On the control software side, there currently exist multiple commercial and free software packages for interfacing with quantum hardware, such as \textsf{QCoDeS}\cite{qcodes}, the related \textsf{pyCQED}\cite{pycqed}, \textsf{qKIT}\cite{qkit} and \textsf{Labber}\cite{LabberQuantum}. However, many laboratories use software platforms developed in-house, often due to the concurrent development of custom-built, highly specialized electronics and FPGA circuits (many of these developments are not always published, but readers may consult Refs. \onlinecite{Asaad2016,LiuPhD2016,Ryan2017} for three examples). There is currently also a large ongoing development of quantum circuit simulation and compiling software packages. Packages such as \textsf{Qiskit}\cite{qiskit}, \textsf{Forest} (with \textsf{pyQUIL}\cite{Smith2016}), \textsf{ProjectQ}\cite{Steiger2016}, \textsf{Cirq}\cite{Cirq}, \textsf{OpenFermion}\cite{McClean2017}, the Microsoft Quantum Development kit\cite{MS_QDK} provide higher-level programming languages to compile and/or optimize quantum algorithms. For a recent review and comparison of these different software suites, we refer to Ref.\onlinecite{LaRose2019} and Ref.\onlinecite{Chong2017} for a general review on advances in designing quantum software. Since the connectivity and gate set of quantum processors can differ, details of the gate compilation implementation is an important non-trivial problem for larger-scale processors. We note that some of these software packages already interface directly with quantum processors that are available online, supplied, for example, via Rigetti Computing or the IBM Quantum Experience.

\underline{Quantum error correction}: While the qubit lifetimes and gate fidelity have improved dramatically in the last decades, there remains a need for error correction to reach large-scale processors. While certain strategies exist to extend the computational reach of current state-of-the-art physical qubits\cite{Kandala2018}, for truly large-scale algorithms addressing practical problems, the quantum data will have to be embedded in an error-correcting scheme. As briefly mentioned in Sections \ref{subsubsec:CZforQEC} and \ref{subsubsec:CRforQEC}, certain components of the surface code quantum error correcting scheme have already been demonstrated in superconducting qubits (see e.g. Refs.\onlinecite{Kelly2015,Riste2015,Takita2016}). However, the demonstration of a logical qubit with greater lifetime than the underlying physical qubits, remains an outstanding challenge. While the surface code is a promising quantum error correcting code due to its relatively lenient fault tolerance threshold, it cannot implement a \emph{universal} gate set in a fault-tolerant manner. This means that the error-corrected gates in the surface code need to be supplemented, for example, with a \textsf{T} gate, to become universal. Such gates can be implemented by a technique known as \emph{magic state distillation}\cite{Bravyi2005}. The process of \emph{gate-teleportation}, a pre-cursor to magic state distillation, has already been demonstrated using FPGA-based classical feedback with planar superconducting qubits\cite{Ryan2017}, but showing distillation and injection into a surface code logical state remains an open challenge. The development of new quantum codes is also a field in rapid development, and the reader may consult a recent review for more details e.g. Ref.\onlinecite{Campbell2017}. Another important step towards large-scale quantum processor architecture is that of remote entanglement, enabling quantum information to be distributed across different nodes of a quantum processing network\cite{Axline2018,Kurpiers2018}.

\underline{Quantum computational supremacy}: Finally, we mention one of the grand challenges for superconducting qubits in the coming years: the demonstration of quantum computational supremacy\cite{Preskill2012}. The basic idea is to demonstrate a calculation, using qubits and algorithmic gates, which is outside the scope of classical computers (assuming some plausible computational complexity conjectures). For a recent review article, the reader is referred to Ref.\cite{Harrow2017}. A first step towards an approach to demonstrating quantum supremacy was recently reported, using 9 tunable transmons\cite{Neill2018}. It is expected that with somewhere between 50-100 qubits\cite{Dalzell2018}, an extension of the protocol from Refs.\onlinecite{Neill2018,Boixo2018}, will allow researchers to sample from a classically intractable distribution, and thereby demonstrate quantum computational supremacy. The success of this program would constitute a phenomenal result for all of quantum computing.

\begin{acknowledgments}
The authors gratefully acknowledge Mollie Kimchi-Schwartz, Jochen Braum\"uller, Niels-Jakob S{\o}e Loft, and David DiVincenzo for careful reading of the manuscript and Youngkyu Sung for use of his time-dependent qubit drive simulation suite and useful feedback from the entire Engineering Quantum Systems group at MIT. The authors also acknowledges fruitful discussion with Anton Frisk Kockum, Anita Fadavi Roudsari, Daryoush Shiri, and Christian Kri{\v{z}}an.

This research was funded in part by the U.S. Army Research Office Grant No. W911NF-14-1-0682; and by the National Science Foundation Grant No. PHY-1720311. P.K. acknowledges partial support by the Wallenberg Centre for Quantum Technology (WACQT) funded by Knut and Alice Wallenberg Foundation. M.K. gratefully acknowledges support from the Carlsberg Foundation. The views and conclusions contained herein are those of the authors and should not be interpreted as necessarily representing the official policies or endorsements of the US Government.
\end{acknowledgments}

\bibliography{Review_bibliography_Final_v4}{}

\end{document}